\documentclass{ar2e}[10pts, onecolumn,final]
\usepackage{ARAstroBib,amsmath,amssymb,graphicx}
\def\lesssim{\mathrel{\hbox{\rlap{\hbox{\lower4pt\hbox{$\sim$}}}\hbox{$<$}}}}
\def\gtrsim{\mathrel{\hbox{\rlap{\hbox{\lower4pt\hbox{$\sim$}}}\hbox{$>$}}}}

\begin{document}
\input epsf.tex    

\input psfig.sty

\jname{Annu.\ Rev.\ Astron.\ Astrophys.}
\jyear{2008}
\jvol{1}

\title{Gravitational waves\\ from merging compact binaries}

\markboth{Scott A.\ Hughes}{Compact binary gravitational waves}

\author{Scott A.\ Hughes
\affiliation{Department of Physics and MIT Kavli Institute,
Massachusetts Institute of Technology, 77 Massachusetts Avenue,
Cambridge, MA 02139}}


\begin{abstract}
Largely motivated by the development of highly sensitive
gravitational-wave detectors, our understanding of merging compact
binaries and the gravitational waves they generate has improved
dramatically in recent years.  Breakthroughs in numerical relativity
now allow us to model the coalescence of two black holes with no
approximations or simplifications.  There has also been outstanding
progress in our analytical understanding of binaries.  We review these
developments, examining merging binaries using black hole perturbation
theory, post-Newtonian expansions, and direct numerical integration of
the field equations.  We summarize these approaches and what they have
taught us about gravitational waves from compact binaries.  We place
these results in the context of gravitational-wave generating systems,
analyzing the impact gravitational wave emission has on their sources,
as well as what we can learn about them from direct gravitational-wave
measurements.
\end{abstract}

\maketitle

\section{Introduction}
\label{sec:intro}

\subsection{History and motivation}

Most physics students learn to solve the Newtonian gravitational
two-body problem early in their studies.  An exact solution for point
masses requires only a few equations and an elliptic integral.
Coupling this simple solution to perturbation theory lets us include
the effect of non-spherical mass distributions and the impact of
additional bodies.  We can then accurately model an enormous range of
astrophysically interesting and important systems.

By contrast, no exact analytic solution describes binaries in general
relativity (GR).  GR is nonlinear (making analytic solutions difficult
to find except for problems that are highly symmetric) and includes
radiation (so that any bound solution will evolve as waves carry off
energy and angular momentum).  Indeed, for systems containing black
holes, GR doesn't so much have a ``two-body'' problem as it has a
``one-spacetime'' problem: One cannot even delineate where the
boundaries (the event horizons) of the ``bodies'' are until the entire
spacetime describing the binary has been constructed.

Such difficulties in describing binary systems in GR bedeviled the
theoretical development of this topic.  Many early discussions
centered on the even more fundamental question of which motions would
generate radiation and which would not.  A particularly clear
formulation of the confusion is expressed in attempts to answer the
following question: {\it If a charge falls freely in the Earth's
gravitational field, does it radiate?}  On one hand, in this
non-relativistic limit, we should expect that our usual intuition
regarding accelerating charges would hold, and a falling charge should
radiate exactly as described in {\cite{jackson}} with an acceleration
$\vec a = -g \vec e_r$.  On the other hand, in GR a falling charge
simply follows a geodesic of the Earth's curved spacetime; it is not
``accelerating'' relative to freely falling frames, and so is not
accelerating in a meaningful sense.  The charge just follows the
trajectory geometry demands it follows.  In this picture, the falling
charge appears to {\it not} radiate.  John Wheeler once asked a group
of relativity theorists to vote on whether the falling charge radiates
or not; their responses were split almost precisely down the middle
(\citealt{kennefick}, p.\ 157)\footnote{The correct answer is now
understood thanks to \cite{db60}: The charge {\it does} radiate,
precisely reproducing the non-relativistic limit.  The intuition that
the charge follows a geodesic is not quite right.  Though the charge
``wants'' to follow a geodesic, the charge {\it plus its associated
field} is extended and nonlocal, and so cannot follow a geodesic.  The
bending of the charge's field as it falls in spacetime enforces the
laws of radiation emission.}.

Similar conceptual issues affect the general two-body problem in
relativity.  As recently as 1976, it was pointed out {\citep{ergh76}}
that there had not yet been a fully self-consistent derivation for the
energy loss from a binary due to gravitational-wave (GW) backreaction.
A major motivation for this criticism was the discovery of the binary
pulsar PSR 1913+16 {\citep{ht75}}.  It was clear that this system
would be a powerful laboratory for testing theories of gravity,
including the role of GW emission.  However, as Ehlers et al.\ spelled
out, the theoretical framework needed for such tests was not in good
shape.  Various approaches to understanding the evolution of binary
systems tended to be inconsistent in the nature of the approximations
they made, often treating the motion of the members of the binary at a
different level of rigor than they treated the solution for the
spacetime.  These discrepancies were most notable when the members of
the binary were strongly self gravitating; a weak-field approach is
ill-suited to a binary containing neutron stars or black holes.  These
calculations generally predicted that the system would, at leading
order, lose energy at a rate related to the third time derivative of
the source's ``quadrupole moment.''  However, the precise definition
of this moment for strong-field sources was not always clear.

The Ehlers et al.\ criticism served as a wake-up call, motivating the
formulation of methods for modeling binary dynamics in a
mathematically rigorous and self-consistent manner.  Several
independent approaches were developed; a concise and cogent summary of
the principles and the leading results is given by {\cite{damour83}}.
For the purpose of this review, a major lesson of the theoretical
developments from this body of work is that the so-called ``quadrupole
formula'' describing the loss of orbital energy and angular momentum
due to GW emission is valid.  Somewhat amazingly, one finds that the
equations of motion are {\it insensitive} to much of the detailed
structure of a binary's members.  Features such as the members' size
and compactness can be absorbed into the definition of their masses.
This ``principle of effacement'' {\citep{damour_dt83}} tells us that
the motions of two bodies with masses $m_1$ and $m_2$ can be predicted
independent of whether those bodies are stars, neutron stars, or black
holes\footnote{Other aspects of the members' structure cannot be so
simply absorbed by the principle of effacement.  For example, at a
certain order, the spins of a binary's members impact its motion.
Spin's effects cannot be absorbed into the definition of mass, but
affect the binary's dynamics directly.  See {\cite{th85}} for further
detailed discussion.}.  Over 30 years of study have since found
extraordinary agreement between prediction and observation for the
evolution of PSR 1913+16's orbit {\citep{wt05}}.  Additional
inspiraling systems have been discovered; in all cases for which we
have enough data to discern period evolution, the data agree with
theory to within measurement precision (\citealt{stairs98},
\citealt{nice05}, \citealt{jacoby06}, \citealt{kramerstairs08},
\citealt{bhat08}).  At least one additional recently discovered system
is likely to show a measurable inspiral in the next few years
{\citep{kasian08}}.  These measurements have validated our theory of
GW generation, and are among our most powerful tests of GR.

Measuring GWs with the new generations of detectors will require even
more complete models for their waveforms, and hence complete models of
a binary's evolution.  Motivated by this, our theoretical
understanding of merging compact binaries and their GWs has grown
tremendously.  The purpose of this review is to summarize what we have
learned about these systems and their waves, focusing on theory but
connecting it to current and planned observations.  We examine the
analytic and numerical toolkits that have been developed to model
these binaries, discuss the different regimes in which these tools can
be used, and summarize what has been learned about their evolution and
waves.  We begin by first examining compact binaries as astrophysical
objects, turning next to how they are treated within the theory of GR.

\subsection{Compact binaries: The astrophysical view}
\label{sec:astroview}

From the standpoint of GR, all compact binary systems are largely the
same until their members come into contact, modulo the value of
parameters such as the members' masses and spins and the orbital
period at a given moment.  This means that we only need one
theoretical framework to model any compact binary that we encounter in
nature.  From the standpoint of astrophysics, though, all compact
binary systems are {\it not} the same: A $1.4\,M_\odot - 1.4\,M_\odot$
neutron star binary forms in very different processes than those which
form a $10^6\,M_\odot - 10^6\,M_\odot$ black hole binary.  In this
section, we summarize the astrophysical wisdom regarding the various
compact binary systems that should be strong generators of GWs.

Compact binaries are organized most naturally by their masses.  At the
low end we have {\it stellar-mass} binaries, which include the binary
pulsars discussed in the previous section.  The data on binaries in
this end are quite solid, thanks to the ability to tie models for the
birth and evolution of these systems to observations.  At least some
fraction of short gamma-ray bursts are likely to be associated with
the mergers of neutron star-neutron star (NS-NS) or black hole-neutron
star (BH-NS) systems (\citealt{eichler89}; \citealt{fox05}).
Gamma-ray telescopes may already be telling us about compact binary
merger many times per year {\citep{nakar06}}.

There is also evidence that nature produces {\it supermassive}
binaries, in which the members are black holes with $M \sim 10^6 -
10^8\,M_\odot$ such as are found at the centers of galaxies.  As
described in more detail below, theoretical arguments combining
hierarchical galaxy growth scenarios with the hypothesis that most
galaxies host black holes generically predict the formation of such
binaries.  We have now identified quite a few systems with properties
indicating that they may host such binaries.  The evidence includes
active galaxies with double cores {\citep{komossa03, maness04,
rodriguez06}}; systems with doubly-peaked emission lines
({\citealt{zhou04}}, {\citealt{gerke07}}); helical radio jets,
interpreted as the precession or modulation of a jet due to binarity
({\citealt{bbr80}}, {\citealt{cw95}}, {\citealt{lr05}}); and systems
that appear to be periodic or semi-periodic, such as the blazar OJ287
{\citep{valtonen08}}.  There are also sources which suggest the system
hosted a binary that recently merged, leading to the spin flip of a
radio jet {\citep{me02}} or to the interruption and later restarting
of accretion activity {\citep{liu03}}.  As surveys go deeper and
resolution improves, we may expect the catalog of candidate
supermassive black hole binaries to expand.

Turn now from the observational evidence to theoretical models.  If we
assume that our galaxy is typical, and that the inferred density of
NS-NS systems in the Milky Way should carry over to similar galaxies
(correcting for factors such as typical stellar age and the proportion
of stars that form neutron stars), then we can estimate the rate at
which binary systems merge in the universe.  {\citet{nps91}} and
{\citet{phinney91}} first performed such estimates, finding a
``middle-of-the-road'' estimate that about 3 binaries per year merge
to a distance of 200 Mpc.  More recent calculations based on later
surveys and observations of NS-NS systems have amended this number
somewhat; the total number expected to be measured by advanced
detectors is in the range of several tens per year (see, e.g.,
{\citet{kalogera07}} for a detailed discussion of methodology, and
{\citet{kim06}} for a summary).

Population synthesis gives us a second way to model the distribution
of stellar mass compact binaries.  Such calculations combine data on
the observed distribution of stellar binaries with models for how
stars evolve.  Begin with a pair of main sequence stars.  The more
massive star evolves into a giant, transfers mass onto its companion,
and then goes supernova, leaving a neutron star or black hole.  After
some time, the companion also evolves into a giant and transfers mass
onto its compact companion (and may be observable as a high-mass x-ray
binary).  In almost all cases, the compact companion is swallowed by
the envelope of the giant star, continuing to orbit the giant's core.
The orbiting compact body can then unbind the envelope, leaving the
giant's core behind to undergo a supernova explosion and form a
compact remnant.  See {\cite{tvdh03}}, especially Fig.\ 16.12, for
further discussion.

An advantage of population synthesis is that one can estimate the rate
of formation and merger for systems which we cannot at present
observe, such as stellar mass black hole-black hole (BH-BH) binaries,
or for which we have only circumstantial evidence, such as neutron
star-black hole (NS-BH) binaries (which presumably form some fraction
of short gamma ray bursts).  A disadvantage is that the models of
stellar evolution in binaries have many uncertainties.  There are
multiple branch points in the scenario described, such as whether the
binary remains bound following each supernova, and whether the binary
survives common envelope evolution.  As a consequence, the predictions
of calculations based on population synthesis can be quite diverse.
Though different groups generally agree well with the rates for NS-NS
systems (by design), their predictions for NS-BH and BH-BH systems
differ by quite a bit (\citealt{ypz98}, \citealt{pp99}).  New data are
needed to clear the theoretical cobwebs.

Binaries can also form dynamically through many-body interactions in
dense environments, such as globular clusters.  In such a cluster, the
most massive bodies will tend to sink through mass segregation
{\citep{spitzer69}}; as such, the core of the cluster will become
populated with the heaviest bodies, either stars which will evolve
into compact objects, or the compact objects themselves.  As those
objects interact with one another, they will tend to form massive
binaries; calculations show that the production of BH-BH binaries is
particularly favored.  It is thus likely that globular clusters will
act as ``engines'' for the production of massive compact binaries
(\citealt{pzm00}, \citealt{oor07}, \citealt{mwdg08}).

As mentioned above, the hierarchical growth scenario for galaxies,
coupled with the hypothesis that most galactic bulges host large black
holes (\citealt{kg01}, \citealt{ferrarese02}) generically predicts the
formation of supermassive binaries, especially at high redshifts when
mergers were common.  The first careful discussion of this scenario
was by {\citet{bbr80}}.  In order for the black holes to get close
enough to merge with one another due to GW emission, the black holes
hosted by the merging galaxies must sink, via dynamical friction, into
the center of the newly merged object.  The binary thus formed will
typically be very widely separated, and only harden through
interactions with field stars (ejecting them from the center).  For
some time, it was unclear whether there would be enough stars to bring
the holes close enough that they would be strong GW emitters.  It is
now thought that, at least on the low end of the black hole mass
function ($M \lesssim 10^6-10^7\,M_\odot$), this so-called ``last
parsec problem'' is not such a problem.  Quite a few mechanisms have
been found to carry the binary into the regime where GWs can merge the
binary (\citealt{an02}, \citealt{mm05}).  It is now fairly common to
assume that some mechanism will bring a binary's members into the
regime where they can merge.

Much theoretical activity in recent years has thus focused on the
coevolution of black holes and galaxies in hierarchical scenarios
(\citealt{mhn01}, \citealt{yt02}, \citealt{vhm03}).  Galaxy mergers
appear to be a natural mechanism to bring ``fuel'' to one or both
black holes, igniting quasar activity; the formation of a binary may
thus be associated with the duty cycle of quasars (\citealt{hco04},
\citealt{hckh08}, \citealt{dcshs08}).  Such scenarios typically find
that most black hole mergers come at fairly high redshift ($z \gtrsim
3$ or so), and that the bulk of a given black hole's mass is due to
gas it has accreted over its growth.

A subset of binaries in the supermassive range are of particular
interest to the relativity theorist.  These binaries form by the
capture of a ``small'' ($1 - 100\,M_\odot$) compact object onto an
orbit around the black hole in a galactic center.  Such binaries form
dynamically through stellar interactions in the core (\citealt{sr97},
\citealt{ha05}); the formation rate predicted by most models is
typically $\sim 10^{-7}$ extreme mass ratio binaries per galaxy per
year \citep{ha05}.  If the inspiraling object is a white dwarf or
star, it could tidally disrupt as it comes close to the massive black
hole, producing an x-ray or gamma-ray flare (\citealt{rbb05},
\citealt{mhk08}, \citealt{rrh08}).  If the inspiraling object is a
neutron star or black hole, it will be swallowed whole by the large
black hole.  As such, it will almost certainly be electromagnetically
quiet; however, as will be discussed in more detail below and in Sec.\
{\ref{sec:pert}}, its GW signature will be loud, and is a particularly
interesting target for GW observers.

\subsection{Compact binaries: The relativity view}
\label{sec:relview}

Despite the diverse astrophysical paths to forming a compact binary,
the end result always looks more-or-less the same from the standpoint
of gravity.  We now briefly outline the general features of binary
evolution in GR.  As described near the beginning of Sec.\
{\ref{sec:intro}}, in GR a binary is not so much described by ``two
bodies'' as by ``one spacetime.''  The methods used to describe this
spacetime depend on the extent to which the two-body description is
useful.

Although it is something of an oversimplification, it is useful to
divide the evolution of a binary into two or three broad epochs,
following {\citet{fh98}}.  First is the binary's {\it inspiral}, in
which its members are widely separated and can be readily defined as a
pair of distinct bodies.  During the inspiral, the binary's mean
separation gradually decreases due to the secular evolution of its
orbital energy and angular momentum by the backreaction of
gravitational radiation.  The bodies eventually come together, merging
into a single highly dynamical and asymmetric object.  We call the
{\it merger} the final transition from two bodies into one.  If the
final state of the system is a black hole, then its last dynamics are
given by a {\it ringdown} as that black hole settles down from the
highly distorted post-merger state into a Kerr black hole as required
by the ``no hair'' theorems of GR {\citep{carter71,robinson75}}.

How we solve the equations of GR to model a binary and its GWs depends
upon the epoch that describes it.  When the binary is widely
separated, the {\it post-Newtonian} (pN) expansion of GR works very
well.  In this case, the Newtonian potential $\phi \equiv GM/rc^2$
(where $M = m_1 + m_2$ is the total mass of a binary, and $r$ is the
orbital separation) can be taken to be a small parameter.  Indeed, we
must {\it always} have $r \gtrsim (\mbox{a few}) \times GM/c^2$: The
closest the members of the binary can come is set by their physical
size, which has a lower bound given by the radius they would have if
they were black holes.  The pN expansion is what we get by iterating
GR's field equations from the Newtonian limit to higher order in
$\phi$.  We review the basic principles of the pN expansion and
summarize important results in Sec.\ {\ref{sec:pn}}.

Some binaries, such as the extreme mass ratio captures described in
Sec.\ {\ref{sec:astroview}}, will have $m_1 \ll m_2$.  For these
cases, the reduced mass ratio $\eta \equiv \mu/M = m_1 m_2/(m_1 +
m_2)^2$ can define a perturbative expansion.  In this limit, one can
treat the binary's spacetime as an exact background solution with mass
$M$ (e.g., that of a black hole) perturbed by a smaller mass $\mu$.
By expanding GR's field equations around the exact background, one can
typically derive tractable equations describing the perturbation and
its evolution; those perturbations encode the dynamical evolution of
the binary and its evolution.  We discuss perturbative approaches to
binary modeling in Sec.\ {\ref{sec:pert}}.

For some binaries, {\it no} approximation scheme is useful.  Consider,
for example, the last moments of two black holes coming together and
fusing into a single body.  In these moments, the spacetime can be
highly dynamical and asymmetric; no obvious small parameter organizes
our thinking about the spacetime.  Instead, one must simply solve
Einstein's field equations as exactly as possible using numerical
methods.  The essential question this field of {\it numerical
relativity} asks is how one can take a ``slice'' of spacetime (that
is, a single 3-dimensional moment of time) and use the field equations
to understand how that spacetime evolves into the future.  This
requires us to explicitly split ``spacetime'' into ``space'' and
``time.''  Progress in numerical relativity has exploded in recent
years.  Since 2005, practitioners have moved from being able to just
barely model a single binary orbit for a highly symmetric system to
modeling multiple orbits and the final coalescence for nearly
arbitrary binary masses and spins.  We summarize the major principles
of this field and review the explosion of recent activity in Sec.\
{\ref{sec:numrel}}.

Roughly speaking, for comparable mass binaries, pN methods describe
the inspiral, and numerical relativity describes the merger. The line
dividing these regimes is fuzzy.  A technique called the {\it
effective one-body} (EOB) approximation blurs it even further by
making it possible to extend pN techniques beyond their naive range of
validity {\citep{damour_eob08}}, at least when both members of the
binary are black holes.  [When at least one of the members is a
neutron star, at some point the nature of the neutron star fluid will
have an impact.  Detailed numerical modeling will then surely be
critical; see \cite{su06}, \cite{etienne08}, \cite{bgr08}, and
\cite{skyt09} for examples of recent progress.]  Detailed tests show
that using EOB methods greatly extends the domain for which analytical
waveform models can accurately model numerical relativity waveforms
(\citealt{betal07}, \citealt{dnhhb08}).  A brief discussion of these
techniques is included in Secs.\ {\ref{sec:pn}} and
{\ref{sec:numrel}}.

Finally, it's worth noting that the ringdown waves that come from the
last dynamics of a newly-born black hole can also be modeled using
perturbation theory.  The spacetime is accurately modeled as a black
hole plus a small deviation.  Perturbation theory teaches us that
black holes ``ring'' in a series of modes, with frequencies and
damping times that are controlled by the mass and the spin of the
black hole {\citep{leaver85}}.  Any deviation from an exact black hole
solution is carried away by such modes, enforcing the black hole
no-hair theorems (\citealt{price72a,price72b}).  We will not say much
about the ringdown in this review, except to note that the last waves
which come from numerical relativity simulations have been shown to
agree excellently with these perturbative calculations.

\subsection{Notation and conventions}
\label{sec:notation}

The underlying theory of GWs is general relativity (GR); we review its
most crucial concepts in Sec.\ {\ref{sec:gr}}.  Because multiple
conventions are often used in the GR literature, we first describe our
notation and conventions.

Throughout this review, Greek indices denote {\it spacetime}
components of tensors and vectors.  These indices run over
$(0,1,2,3)$, with $0$ generally denoting time $t$, and $(1,2,3)$
denoting spatial directions.  Spacetime vectors are sometimes written
with an overarrow:
\begin{equation}
\vec A = \{A^\mu\} \doteq (A^0, A^1, A^2, A^3)\;.
\label{eq:vector_notation}
\end{equation}
Equation (\ref{eq:vector_notation}) should be read as ``The vector
$\vec A$ has components $A^\mu$ whose values in a specified coordinate
system are $A^0$, $A^1$, $A^2$, and $A^3$.''  Lowercase Latin indices
denote {\it spatial} components.  Spatial vectors are written
boldface:
\begin{equation}
{\bf a} = \{a^i\} \doteq (a^1, a^2, a^3)\;.
\label{eq:spatial_vector_notation}
\end{equation}
We use the Einstein summation convention throughout, meaning that
repeated adjacent indices in superscript and subscript positions are
to be summed:
\begin{equation}
A^\mu B_\mu \equiv \sum_{\mu = 0}^3 A^\mu B_\mu\;.
\label{eq:einstein_summation}
\end{equation}
Indices are raised and lowered using the metric of spacetime
(discussed in more detail in Sec.\ {\ref{sec:gr}}) as a raising or
lowering operator:
\begin{equation}
A^\mu B_\mu = g_{\mu\nu} A^\mu B^\nu = A_\nu B^\nu = g^{\mu\nu} A_\mu
B_\nu\;.
\label{eq:raiselower}
\end{equation}

When we discuss the linearized limit of GR (particularly in Sec.\
{\ref{sec:gwbasics}}), it is useful to work in coordinates such that
the spacetime metric can be written as that of special relativity plus
a small perturbation:
\begin{equation}
g_{\mu\nu} = \eta_{\mu\nu} + h_{\mu\nu}\;,
\end{equation}
where $\eta_{\mu\nu} = {\rm diag}(-1,1,1,1)$.  This means that the
spatial part of the background metric is the Kronecker delta
$\delta_{ij}$.  For certain calculations in linearized theory, it is
useful to abuse the Einstein summation convention and sum over
repeated adjacent {\it spatial} indices regardless of position:
\begin{equation}
\sum_{i = 1}^3 a_i b^i = a_i b^i = a_i b_i = a^i b^i\;.
\end{equation}
This is allowable because using the Kronecker delta for the spatial
part of the metric means $a^i = a_i$ to this order.

Throughout this review, we abbreviate the partial derivative
\begin{equation}
\frac{\partial}{\partial x^\mu} \equiv \partial_\mu\;.
\end{equation}
With this notation defined, we can write $\partial^\mu =
g^{\mu\nu}\partial_\nu$.

Finally, it is common in GR research to use units in which the
gravitational constant $G$ and the speed of light $c$ are set to 1.
This has the advantage that mass, space, and time have the same units,
but can be confusing when applied to astrophysical systems.  In
deference to the astronomical audience of this review, we have put
$G$s and $c$s back into the relativity formulas.  An exception to the
rule that $G = c = 1$ everywhere is {\citet{shapteuk}}, especially
Chap.\ 15.

\section{Synopsis of general relativity}
\label{sec:gr}

GR describes gravity as geometry.  The foundation of this is the {\it
metric}, which provides a notion of spacetime distance.  Suppose event
A occurs at coordinate $x^\alpha$, and event B at $x^\alpha +
dx^\alpha$.  The proper spacetime separation, $ds$, between A and B is
given by
\begin{equation}
ds^2 = g_{\alpha\beta} dx^\alpha dx^\beta\;.
\label{eq:metric}
\end{equation}
The metric $g_{\alpha\beta}$ translates the information in
coordinates, which can be arbitrary, into a ``proper'' quantity, which
can be measured.  In the limit of special relativity,
$g_{\alpha\beta}$ becomes $\eta_{\alpha\beta}$ (defined in Sec.\
{\ref{sec:notation}}).  The general spacetime metric is determined by
the distribution of mass and energy; we describe how to compute it
below.  It will sometimes be useful to work with the inverse metric
$g^{\alpha\beta}$, defined by
\begin{equation}
g^{\alpha\beta}g_{\beta\gamma} = {\delta^\alpha}_\gamma\;.
\label{eq:metric_inverse}
\end{equation}
The metric also takes inner products between vectors and tensors:
\begin{equation}
\vec A\cdot \vec B = g_{\alpha\beta} A^\alpha B^\beta\;.
\end{equation}
$\vec A$ is {\it timelike} if $\vec A\cdot\vec A < 0$, {\it spacelike}
if $\vec A\cdot\vec A > 0$, and {\it lightlike} or {\it null} if $\vec
A\cdot\vec A = 0$.

Consider a {\it worldline} or spacetime trajectory $z^\mu(\tau)$,
where $\tau$ is ``proper time'' (time as measured by an observer on
that worldline).  The vector $u^\mu \equiv dz^\mu/d\tau$ is the
tangent to the worldline.  If $u^\mu$ is timelike, it is the
4-velocity of an observer following the worldline, and is normalized
$u^\mu u_\mu = -1$.  Suppose the worldline extends from A to B.
The total spacetime separation between these points is
\begin{equation}
s = \int_{\rm A}^{\rm B} d\tau \sqrt{g_{\alpha\beta} u^\alpha u^\beta}\;.
\end{equation}
We now extremize $s$: fix the endpoints, allow quantities under the
integral to vary, but require the variation to be stationary (so that
$\delta s = 0$).  The $u^\alpha$ which extremizes $s$ is given by the
{\it geodesic equation}:
\begin{equation}
\frac{du^\alpha}{d\tau} + {\Gamma^\alpha}_{\beta\gamma} u^\beta
u^\gamma = 0\;.
\label{eq:geodesic}
\end{equation}
We have introduced the ``connection'' ${\Gamma^\alpha}_{\beta\gamma}$;
it is built from the metric by
\begin{equation}
{\Gamma^\alpha}_{\beta\gamma} = \frac{1}{2} g^{\alpha\mu}\left(
\partial_\gamma g_{\mu\beta} + \partial_\beta g_{\gamma\mu} -
\partial_\mu g_{\beta\gamma}\right)\;.
\label{eq:connection}
\end{equation}
Geodesics are important for our discussion because {\it freely falling
bodies follow geodesics of spacetime in GR.}  Geodesics express the
rule that ``spacetime tells bodies how to move.''

Timelike geodesics describe material bodies.  {\it Null} geodesics,
for which $u^\mu u_\mu = 0$, describe massless bodies or light rays.
Our discussion above describes null geodesics, with one modification:
We cannot parameterize a null worldline with $\tau$, as proper time is
not meaningful for a ``speed of light'' trajectory.  Instead, one uses
an {\it affine parameter} $\lambda$ which uniformly ``ticks'' along
that trajectory.  A convenient choice is to set $u^\alpha \equiv
dx^\alpha/d\lambda$ to be the 4-momentum of our radiation or massless
body.  With this choice, our discussion describes null trajectories
just as well as timelike ones.

The connection also defines the {\it covariant derivative} of a vector
or tensor:
\begin{eqnarray}
\nabla_\alpha A^\beta &=& \partial_\alpha A^\beta + A^\mu
{\Gamma^\beta}_{\alpha\mu}\;,
\nonumber\\
\nabla_\alpha A_\beta &=& \partial_\alpha A_\beta - A_\mu
{\Gamma^\mu}_{\alpha\beta}\;,
\nonumber\\
\nabla_\alpha {A^\beta}_\gamma &=& \partial_\alpha {A^\beta}_\gamma +
{A^\mu}_\gamma {\Gamma^\beta}_{\alpha\mu} -
{A^\beta}_\mu {\Gamma^\mu}_{\alpha\gamma}\;.
\label{eq:covar}
\end{eqnarray}
The pattern continues as we add indices.  This derivative follows by
comparing vectors and tensors that are slightly separated by {\it
parallel transporting} them together to make the comparison; the
connection encodes the twists of our curved geometry.  Using the
covariant derivative, the geodesic equation can be written
\begin{equation}
u^\beta \nabla_\beta u^\alpha = 0\;.
\label{eq:geodesic2}
\end{equation}

In curved spacetime, nearby geodesics diverge.  Because a geodesic
describes a freely-falling body, the rate of at which geodesics
diverge describes {\it tides}.  Let $\xi^\alpha$ be the displacement
between geodesics.  Then the rate of divergence is given by
\begin{equation}
\frac{D^2\xi^\alpha}{d\tau^2} = {R^\alpha}_{\beta\gamma\delta}u^\beta
u^\gamma \xi^\delta\;.
\end{equation}
We have introduced the {\it Riemann curvature tensor},
\begin{equation}
{R^\alpha}_{\beta\gamma\delta} =
\partial_\gamma{\Gamma^\alpha}_{\beta\delta} -
\partial_\delta{\Gamma^\alpha}_{\beta\gamma} +
{\Gamma^\alpha}_{\mu\gamma}{\Gamma^\mu}_{\beta\delta} -
{\Gamma^\alpha}_{\mu\delta}{\Gamma^\mu}_{\beta\gamma}\;.
\label{eq:Riemann}
\end{equation}
Some variants of Riemann are important.  First, there is the Ricci
curvature:
\begin{equation}
R_{\alpha\beta} = {R^\mu}_{\alpha\mu\beta}\;.
\label{eq:ricci}
\end{equation}
Ricci is the trace of Riemann.  Taking a further trace gives us the
Ricci scalar,
\begin{equation}
R = {R^\mu}_{\mu}\;.
\end{equation}
The Ricci tensor and Ricci scalar combine to produce the Einstein
curvature:
\begin{equation}
G_{\alpha\beta} = R_{\alpha\beta} - \frac{1}{2}g_{\alpha\beta}R\;.
\label{eq:einstein_curve}
\end{equation}

The Riemann curvature satisfies the {\it Bianchi identity},
\begin{equation}
\nabla_\gamma R_{\alpha\beta\mu\nu} +
\nabla_\beta R_{\gamma\alpha\mu\nu} +
\nabla_\alpha R_{\beta\gamma\mu\nu} = 0\;.
\label{eq:bianchi}
\end{equation}
By tracing over certain combinations of indices, the Bianchi identity
implies
\begin{equation}
\nabla^\alpha G_{\alpha\beta} = 0\;,
\label{eq:bianchi_contr}
\end{equation}
a result sometimes called the ``contracted'' Bianchi identity.

So far, we have mostly described the mathematics of curved geometry.
We must also introduce tools to describe matter and fields.  The most
important tool is the stress-energy tensor:
\begin{equation}
T^{\mu\nu} \equiv \mbox{Flux of momentum $p^\mu$ in the $x^\nu$
direction.}
\label{eq:Tmunu_def}
\end{equation}
An observer who uses coordinates $(t,x^i)$ to make local measurements
interprets the components of this tensor as
\begin{eqnarray}
T^{tt} &\equiv& \mbox{Local energy density}
\label{eq:energy_density}
\\
T^{ti} &\equiv& \mbox{Local energy flux (times $c$)}
\label{eq:energy_flux}
\\
T^{it} &\equiv& \mbox{Local momentum density (times $c$)}
\label{eq:momentum_density}
\\
T^{ij} &\equiv& \mbox{Local momentum flux (times $c^2$); $T^{ii}$
acts as pressure.}
\label{eq:momentum_flux}
\end{eqnarray}
[The factors of $c$ in Eqs.\ (\ref{eq:energy_flux}) --
(\ref{eq:momentum_flux}) ensure that the components of $T^{\mu\nu}$
have the same dimension.]  Local conservation of energy and momentum
is expressed by
\begin{equation}
\nabla_\mu T^{\mu\nu} = 0\;.
\label{eq:local_en_cons}
\end{equation}
In GR, we generally lose the notion of {\it global} energy
conservation: We cannot integrate Eq.\ (\ref{eq:local_en_cons}) over
an extended region to ``add up'' the total energy and momentum.  This
is because $\nabla_\mu T^{\mu\nu}$ is a spacetime vector, and in
curved spacetime one cannot unambiguously compare widely separated
vectors.

Einstein's hypothesis is that stress energy is the source of spacetime
curvature.  If $T^{\mu\nu}$ is our source, then the curvature must
likewise be divergence free.  The contracted Bianchi identity
(\ref{eq:bianchi_contr}) shows us that the Einstein tensor is the
curvature we need.  This logic yields the {\it Einstein field
equation}:
\begin{equation}
G_{\mu\nu} = \frac{8\pi G}{c^4} T_{\mu\nu}\;.
\label{eq:einstein}
\end{equation}
The factor $8\pi G/c^4$ guarantees that this equation reproduces
Newtonian gravity in an appropriate limit.  Note its value:
\begin{equation}
\frac{G}{c^4} = 8.26\times 10^{-50}\frac{{\rm cm}^{-2}}{{\rm erg/cm}^3}\;.
\end{equation}
It takes an {\it enormous} amount of energy density to produce
spacetime curvature (measured in inverse length squared).  Note that
the reciprocal of this quantity, times $c$, has the dimensions of
power:
\begin{equation}
\frac{c^5}{G} = 3.63 \times 10^{59}\,{\rm erc/sec}\;.
\end{equation}
This is the scale for the power that is generated by a GW source.

\section{Gravitational-wave basics}
\label{sec:gwbasics}

We now give a brief description of how gravitational waves arise in
GR.  Our purpose is to introduce the main ideas of this field, and
also provide some results against which the more complete calculations
we discuss later can be compared.

\subsection{Leading waveform}
\label{sec:waveform}

Begin with ``weak'' gravity, so that spacetime is nearly that of
special relativity,
\begin{equation}
g_{\alpha\beta} = \eta_{\alpha\beta} + h_{\alpha\beta}\;.
\end{equation}
Take the correction to flat spacetime to be small, so that we can
linearize about $\eta_{\alpha\beta}$.  Consider, for example, raising
and lowering indices:
\begin{equation}
h^{\alpha\beta} \equiv g^{\alpha\mu}g^{\beta\nu} h_{\mu\nu} =
\eta^{\alpha\mu}\eta^{\beta\nu} h_{\mu\nu} + {\cal O}(h^2)\;.
\end{equation}
Because we only keep quantities to first order in $h$, we will
consistently use the flat metric to raise and lower indices for
quantities related to the geometry.

Applying this logic repeatedly, we build the linearized Einstein
tensor:
\begin{equation}
G_{\alpha\beta} = \frac{1}{2} \left(\partial_\alpha\partial^\mu
h_{\mu\beta} + \partial_\beta\partial^\mu h_{\mu\alpha} -
\partial_\alpha\partial_\beta h - \Box h_{\alpha\beta} +
\eta_{\alpha\beta}\Box h - \eta_{\alpha\beta}\partial^\mu\partial^\nu
h_{\mu\nu}\right)\;,
\label{eq:lin_einstein1}
\end{equation}
where $h \equiv \eta^{\alpha\beta}h_{\alpha\beta}$ is the trace of
$h_{\alpha\beta}$, and $\Box \equiv \eta^{\alpha\beta}\partial_\alpha
\partial_\beta$ is the flat spacetime wave operator.

Equation (\ref{eq:lin_einstein1}) is rather messy.  We clean it up in
two steps.  The first is pure sleight of hand: We introduce the {\it
trace-reversed} metric perturbation
\begin{equation}
{\bar h}_{\alpha\beta} \equiv h_{\alpha\beta} -
\frac{1}{2}\eta_{\alpha\beta} h\;.
\end{equation}
With this definition, Eq.\ (\ref{eq:lin_einstein}) becomes
\begin{equation}
G_{\alpha\beta} = \frac{1}{2} \left(\partial_\alpha\partial^\mu \bar
h_{\mu\beta} + \partial_\beta\partial^\mu \bar h_{\mu\alpha} - \Box
\bar h_{\alpha\beta} - \eta_{\alpha\beta}\partial^\mu\partial^\nu
h_{\mu\nu}\right)\;.
\label{eq:lin_einstein2}
\end{equation}
Next, we take advantage of the {\it gauge-freedom} of linearized
gravity.  Recall that in electrodynamics, if one adjusts the potential
by the gradient of a scalar, $A_\mu \to A_\mu - \partial_\mu \Lambda$,
then the field tensor $F_{\mu\nu} = \partial_\mu A_\nu - \partial_\nu
A_\mu$ is unchanged.  In linearized GR, a similar operation follows by
adjusting one's coordinates: If one changes coordinates $x^\alpha \to
x^\alpha + \xi^\alpha$ (requiring $\partial_\mu\xi^\alpha \ll 1$),
then
\begin{equation}
h_{\mu\nu} \to h_{\mu\nu} - \partial_\mu\xi_\nu - \partial_\nu\xi_\mu\;.
\label{eq:gauge_lingrav}
\end{equation}
One can easily show [see, e.g., \cite{carroll}, Sec.\ 7.1] that
changing gauge leaves the Riemann tensor (and thus all tensors derived
from it) unchanged.

We take advantage of our gauge freedom to choose $\xi^\alpha$ so that
\begin{equation}
\partial^\mu {\bar h}_{\mu\nu} = 0\;.
\end{equation}
This is called ``Lorenz gauge'' in analogy with the electrodynamic
Lorenz gauge condition $\partial^\mu A_\mu = 0$.  This simplifies our
Einstein tensor considerably, yielding
\begin{equation}
G_{\alpha\beta} = -\frac{1}{2}\Box{\bar h}_{\alpha\beta}\;.
\label{eq:lin_einstein}
\end{equation}
The Einstein equation for linearized gravity thus takes the simple
form
\begin{equation}
\Box{\bar h}_{\alpha\beta} = -\frac{16\pi G}{c^4} T_{\alpha\beta}\;.
\label{eq:lin_efe}
\end{equation}

Any linear equation of this form can be solved using a radiative
Green's function [e.g., \cite{jackson}, Sec.\ 12.11].  Doing so yields
\begin{equation}
{\bar h}_{\alpha\beta}({\bf x}, t) =
    \frac{4G}{c^4}\int\frac{T_{\alpha\beta}({\bf x}', t - |{\bf x} -
    {\bf x}'|/c)}{|{\bf x} - {\bf x}'|} d^3x'\;.
\label{eq:lin_soln}
\end{equation}
In this equation, ${\bf x}$ is the ``field point,'' where $\bar
h_{\alpha\beta}$ is evaluated, and ${\bf x}'$ is a ``source point,''
the coordinate that we integrate over the source's spatial extent.
Notice that the solution at $t$ depends on what happens to the source
at {\it retarded time} $t - |{\bf x} - {\bf x'}|/c$.  Information must
causally propagate from ${\bf x'}$ to ${\bf x}$.

Equation (\ref{eq:lin_soln}) is formally an exact solution to the
linearized Einstein field equation.  However, it has a serious
problem: It gives the impression that {\it every component} of the
metric perturbation is radiative.  This is an unfortunate consequence
of our gauge.  Just as one can choose a gauge such that an isolated
point charge has an oscillatory potential, the Lorenz gauge we have
used makes {\it all} components of the metric appear radiative, even
if they are static\footnote{In the electromagnetic case, it is
unambiguous which {\it field} components are radiative and which are
static.  Similarly, one can always tell which Riemann {\it curvature}
components are radiative and which are static.  {\cite{eddington22}}
appears to have been the first to use the curvature tensor to
categorize the gravitational degrees of freedom in this way.}.

Fortunately, it is not difficult to see that only a subset of the
metric represents the truly radiative degrees of freedom in {\it all}
gauges.  We will only quote the result here; interested readers can
find the full calculation in \citet{fh05}, Sec.\ 2.2: {\it Given a
solution $h_{\alpha\beta}$ to the linearized Einstein field equations,
only the {\bf spatial}, {\bf transverse}, and {\bf traceless}
components $h^{\rm TT}_{ij}$ describe the spacetime's gravitational
radiation in a gauge-invariant manner.}  (The other components can be
regarded as ``longitudinal'' degrees of freedom, much like the Coulomb
potential of electrodynamics.)  Traceless means
\begin{equation}
\delta_{ij} h^{\rm TT}_{ij} = 0\;;
\label{eq:traceless}
\end{equation}
``transverse'' means
\begin{equation}
\partial_i h^{\rm TT}_{ij} = 0\;.
\label{eq:transverse}
\end{equation}
This condition tells us that $h^{\rm TT}_{ij}$ is orthogonal to the
direction of the wave's propagation.  Expanding $h^{\rm TT}_{ij}$ in
Fourier modes shows that Eq.\ (\ref{eq:transverse}) requires $h^{\rm
TT}_{ij}$ to be orthogonal (in space) to each mode's wave vector ${\bf
k}$.

Conditions (\ref{eq:traceless}) and (\ref{eq:transverse}) make it
simple to construct $h^{\rm TT}_{ij}$ given some solution $h_{ij}$ to
the linearized field equations.  Let $n_i$ denote components of the
unit vector along the wave's direction of propagation.  The tensor
\begin{equation}
P_{ij} = \delta_{ij} - n_in_j
\end{equation}
projects spatial components orthogonal to ${\bf n}$.  It is then
simple to verify that
\begin{equation}
h^{\rm TT}_{ij} = h_{kl}\left(P_{ki}P_{lj} -
\frac{1}{2}P_{kl}P_{ij}\right)
\label{eq:projected_hTT}
\end{equation}
represents the transverse and traceless metric perturbation.

The simplest example solution to Eq.\ (\ref{eq:projected_hTT}) can be
built by going back to Eq.\ (\ref{eq:lin_soln}) and focusing on the
spatial components:
\begin{equation}
{\bar h}_{ij} = \frac{4G}{c^4}\int\frac{T_{ij}({\bf x}', t - |{\bf x}
  - {\bf x}'|/c)}{|{\bf x} - {\bf x}'|} d^3x'\;.
\end{equation}
Consider a very distant source, putting $|{\bf x} - {\bf x}'| \simeq
R$:
\begin{equation}
{\bar h}_{ij} \simeq \frac{4}{R}\frac{G}{c^4}\int T_{ij}({\bf x}', t -
R/c) d^3x'\;.
\end{equation}
To proceed, we invoke an identity.  Using the fact that $\nabla^\mu
T_{\mu\nu} = 0$ goes to $\partial^\mu T_{\mu\nu} = 0$ in linearized
theory and in our chosen coordinates, we have
\begin{equation}
\partial^t T_{tt} + \partial^j T_{jt} = 0\;,\qquad
\partial^t T_{tj} + \partial^i T_{ij} = 0\;.
\end{equation}
Combine these identities with the fact that $\partial^t =
-\partial_t$; use integration by parts to convert volume integrals to
surface integrals; discard those integrals by taking the surface
outside our sources.  We then find
\begin{equation}
\int T_{ij}({\bf x'}, t)\,d^3x' = \frac{1}{2}\frac{d^2}{dt^2}\int
x^{i'}x^{j'} T_{tt}({\bf x}', t)\,d^3x' \equiv \frac{1}{2}
\frac{d^2}{dt^2} I_{ij}(t)\;.
\end{equation}
We have introduced here the {\it quadrupole moment} $I_{ij}$.  This
allows us to at last write the transverse-traceless waveform as
\begin{equation}
h^{\rm TT}_{ij} =
\frac{2}{R}\frac{G}{c^4}\frac{d^2I_{kl}}{dt^2}\left(P_{ik}P_{jl} -
\frac{1}{2}P_{kl}P_{ij}\right)\;.
\label{eq:quadrupole1}
\end{equation}
It is straightforward to show that the trace $I \equiv I_{ii}$ does
not contribute to Eq.\ (\ref{eq:quadrupole1}), so it is common to use
the ``reduced'' quadrupole moment,
\begin{equation}
{\cal I}_{ij} = I_{ij} - \frac{1}{3}\delta_{ij}I\;.
\end{equation}
The waveform then takes the form in which it is usually presented,
\begin{equation}
h^{\rm TT}_{ij} =
\frac{2}{R}\frac{G}{c^4}\frac{d^2{\cal I}_{kl}}{dt^2}\left(P_{ik}P_{jl} -
\frac{1}{2}P_{kl}P_{ij}\right)\;,
\label{eq:quadrupole}
\end{equation}
the {\it quadrupole formula} for GW emission.

A more accurate approximation than $|{\bf x} - {\bf x}'| \simeq R$ is
\begin{equation}
|{\bf x} - {\bf x}'| \simeq R - {\bf n}\cdot{\bf x}'\;,
\end{equation}
where ${\bf n}$ is the unit vector pointing from ${\bf x}'$ to ${\bf
x}$.  Revisiting the calculation, we find that Eq.\
(\ref{eq:quadrupole}) is the first term in a {\it multipolar
expansion} for the radiation.  Detailed formulae and notation can be
found in {\citet{thorne80}}, Sec.\ IVA.  Schematically, the resulting
waveform can be written
\begin{equation}
h^{\rm TT} = \frac{1}{R}\frac{G}{c^2}\sum_{l = 2}^\infty \left\{
\frac{{\cal A}_l}{c^l}\frac{d^l{\cal I}_l}{dt^l} + \frac{{\cal
B}_l}{c^{l+1}}\frac{d^l{\cal S}_l}{dt^l}\right\}^{\rm STF}\;.
\label{eq:multipolar_form}
\end{equation}
We have hidden various factorials of $l$ in the coefficients ${\cal
A}_l$ and ${\cal B}_l$; the superscript ``STF'' means to symmetrize
the result on any free indices and remove the trace.

The symbol ${\cal I}_l$ stands for the $l$th {\it mass moment} of the
source.  \citet{thorne80} precisely defines ${\cal I}_l$; for our
purposes, it is enough to note that it represents an integral of $l$
powers of length over the source's mass density $\rho$:
\begin{equation}
{\cal I}_l \sim \int \rho(x') (x')^l\,d^3x'\;.
\end{equation}
The mass moment plays a role in gravitational radiation similar to
that played by the electric charge moment in electrodynamics.  The
symbol ${\cal S}_l$ describes the $l$th {\it mass-current moment}; it
represents an integral of $l$ powers of length over the source's
mass-current ${\bf J} = \rho{\bf v}$, where ${\bf v}$ describes a
source's internal motions:
\begin{equation}
{\cal S}_l \sim \int \rho(x') v(x') (x')^l\,d^3x'\;.
\end{equation}
${\cal S}_l$ plays a role similar to the magnetic moment.

The similarity of the multipolar expansion (\ref{eq:multipolar_form})
for GWs to that for electromagnetic radiation should not be a
surprise; after all, our linearized field equation (\ref{eq:lin_efe})
is very similar to Maxwell's equation for the potential $A_\mu$
(modulo an extra index and a factor of four).  One should be cautious
about taking this analogy too far.  Though electromagnetic intuition
applied to linearized theory works well to compute the GWs from a
source, one must go beyond this linearized form to understand deeper
aspects of this theory.  In particular, one must go to higher order in
perturbation theory to see how energy and angular momentum are carried
from a radiating source.  We now sketch how this works in GR.

\subsection{Leading energy loss}
\label{sec:energyloss}

Electromagnetic radiation carries a flux of energy and momentum
described by the Poynting vector, ${\bf S} = (c/4\pi){\bf E}\times{\bf
B}$.  Likewise, electromagnetic fields generate stresses proportional
to $(|{\bf E}|^2 + |{\bf B}|^2)/8\pi$.  The lesson to take from this
is that the energy content of radiation should be {\it quadratic} in
wave amplitude.  Properly describing the energy content of radiation
requires second-order perturbation theory.  In this section, we will
discuss the key concepts and ideas in this analysis, which was first
given by {\cite{isaacson68}}.

Begin by writing the spacetime
\begin{equation}
g_{\alpha\beta} = {\hat g}_{\alpha\beta} + \epsilon h_{\alpha\beta}
+ \epsilon^2 j_{\alpha\beta}\;.
\end{equation}
We have introduced a parameter $\epsilon$ whose formal value is 1; we
use it to gather terms that are of the same order.  Note that now we
do not restrict the background to be flat.  This introduces a
conceptual difficulty: Measurements directly probe only the {\it
total} spacetime $g_{\alpha\beta}$ (or a derived surrogate like
curvature), so how do we distinguish background $\hat g_{\alpha\beta}$
from perturbation?  The answer is to use {\it separation of
lengthscales}.  Our background will only vary on ``long'' lengthscales
and timescales ${\cal L}, {\cal T}$; our perturbation varies on
``short'' lengthscales and timescales $\lambda, \tau$.  We require
${\cal L} \gg \lambda$ and ${\cal T} \gg \tau$.  Let $\langle f
\rangle$ denote a quantity $f$ averaged over long scales; this
averaging is well-defined even for tensors up to errors ${\cal
O}(\lambda^2/{\cal L}^2)$.  Then, to first order in $\epsilon$,
\begin{eqnarray}
\hat g_{\alpha\beta} &=& \langle g_{\alpha\beta} \rangle
\\
h_{\alpha\beta} &=& g_{\alpha\beta} - \hat g_{\alpha\beta}\;.
\end{eqnarray}
The second-order contribution will be of order $h^2$, and (as we'll
see below) will have contributions on both long and short scales.

Begin by expanding the Einstein tensor in $\epsilon$.  The result can
be written
\begin{equation}
G_{\mu\nu}(g_{\alpha\beta}) = G^0_{\mu\nu}({\hat g}_{\alpha\beta}) +
\epsilon G^1_{\mu\nu}(h_{\alpha\beta}) + \epsilon^2
[G^2_{\mu\nu}(h_{\alpha\beta}) + G^1_{\mu\nu}(j_{\alpha\beta})]\;.
\label{eq:einstein_expanded}
\end{equation}
For simplicity, take the spacetime to be vacuum --- there are no
non-gravitational sources of stress and energy in the problem.  We
then require Einstein's equation, $G_{\mu\nu} = 0$, to hold at each
order.  The zeroth order result,
\begin{equation}
G^0_{\mu\nu}({\hat g}_{\alpha\beta}) = 0\;,
\end{equation}
is just a statement that we assume our background to be a vacuum
solution.

Expanding $G^1_{\mu\nu}(h_{\alpha\beta}) = 0$, we find
\begin{equation}
-\frac{1}{2}{\hat\Box}{\bar h}_{\alpha\beta} - {\hat
 R}_{\alpha\mu\beta\nu}{\bar h}^{\mu\nu} = 0\;,
\label{eq:waveeqn_curvedbackground}
\end{equation}
where $\hat\Box \equiv {\hat g}^{\mu\nu} {\hat\nabla}_\mu
{\hat\nabla}_\nu$ is the wave operator for ${\hat g}_{\mu\nu}$ (with
$\hat\nabla_\mu$ the covariant derivative on $\hat g_{\mu\nu}$),
${\hat R}_{\alpha\mu\beta\nu}$ is the Riemann curvature built from
$\hat g_{\mu\nu}$, and ${\bar h}_{\mu\nu} = h_{\mu\nu} - (1/2){\hat
g}_{\mu\nu} h$.  Equation (\ref{eq:waveeqn_curvedbackground}) is just
the wave equation for radiation propagating on a curved background.
The coupling between $h$ and the background Riemann tensor describes a
correction to the ``usual'' geometric optics limit.

Next, consider second order:
\begin{equation}
G^1_{\mu\nu}(j_{\alpha\beta}) = -G^2_{\mu\nu}(h_{\alpha\beta})\;.
\end{equation}
To make sense of this, invoke separation of scales.  The second-order
perturbation has contributions on both scales:
\begin{equation}
j_{\alpha\beta} = \langle j_{\alpha\beta} \rangle + \delta
j_{\alpha\beta}\;,
\end{equation}
where $\langle j_{\alpha\beta}\rangle$ varies on long scales and
$\delta j_{\alpha\beta}$ is oscillatory and varies on short scales.
The second-order metric can now be written
\begin{equation}
g_{\alpha\beta} = g^{\cal L}_{\alpha\beta} + \epsilon h_{\alpha\beta}
+ \epsilon^2 \delta j_{\alpha\beta}\;,
\end{equation}
where
\begin{equation}
g^{\cal L}_{\alpha\beta} \equiv \hat g_{\alpha\beta} + \epsilon^2
\langle j_{\alpha\beta} \rangle\;,
\end{equation}
is a ``corrected'' background which includes all pieces that vary on
long scales.

Now return to the Einstein equation (\ref{eq:einstein_expanded}), but
consider its {\it averaged} value.  Thanks to linearity, we can take
the averaging inside the operator $G^1$:
\begin{equation}
\langle G^1_{\mu\nu}(h_{\alpha\beta})\rangle = G^1_{\mu\nu}(\langle
h_{\alpha\beta}\rangle) = 0\;.
\end{equation}
We have used here $\langle h_{\alpha\beta} \rangle = 0$. Putting all
this together, we find
\begin{equation}
G^0_{\mu\nu}({\hat g}_{\alpha\beta}) + \epsilon^2 [\langle
G^2_{\mu\nu}(h_{\alpha\beta})\rangle + G^1_{\mu\nu}(\langle
j_{\alpha\beta}\rangle )] = 0\;.
\end{equation}
Let us rewrite this as
\begin{equation}
G_{\mu\nu}({\hat g}_{\alpha\beta} + \epsilon^2 \langle j_{\alpha\beta}
\rangle) = -\epsilon^2 \langle
G^2_{\mu\nu}(h_{\alpha\beta})\rangle\;,
\end{equation}
or, putting $\epsilon = 1$,
\begin{equation}
G_{\mu\nu}(g^{\cal L}_{\alpha\beta}) = -\langle
G^2_{\mu\nu}(h_{\alpha\beta})\rangle\;.
\label{eq:2ndorder_averaged_einstein}
\end{equation}
Equation (\ref{eq:2ndorder_averaged_einstein}) says that the
second-order averaged Einstein tensor acts as a source for the long
lengthscale background spacetime.  This motivates the definition
\begin{equation}
T_{\mu\nu}^{\rm GW} = -\frac{c^4}{8\pi G}
\left\langle G^2_{\mu\nu}(h_{\alpha\beta})\right\rangle\;.
\end{equation}
Choosing a gauge so that $\hat\nabla^\mu\bar h_{\mu\nu} = 0$,
$T_{\mu\nu}^{\rm GW}$ takes a very simple form:
\begin{equation}
T_{\mu\nu}^{\rm GW} = \frac{c^4}{32\pi G} \langle \hat\nabla_\mu
h_{\alpha\beta} \hat\nabla_\nu h^{\alpha\beta} \rangle\;.
\label{eq:isaacson_tmunu}
\end{equation}
This quantity is known as the Isaacson stress-energy tensor
\citep{isaacson68}.

\subsection{The ``Newtonian, quadrupole'' waveform}
\label{sec:newtonian_waves}

A useful exercise is to consider a binary with Newtonian orbital
dynamics that radiates GWs according to Eq.\ (\ref{eq:quadrupole}).
Further, allow the binary to slowly evolve by energy and angular
momentum carried off in accordance with Eq.\
(\ref{eq:isaacson_tmunu}).

Begin by considering such a binary with its members in circular orbit
of separation $R$.  This binary is characterized by orbital energy
\begin{equation}
E^{\rm orb} = \frac{1}{2}m_1 v_1^2 + \frac{1}{2}m_2 v_2^2 -
\frac{Gm_1m_2}{R} = -\frac{G\mu M}{2R}\;,
\end{equation}
(where $M = m_1 + m_2$ and $\mu = m_1 m_2/M$) and orbital frequency
\begin{equation}
\Omega_{\rm orb} = \sqrt{\frac{GM}{R^3}}\;.
\end{equation}

Next consider the energy that GWs carry from the binary to distant
observers.  When evaluated far from a source, Eq.\
(\ref{eq:isaacson_tmunu}) gives a simple result for energy flux:
\begin{equation}
\frac{dE}{dAdt} = \frac{c^4}{32\pi G}\langle \partial_t h^{\rm TT}_{ij}
\partial_k h^{\rm TT}_{ij}\rangle n^k\;.
\end{equation}
Plugging in Eq.\ (\ref{eq:quadrupole}) and integrating over the
sphere, we find
\begin{equation}
\frac{dE}{dt}^{\rm GW} = \int dA\,\frac{dE}{dAdt} =
\frac{G}{5c^5}\left\langle \frac{d^3{\cal I}_{ij}}{dt^3} \frac{d^3{\cal
I}_{ij}}{dt^3}\right\rangle\;.
\label{eq:quadrupole_edot}
\end{equation}
For the Newtonian binary,
\begin{equation}
{\cal I}_{ij} = \mu \left(x_i x_j -
\frac{1}{3}R^2\delta_{ij}\right)\;;
\end{equation}
we choose coordinates for this binary such that the components of the
separation vector are $x_1 = R\cos\Omega_{\rm orb} t$, $x_2 =
R\sin\Omega_{\rm orb} t$, $x_3 = 0$.  Inserting into Eq.\
(\ref{eq:quadrupole_edot}), we find
\begin{equation}
\frac{dE}{dt}^{\rm GW} = \frac{32}{5}\frac{G}{c^5} \mu^2 R^4
\Omega^6\;.
\label{eq:circbin_edot}
\end{equation}
We now assert that the binary evolves quasi-statically, meaning that
any radiation carried off by GWs is accounted for by the evolution of
its orbital energy:
\begin{equation}
\frac{dE}{dt}^{\rm orb} + \frac{dE}{dt}^{\rm GW} = 0\;.
\label{eq:energy_balance}
\end{equation}
We evaluate $dE^{\rm orb}/dt$ by allowing the orbital radius to
slowly change in time,
\begin{equation}
\frac{dE}{dt}^{\rm orb} = \frac{dE}{dR}^{\rm orb}\frac{dR}{dt}\;.
\label{eq:eorb_dot}
\end{equation}
Combining Eqs.\ (\ref{eq:circbin_edot}), (\ref{eq:energy_balance}),
and (\ref{eq:eorb_dot}), we find
\begin{equation}
R(t) = \left[\frac{256 G^3 \mu M^2(t_c - t)}{5c^5}\right]^{1/4}\;.
\label{eq:rorb_of_time}
\end{equation}
This in turn tells us that the orbital frequency changes according to
\begin{equation}
\Omega_{\rm orb}(t) = \left[\frac{5c^5}{256(G{\cal M})^{5/3}(t_c -
t)}\right]^{3/8}\;.
\label{eq:Newt_quad}
\end{equation}
We have introduced the binary's {\it chirp mass} ${\cal M} \equiv
\mu^{3/5}M^{2/5}$, so called because it sets the rate at which the
binary sweeps upward in frequency, or ``chirps.''  We have also
introduced the coalescence time $t_c$, which formally describes when
the separation goes to zero (equivalently, when the frequency goes to
infinity).  By rearranging Eq.\ (\ref{eq:rorb_of_time}), we find the
time remaining for a circular binary of radius $R$ to coalesce due to
GW emission:
\begin{eqnarray}
T_{\rm remaining} &=& \frac{5}{256}\frac{c^5}{G^3}\frac{R^4}{\mu M^2}
\nonumber\\
&=& 3\times10^6\,{\rm years}\left(\frac{2.8\,M_\odot}{M}\right)^3
\left(\frac{R}{R_\odot}\right)^4
\nonumber\\
&=& 2\,{\rm months}\left(\frac{2\times10^6\,M_\odot}{M}\right)^3
\left(\frac{R}{{\rm AU}}\right)^4
\nonumber\\
&=& 3\times10^8{\rm years}\left(\frac{2\times10^6\,M_\odot}{M}\right)^3
\left(\frac{R}{0.001\,{\rm pc}}\right)^4\;.
\label{eq:coal_time}
\end{eqnarray}
The fiducial numbers we show here are for equal mass binaries, so $\mu
= M/4$.

Given the substantial eccentricity of many binaries, restriction to
circular orbits may not seem particularly realistic.  Including
eccentricity means that our binary will have two evolving parameters
(semi-major axis $a$ and eccentricity $e$) rather than just one
(orbital radius $R$).  To track their evolution, we must separately
compute the radiated energy and angular momentum
(\citealt{pm63,peters64}):
\begin{equation}
\frac{dE}{dt}^{\rm GW} = \frac{G}{5c^5}\langle\frac{d^3{\cal
I}_{ij}}{dt^3} \frac{d^3{\cal I}_{ij}}{dt^3}\rangle =
\frac{32}{5}\frac{G}{c^5} \mu^2 a^4 \Omega^6 f(e)\;,
\label{eq:edot_eccentric}
\end{equation}
\begin{equation}
\frac{dL_z}{dt}^{\rm GW} =
\frac{2G}{5c^5}\epsilon_{zjk}\langle\frac{d^2{\cal I}_{jm}}{dt^2}
\frac{d^3{\cal I}_{km}}{dt^3}\rangle =
\frac{32}{5}\frac{G}{c^5} \mu^2 a^4 \Omega^5 g(e)\;.
\label{eq:ldot_eccentric}
\end{equation}
Because the binary is in the $x-y$ plane, the angular momentum is purely
along the $z$ axis.  The eccentricity corrections $f(e)$ and $g(e)$
are given by
\begin{equation}
f(e) = \frac{1 + \frac{73}{24}e^2 + \frac{37}{96}e^4}{(1 - e^2)^{7/2}}\;,
\qquad
g(e) = \frac{1 + \frac{7}{8}e^2}{(1 - e^2)^2}\;.
\end{equation}
Using standard definitions relating the semi-major axis and
eccentricity to the orbit's energy $E$ and angular momentum $L_z$,
Eqs.\ (\ref{eq:edot_eccentric}) and (\ref{eq:ldot_eccentric}) imply
\citep{peters64}
\begin{equation}
\frac{da}{dt} = -\frac{64}{5}\frac{G^3}{c^5}\frac{\mu M^2}{a^3} f(e)\;,
\label{eq:adot}
\end{equation}
\begin{equation}
\frac{de}{dt} = -\frac{304}{15}e\frac{G^3}{c^5}\frac{\mu M^2}{a^4}
\frac{1 + \frac{121}{304}e^2}{(1 - e^2)^{5/2}}\;.
\label{eq:edot}
\end{equation}
It is then simple to compute the rate at which an eccentric binary's
orbital period changes due to GW emission.  This result is compared
with data in investigations of GW generating binary pulsars.  Because
eccentricity tends to enhance a system's energy and angular momentum
loss, the timescales given in Eq.\ (\ref{eq:coal_time}) can be
significant {\it under}estimates of a binary's true coalescence time.

Notice that a binary's eccentricity decreases: GWs tend to circularize
orbits.  Many binaries are expected to be essentially circular by the
time their waves enter the sensitive band of many GW detectors; the
circular limit is thus quite useful.  Exceptions are binaries which
form, through capture processes, very close to merging.  The extreme
mass ratio inspirals discussed in Sec.\ {\ref{sec:pert}} are a
particularly interesting example of this.

We conclude this section by writing the gravitational waveform
predicted for quadrupole emission from the Newtonian, circular binary.
Evaluating Eq.\ (\ref{eq:quadrupole}), we find that $h_{ij}$ has two
polarizations.  These are traditionally labeled ``plus'' and ``cross''
from the lines of force associated with their tidal stretch and
squeeze.  Taking our binary to be a distance $D$ from Earth, its
waveform is written
\begin{eqnarray}
h_+ &=& -\frac{2G{\cal M}}{c^2D}\left(\frac{\pi G{\cal
M}f}{c^3}\right)^{2/3}(1 + \cos^2\iota) \cos2\Phi_N(t)\;,
\nonumber\\
h_\times &=& -\frac{4G{\cal M}}{c^2D}\left(\frac{\pi G{\cal
M}f}{c^3}\right)^{2/3}\cos\iota \sin2\Phi_N(t)\;,
\label{eq:h_NQ}
\end{eqnarray}
where the phase
\begin{equation}
\Phi_N(t) = \int \Omega_{\rm orb}\,dt = \Phi_c - \left[\frac{c^3(t_c
- t)}{5G{\cal M}}\right]^{5/8}\;,
\label{eq:phi_NQ}
\end{equation}
and where $f = (1/\pi)d\Phi_N/dt$ is the GW frequency.  The system's
inclination $\iota$ is just the projection of its orbital angular
momentum, ${\bf L}$, to the wave's direction of propagation ${\bf n}$:
$\cos\iota = \hat{\bf L}\cdot{\bf n}$ (where $\hat{\bf L} = {\bf
L}/|{\bf L}|$).  We show fiducial values for the GW amplitudes when we
briefly describe GW measurement in Sec.\ {\ref{sec:gwastro}}.

In later sections, we will use Eq.\ (\ref{eq:h_NQ}) as a reference to
calibrate how effects we have neglected so far change the waves.  Note
that $h_+$ and $h_\times$ depend on, and thus encode, the chirp mass,
distance, the position on the sky (via the direction vector ${\bf
n}$), and the orientation of the binary's orbital plane (via $\hat{\bf
L}$).

\subsection{Nonlinear description of waves}
\label{sec:nonlinear}

The nonlinear nature of GR is one of its most important defining
characteristics.  By linearizing, perhaps we are throwing out the baby
with the bathwater, failing to characterize important aspects of
gravitational radiation.  Fortunately, we can derive a wave equation
that fully encodes all nonlinear features of GR.  This was apparently
first derived by {\cite{penrose60}}; {\cite{ryan74}} gives a very nice
discussion of this equation's history and derivation.  Begin by taking
an additional derivative $\nabla^\gamma$ of the Bianchi identity
(\ref{eq:bianchi}), obtaining
\begin{equation}
\Box_g R_{\alpha\beta\mu\nu} = -\nabla^\gamma \nabla_\beta
R_{\gamma\alpha\mu\nu} - \nabla^\gamma \nabla_\alpha
R_{\beta\gamma\mu\nu} \;,
\label{eq:nonlinwave1}
\end{equation}
where $\Box_g \equiv g^{\gamma\delta}\nabla_\gamma\nabla_\delta$ is a
covariant wave operator.  Next, use the fact that the commutator of
covariant derivatives generates a Riemann:
\begin{equation}
\left[\nabla_\gamma,\nabla_\delta\right]p_\mu \equiv
\left(\nabla_\gamma\nabla_\delta -
\nabla_\delta\nabla_\gamma\right)p_\mu = 
-{R^\sigma}_{\mu\gamma\delta}p_{\sigma}\;,
\end{equation}
\begin{equation}
\left[\nabla_\gamma,\nabla_\delta\right]p_{\mu\nu} = 
-{R^\sigma}_{\mu\gamma\delta}p_{\sigma\nu}
-{R^\sigma}_{\nu\gamma\delta}p_{\mu\sigma}
\;.
\end{equation}
Extension to further indices is hopefully obvious.  Manipulating
(\ref{eq:nonlinwave1}) yields a wave equation for Riemann in which
$(\mbox{Riemann})^2$ acts as the source term.  This is the Penrose
wave equation; see \cite{ryan74} for more details.

If spacetime is vacuum [$T_{\mu\nu} = 0$, so that via the Einstein
equation (\ref{eq:einstein}) $R_{\mu\nu} = 0$], the Penrose wave
equation simplifies quite a bit, yielding
\begin{equation}
\Box_g R_{\alpha\beta\mu\nu} =
2R_{\mu\sigma\beta\tau} {{{R_\nu}^\sigma}_\alpha}^\tau -
2R_{\mu\sigma\alpha\tau} {{{R_\nu}^\sigma}_\beta}^\tau +
R_{\mu\sigma\tau\sigma} {R^{\tau\sigma}}_{\alpha\beta}\;.
\label{eq:nonlinwave}
\end{equation}
A variant of Eq.\ (\ref{eq:nonlinwave}) underlies much of black hole
perturbation theory, our next topic.

\section{Perturbation theory}
\label{sec:pert}

Perturbation theory is the first technique we will discuss for
modeling strong-field merging compact binaries.  The basic concept is
to assume that the spacetime is an exact solution perturbed by a small
orbiting body, and expand to first order in the binary's mass ratio.
Some of the most interesting strong-field binaries have black holes,
so we will focus on black hole perturbation theory.  Perturbation
theory analysis of binaries has two important applications.  First, it
can be a limiting case of the pN expansion: Perturbation theory for
binaries with separations $r \gg GM/c^2$ should give the same result
as pN theory in the limit $\mu \ll M$.  We return to this point in
Sec.\ {\ref{sec:pn}}.  Second, perturbation theory is an ideal tool
for extreme mass ratio captures, binaries created by the scattering of
a stellar mass ($m \sim 1 - 100\,M_\odot$) body onto a strong-field
orbit of a massive ($M \sim 10^5 - 10^7\,M_\odot$) black hole.

\subsection{Basic concepts and overview of formalism}
\label{sec:perturb_formalism}

At its most basic, black hole perturbation theory is developed much
like the weak gravity limit described in Sec.\ {\ref{sec:waveform}},
replacing the flat spacetime metric $\eta_{\alpha\beta}$ with the
spacetime of a black hole:
\begin{equation}
g_{\mu\nu} = g_{\mu\nu}^{\rm BH} + h_{\mu\nu}\;.
\end{equation}
For astrophysical scenarios, one uses the Schwarzschild (non-rotating
black hole) or Kerr (rotating) solutions for $g_{\mu\nu}^{\rm BH}$.
It is straightforward (though somewhat tedious) to then develop the
Einstein tensor for this spacetime, keeping terms only to first order
in the perturbation $h$.

This approach works very well when the background is non-rotating,
\begin{equation}
(ds^2)^{\rm BH} = g_{\mu\nu}^{\rm BH} dx^\mu dx^\nu = -\left(1 -
\frac{2\hat M}{r}\right)dt^2 + \frac{dr^2}{\left(1 - 2\hat M/r\right)} +
r^2d\Omega^2\;,
\end{equation}
where $d\Omega^2 = d\theta^2 + \sin^2\theta d\phi^2$ and $\hat M =
GM/c^2$.  We consider this special case in detail; our discussion is
adapted from {\cite{rezzolla03}}.  Because the background is
spherically symmetric, it is useful to decompose the perturbation into
spherical harmonics.  For example, under rotations in $\theta$ and
$\phi$, $h_{00}$ should transform as a scalar.  We thus put
\begin{equation}
h_{00} = \sum_{lm} a_{lm}(t,r) Y_{lm}(\theta,\phi)\;.
\end{equation}
The components $h_{0i}$ transform like components of a 3-vector under
rotations, and can be expanded in vector harmonics; $h_{ij}$ can be
expanded in tensor harmonics.  One can decompose further with parity:
Even harmonics acquire a factor $(-1)^l$ when $(\theta,\phi) \to (\pi
- \theta, \phi + \pi)$; odd harmonics acquire a factor $(-1)^{l+1}$.

By imposing these decompositions, choosing a particular gauge, and
requiring that the spacetime satisfy the vacuum Einstein equation
$G_{\mu\nu} = 0$, we find an equation that governs the perturbations.
Somewhat remarkably, the $t$ and $r$ dependence for all components of
$h_{\mu\nu}$ for given spherical harmonic indices $(l,m)$ can be
constructed from a function $Q(t,r)$ governed by the simple equation
\begin{equation}
\frac{\partial^2 Q}{\partial t^2} - \frac{\partial^2 Q}{\partial
r_*^2} - V(r)Q = 0\;,
\label{eq:schwarz_pert}
\end{equation}
where $r_* = r + 2 \hat M \ln(r/2\hat M - 1)$.  The potential
$V(r)$ depends on whether we consider even or odd perturbations:
\begin{equation}
V_{\rm even}(r) = \left(1 - \frac{2\hat M}{r}\right) \left[\frac{2q(q +
    1)r^3 + 6q^2\hat M r^2 + 18 q\hat M^2 r + 18\hat M^3} {r^3\left(qr
    + 3\hat M\right)^2}\right]\;,
\label{eq:zerilli_pot}
\end{equation}
where $q = (l - 1)(l + 2)/2$; and
\begin{equation}
V_{\rm odd}(r) = \left(1 - \frac{2\hat M}{r}\right)
\left[\frac{l(l+1)}{r^2} - \frac{6\hat M}{r^3}\right]\;.
\label{eq:reggewheeler_pot}
\end{equation}
For even parity, Eq.\ (\ref{eq:schwarz_pert}) is the {\it Zerilli
equation} {\citep{zerilli70}}; for odd, it is the {\it Regge-Wheeler
equation} {\citep{rw57}}.  For further discussion, including how gauge
is chosen and how to construct $h_{\mu\nu}$ from $Q$, see
{\citet{rezzolla03}}.  Finally, note that when the spacetime
perturbation is due to a body orbiting the black hole, these equations
acquire a source term.  One can construct the full solution for the
waves from an orbiting body by using the source-free equation to build
a Green's function, and then integrating over that source.

How does this procedure fare for rotating holes?  The background
spacetime,
\begin{eqnarray}
(ds^2)^{\rm BH} &=& -\left(1 - \frac{2\hat Mr}{\rho^2}\right)dt^2 -
\frac{4 a\hat M r\sin^2\theta}{\rho^2}dt d\phi +
\frac{\rho^2}{\Delta}dr^2 + \rho^2d\theta^2
\nonumber\\
& & + \left(r^2 + a^2 + \frac{2\hat Mr
a^2\sin^2\theta}{\rho^2}\right)d\phi^2\;,
\label{eq:kerr_metric}
\end{eqnarray}
where
\begin{equation}
a = \frac{|\vec S|}{c M}\;,\qquad
\rho^2 = r^2 + a^2\cos^2\theta\;,\qquad
\Delta = r^2 - 2\hat M r + a^2\;,
\end{equation}
is now markedly nonspherical.  [We have used ``Boyer-Lindquist''
coordinates, but the nonspherical nature is independent of coordinate
choice.  The spin parameter $a$ has the dimension of mass, and must
satisfy $a \le M$ in order for Eq.\ (\ref{eq:kerr_metric}) to
represent a black hole.]  So, the decomposition into spherical
harmonics is not useful.  One could in principle simply expand
$G_{\mu\nu} = 0$ to first order in $h_{\mu\nu}$ and obtain a partial
differential equation in $t$, $r$, and $\theta$.  (The metric is
axially symmetric, so we can easily separate the $\phi$ dependence.)
This author is unaware of such a formulation\footnote{Since originally
posting this paper, I was informed by Plamen Fiziev that
{\cite{chandra83}} in fact develops equations to describe Kerr metric
perturbations, noting that they rather complicated and unwieldy, and
are rarely used.}.  One issue is that the gauge used for the
perturbation must be specified; this may be complicated in the general
case.  More important historically, the equations so developed do not
appear to separate.  As we'll see in a moment, a different approach
{\it does} yield separable equations, which were preferred for much of
the history of this field.

Rather than expanding the metric of the black hole, {\cite{teuk73}}
examined perturbations of its curvature:
\begin{equation}
R_{\alpha\mu\beta\nu} =
R^{\rm BH}_{\alpha\mu\beta\nu} +
\delta R_{\alpha\mu\beta\nu}\;.
\end{equation}
The curvature tensor is invariant to first-order gauge
transformations, an attractive feature.  This tensor also obeys the
nonlinear wave equation (\ref{eq:nonlinwave}).  By expanding that
equation to linear order in $\delta R_{\alpha\mu\beta\nu}$, Teukolsky
showed that perturbations to Kerr black holes are governed by the
equation
\begin{eqnarray}
&&
\!\!\!\!\!\!\!
\left[\frac{(r^2 + a^2)^2 }{\Delta} - a^2\sin^2\theta\right]
\partial^2_{t}\Psi - 4\left[r + ia\cos\theta - \frac{\hat M(r^2 -
a^2)}{\Delta}\right]\partial_t\Psi
\nonumber\\
&&
\!\!\!\!\!\!\!
+\frac{4i \hat M a m r}{\Delta}\partial_t\Psi -
\Delta^{2}\partial_r\left(\Delta^{-1}\partial_r\Psi\right)
- \frac{1}{\sin\theta}\partial_\theta
\left(\sin\theta\partial_\theta\Psi\right)
\nonumber\\
&&
\!\!\!\!\!\!\!
- \left[\frac{a^2}{\Delta} - \frac{1}{\sin^2\theta}\right]m^2 \Psi +
4im \left[\frac{a (r - \hat M)}{\Delta} + \frac{i
    \cos\theta}{\sin^2\theta} \right]\Psi - \left(4\cot^2\theta +
2\right) \Psi = {\cal T}\;.  \nonumber\\
\label{eq:teukolsky}
\end{eqnarray}
$\Psi$ is a complex quantity built from a combination of components of
$\delta R_{\alpha\mu\beta\nu}$, and describes spacetime's radiation;
see {\cite{teuk73}} for details.  (We have assumed $\Psi \propto
e^{im\phi}$.)  Likewise, ${\cal T}$ describes a source function built
from the stress-energy tensor describing a small body orbiting the
black hole.  Somewhat amazingly, Eq.\ (\ref{eq:teukolsky}) separates:
putting
\begin{equation}
\Psi = \int d\omega \sum_{lm} R_{lm}(r)S_{lm}(\theta)e^{im\phi -
i\omega t}
\label{eq:teuk_decomp}
\end{equation}
and applying a similar decomposition to the source ${\cal T}$, we find
that $S_{lm}(\theta)$ is a ``spin-weighted spheroidal harmonic'' (a
basis for tensor functions in a non-spherical background), and that
$R_{lm}(r)$ is governed by a simple ordinary differential equation.
$\Psi$ characterizes Kerr perturbations in much the same way that $Q$
[cf.\ Eq.\ (\ref{eq:schwarz_pert})] characterizes them for
Schwarzschild.  It's worth noting that, although the perturbation
equations are often solved numerically, analytic solutions are known
(\citealt{mst96}, \citealt{fiziev09}), and can be used to dramatically
improve one's scheme for solving for black hole perturbations.

Whether one uses this separation or solves Eq.\ (\ref{eq:teukolsky})
directly, solving for perturbations of black holes is now a
well-understood enterprise.  We now discuss how one uses these
solutions to model compact binaries.

\subsection{Binary evolution in perturbation theory}
\label{sec:evolve_perturbation}

How do we describe the motion of a small body about a black hole?  The
most rigorous approach is to enforce $\nabla^\mu T_{\mu\nu} = 0$,
where $T_{\mu\nu}$ describes the small body in the spacetime of the
large black hole.  If we neglect the small body's perturbation to the
spacetime, this exercise produces the geodesic equation $u^\mu
\nabla_\mu u^\nu = 0$, where $u^\mu$ is the small body's 4-velocity.
Geodesic black hole orbits have been studied extensively; see, for
example, {\cite{mtw}}, Chapter 33.  A key feature of these orbits is
that they are characterized (up to initial conditions) by three
conserved constants: energy $E$, axial angular momentum $L_z$, and
``Carter's constant'' $Q$.  If the black hole does not rotate,
Carter's constant is related to the orbit's total angular momentum:
$Q(a = 0) = {\bf L}\cdot{\bf L} - L_z^2$.  When the black hole rotates
rapidly, $Q$ is not so easy to interpret, but the idea that it is
essentially the rest of the orbit's angular momentum can be useful.

Now take into account perturbations from the small body.  Enforcing
$\nabla^\mu T_{\mu\nu} = 0$, we find that the small body follows a
``forced'' geodesic,
\begin{equation}
u^\mu \hat\nabla_\mu u^\nu = f^\nu\;,
\label{eq:selfforce_eom}
\end{equation}
where $\hat\nabla_\mu$ is the covariant derivative in the background
black-hole spacetime.  The novel feature of Eq.\
(\ref{eq:selfforce_eom}) is the {\it self force} $f^\nu$, a correction
to the motion of order the small body's spacetime perturbation.  The
self force is so named because it arises from the body's interaction
with its own spacetime correction.

Self forces have a long pedigree.  \citet{dirac38} showed that a self
force arises from the backreaction of an electromagnetic charge on
itself, and causes radiative damping.  Computing the gravitational
self force near a black hole is an active area of current research.
It is useful to break the self force into a {\it dissipative} piece,
$f^\nu_{\rm diss}$, which is asymmetric under time reversal, and a
{\it conservative} piece, $f^\nu_{\rm cons}$, which is symmetric.
These contributions have very different impact on the orbit.
Dissipation causes the ``conserved'' quantities $(E, L_z, Q)$ to
decay, driving inspiral of the small body.  {\cite{qw99}} have shown
that the rate at which $E$ and $L_z$ change due to $f^\nu_{\rm diss}$
is identical to what is found when one computes the fluxes of energy
and angular momentum encoded by the Isaacson tensor
(\ref{eq:isaacson_tmunu}).

The conservative self force, by contrast, does not cause orbit decay.
``Conserved'' constants are still conserved when we include this
force; but, the orbits are different from background geodesics.  This
reflects the fact that, even neglecting dissipation, the small body's
motion is determined by the full spacetime, not just the background
black hole.  When conservative effects are taken into account, one
finds that the orbital frequencies are shifted by an amount
\begin{equation}
\delta\Omega_x \sim \Omega_x \times (\mu/M)
\end{equation}
[where $x \in (\phi,\theta,r)$].  Because the GWs have spectral
support at harmonics of the orbital frequencies, these small but
non-negligible frequency shifts are directly encoded in the waves that
the binary generates.  Good discussion and a toy model can be found in
{\cite{ppn05}}.

To date, not a large amount of work is published regarding self forces
and conservative effects for Kerr orbits.  There has, however, been
enormous progress for the case of orbits around non-rotating holes.
{\cite{bs07}} have completed an analysis of the full self force for
circular orbits about a Schwarzschild black hole; generalization to
eccentric orbits is in progress (L.\ Barack, private communication).
An independent approach developed by {\cite{det08}} has been found to
agree with Barack and Sago extremely well; see {\cite{sbd08}} for
detailed discussion of this comparison.

\subsection{Gravitational waves from extreme mass ratio binaries}
\label{sec:emri_waves}

In this section, we discuss the properties of GWs and GW sources as
calculated using perturbation theory.  As discussed in Sec.\
(\ref{sec:relview}), these waves most naturally describe extreme mass
ratio capture sources.  There is also an important overlap with the pN
results discussed in Sec.\ {\ref{sec:pn}}: By specializing to
circular, equatorial orbits and considering the limit $r \gg GM/c^2$,
results from perturbation theory agree with pN results for $\mu/M \ll
1$.

Our goal here is to highlight features of the generic Kerr inspiral
waveform.  As such, we will neglect the conservative self force, which
is not yet understood for the Kerr case well enough to be applied to
these waves.  When conservative effects are neglected, the binary can
be regarded as evolving through a sequence of geodesics, with the
sequence determined by the rates at which GWs change the ``constants''
$E$, $L_z$, and $Q$.  Modeling compact binaries in this limit takes
three ingredients: First, a description of black hole orbits; second,
an algorithm to compute GWs from the orbits, and to infer how the
waves' backreaction evolves us from orbit to orbit; and third, a
method to integrate along the orbital sequence to build the full
waveform.  A description of this method is given in {\cite{hetal05}};
we summarize the main results of these three ingredients here.

\subsubsection{Black hole orbits.}
\label{sec:bhorbits}

\begin{figure}[t]
\includegraphics[width=5.3in]{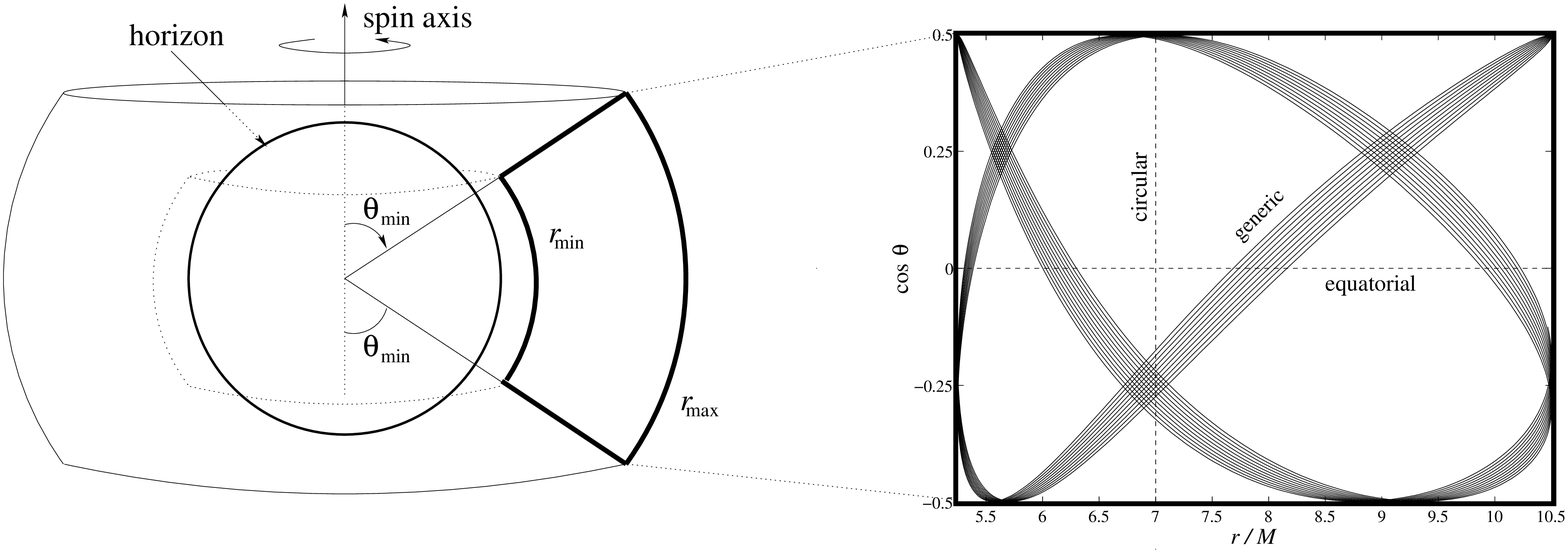}
\caption{The geometry of a generic Kerr black hole orbit [taken from
{\cite{dh06}}].  This orbit is about a black hole with spin parameter
$a = 0.998M$ (recall $a \le M$, so this represents a nearly maximally
spinning black hole).  The range of its radial motion is determined by
$p = 7GM/c^2$ ($G$ and $c$ are set to 1 in the figure) and $e = 1/3$;
$\theta$ ranges from $60^\circ$ to $120^\circ$.  The left panel shows
the torus in coordinate space this torus occupies.  The right panel
illustrates how a generic orbit ergodically fills this torus.}
\label{fig:torus}
\end{figure}

\begin{figure}[t]
\includegraphics[width=5in]{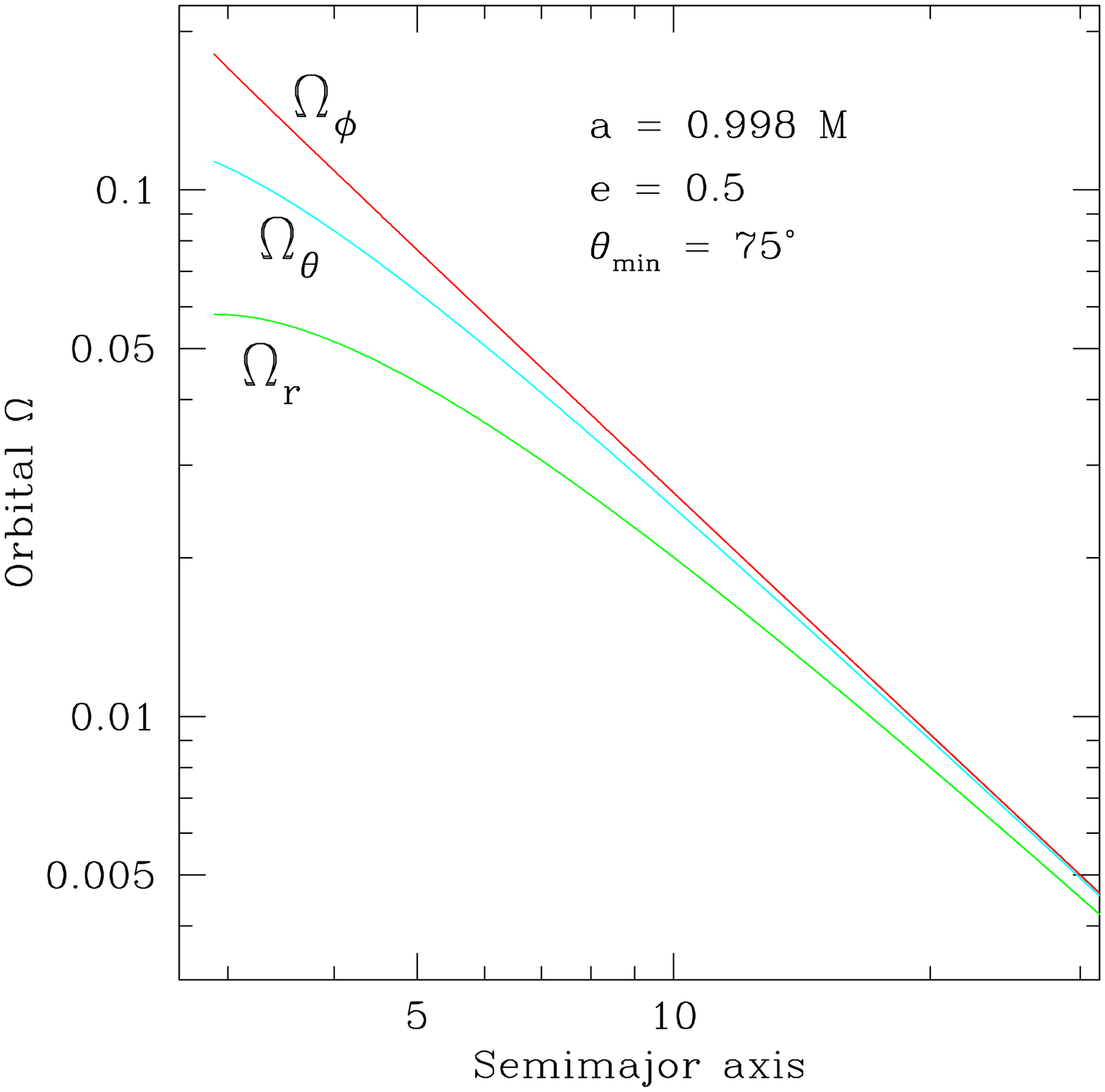}
\caption{Orbital frequencies for generic Kerr black hole orbits.  We
vary the orbits' semilatus rectum $p$, but fix eccentricity $e = 0.5$
and inclination parameter $\theta_{\rm min} = 75^\circ$.  Our results
are plotted as a function of semimajor axis $A = p/\sqrt{1 - e^2}$.
All three frequencies asymptote to the Keplerian value $\Omega =
\sqrt{GM/A^3}$ in the weak field, but differ significantly from each
other in the strong field.}
\label{fig:freqs}
\end{figure}

Motion in the vicinity of a black hole can be conveniently written in
the Boyer-Lindquist coordinates of Eq.\ (\ref{eq:kerr_metric}) as
$r(t)$, $\theta(t)$, and $\phi(t)$.  Because $t$ corresponds to time
far from the black hole, this gives a useful description of the motion
as measured by distant observers.  {\it Bound} black hole orbits are
confined to a region near the hole.  They have $r_{\rm min} \le r(t)
\le r_{\rm max}$ and $\theta_{\rm min} \le \theta(t) \le \pi -
\theta_{\rm min}$.  Bound orbits thus occupy a torus in the 3-space
near the hole's event horizon; an example is shown in Fig.\
{\ref{fig:torus}}, taken from {\cite{dh06}}.  Selecting the orbital
constants $E$, $L_z$, and $Q$ fully determines $r_{\rm min/max}$ and
$\theta_{\rm min}$.  It is useful for some discussions to
reparameterize the radial motion, defining an eccentricity $e$ and a
semi-latus rectum $p$ via
\begin{equation}
r_{\rm min} = \frac{p}{1 + e}\;,\qquad
r_{\rm max} = \frac{p}{1 - e}\;.
\end{equation}
For many bound black hole orbits, $r(t)$, $\theta(t)$, and $\phi(t)$
are periodic ({\citealt{schmidt02}}; see also {\citealt{dh04}}).
(Exceptions are orbits which plunge into the black hole; we discuss
these below.)  Near the hole, the time to cover the full range of $r$
becomes distinct from the time to cover the $\theta$ range, which
becomes distinct from the time to cover $2\pi$ radians of azimuth.
One can say that spacetime curvature splits the Keplerian orbital
frequency $\Omega$ into $\Omega_r$, $\Omega_\theta$, and
$\Omega_\phi$.  Figure {\ref{fig:freqs}} shows these three
frequencies, plotted as functions of semi-major axis $A$ for fixed
values of $e$ and $\theta_{\rm min}$.  Notice that all three approach
$\Omega \propto A^{-3/2}$ for large $A$.

\subsubsection{Gravitational radiation from orbits.}
\label{sec:orbitwaves}

\begin{figure}[t]
\includegraphics[width=5in]{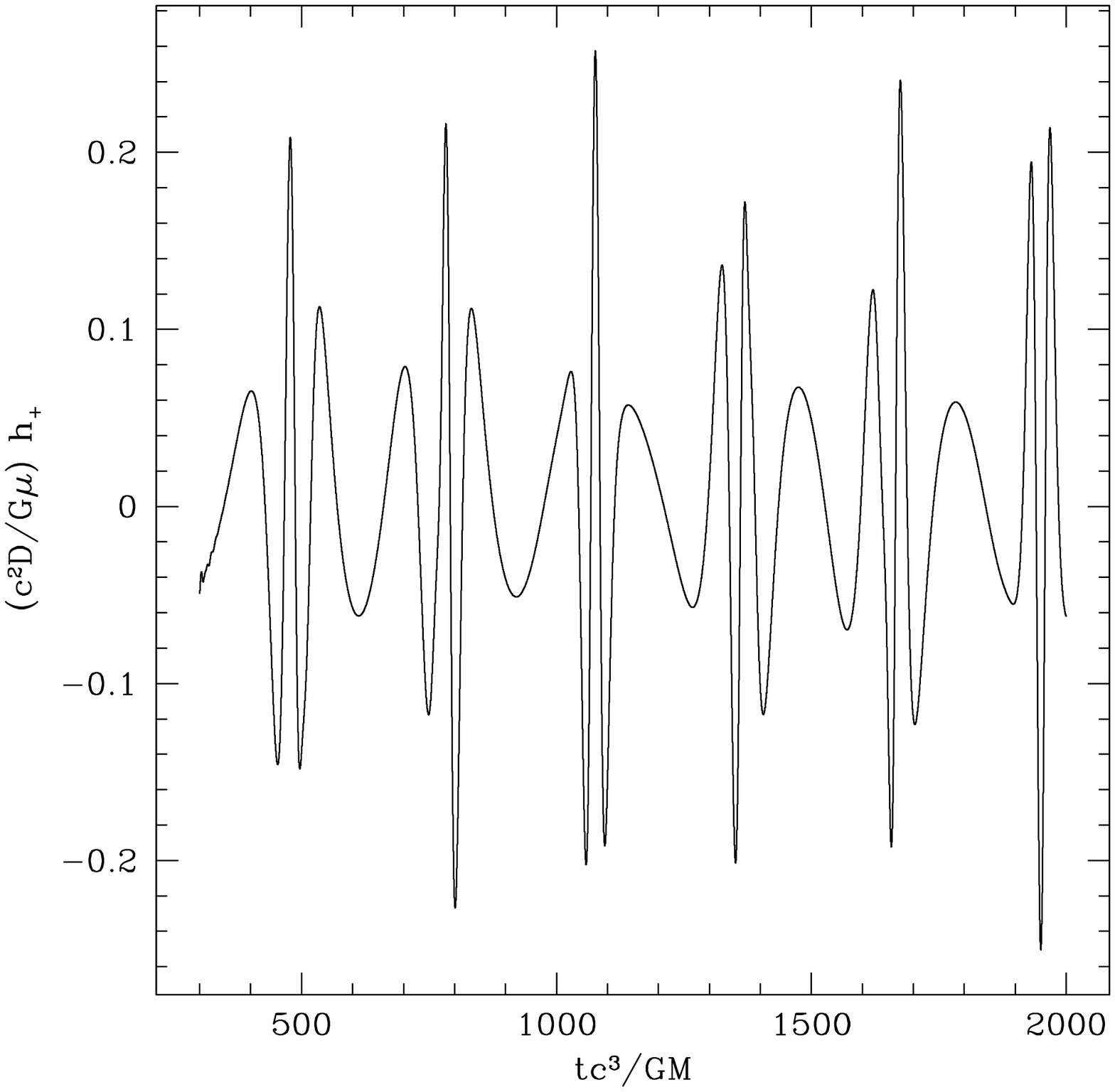}
\caption{Waveform arising from a generic geodesic black hole orbit,
neglecting orbital evolution due to backreaction.  This orbit has $p =
8GM/c^2$ and $e = 0.5$, corresponding to motion in the range
$16GM/3c^2 \le r(t) \le 16GM/c^2$; it also has $\theta_{\rm min} =
60^\circ$.  The large black hole has a spin parameter $a = 0.9M$.
Note that the wave has structure at several timescales, corresponding
to the three frequencies $\Omega_r$, $\Omega_\theta$, and
$\Omega_\phi$ (cf.\ Fig.\ {\ref{fig:freqs}}).}
\label{fig:genwave}
\end{figure}

Because their orbits are periodic, GWs from a body orbiting a black
hole will have support at harmonics of the orbital frequencies.  One
can write the two polarizations
\begin{equation}
h_+ + i h_\times = \sum H_{mkn}
e^{i\omega_{mkn}t}\;,\qquad\mbox{where}
\label{eq:fd_waveform}
\end{equation}
\begin{equation}
\omega_{mkn} = m\Omega_\phi + k\Omega_\theta + n\Omega_r\;.
\label{eq:harmonics}
\end{equation}
The amplitude $H_{mkn}$ can be found by solving the Teukolsky equation
(\ref{eq:teukolsky}) using the decomposition (\ref{eq:teuk_decomp});
details for the general case can be found in {\citet{dh06}}.  An
example of a wave from a geodesic orbit is shown in Fig.\
{\ref{fig:genwave}}.  Note the different timescales apparent in this
wave; they are due to the three distinct frequencies of the underlying
geodesic orbit (and their harmonics).

The expansion (\ref{eq:fd_waveform}) does not work well for orbits
that plunge into the black hole; those orbits are not periodic, and
cannot be expanded using a set of real frequencies.  A better way to
calculate those waves is to solve the Teukolsky equation
(\ref{eq:teukolsky}) {\it without} introducing the decomposion
(\ref{eq:teuk_decomp}).  Results for waves from plunging orbits in the
language of perturbation theory was first given by {\cite{ndt07}};
{\cite{sundararajan08}} has recently extended the cases that we can
model to full generality.

\begin{figure}[t]
\includegraphics[width=5in]{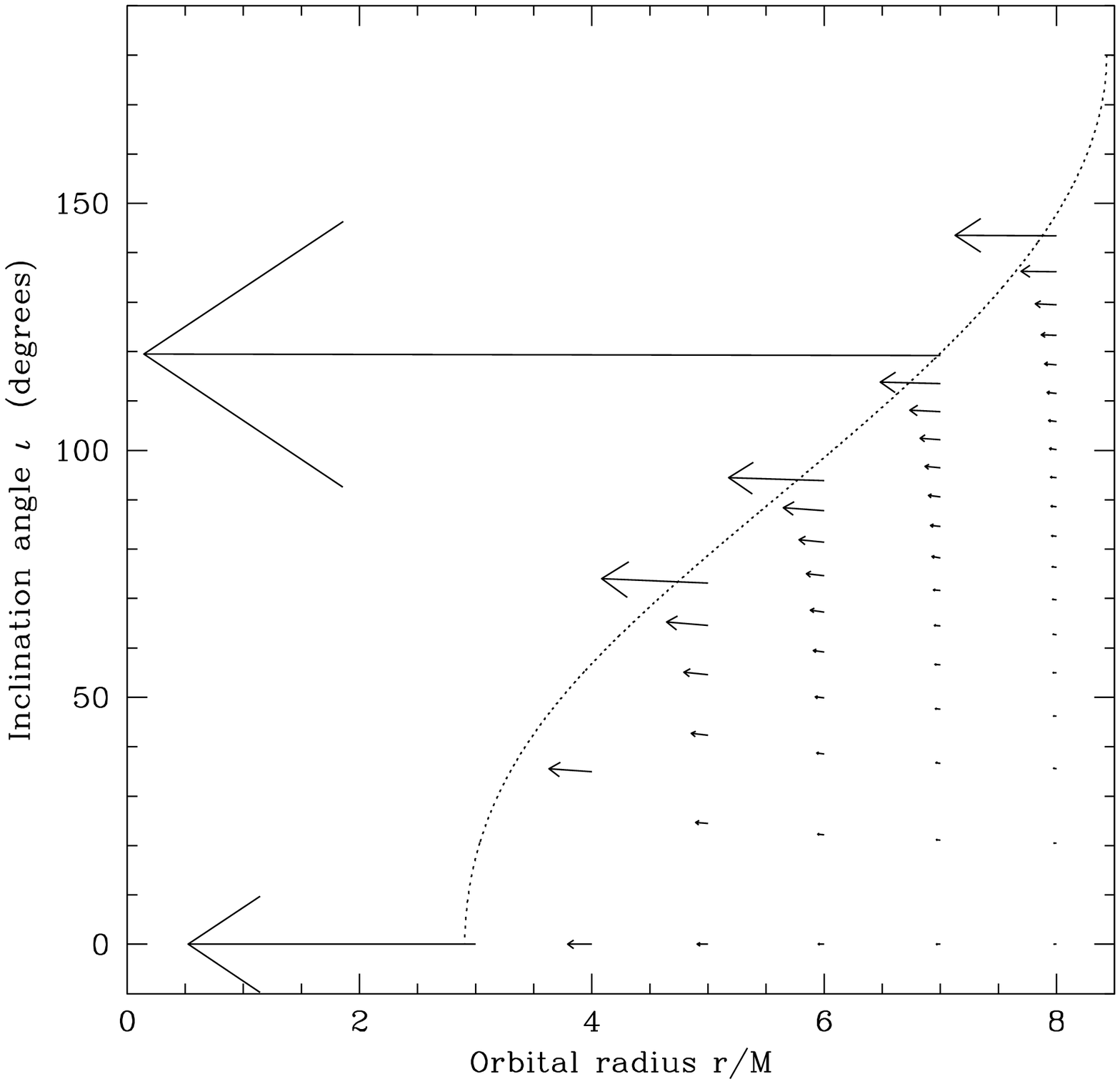}
\caption{The evolution of circular orbits ($e = 0$) about a black hole
with $a = 0.8M$; taken from {\cite{hughes00}}.  The inclination angle
$\iota$ is given by $\iota \simeq \pi/2 - \theta_{\rm min}$; the
equality is exact for $\theta_{\rm min} = 0$ and for $a = 0$.  In the
general case, this relation misestimates $\iota$ by $\lesssim 3\%$;
see {\cite{hughes00}} for detailed discussion.  The dotted line is
this hole's ``last stable orbit''; orbits to the left are unstable to
small perturbations, those to the right are stable.  Each arrow shows
how radiation tends to evolve an orbit; length indicates how strongly
it is driven.  These orbits are driven to smaller radius and to (very
slightly) larger inclination.  The extremely long arrow at $\iota
\simeq 120^\circ$, $r = 7 GM/c^2$ lies very close to the last stable
orbit.  As such, a small push from radiation has a large impact.}
\label{fig:circ_evol}
\end{figure}

As mentioned in Sec.\ {\ref{sec:evolve_perturbation}}, it is fairly
simple to compute the flux of energy $\dot E$ and angular momentum
$\dot L_z$ from the Isaacson tensor, Eq.\ (\ref{eq:isaacson_tmunu}),
once the waves are known.  Recent work {\citep{ganz07}} has shown that
a similar result describes $\dot Q$.  Once $\dot E$, $\dot L_z$, and
$\dot Q$ are known, it is straightforward to evolve the orbital
elements $r_{\rm min/max}$ and $\theta_{\rm min}$, specifying the
sequence of orbits through which gravitational radiation drives the
system.  Figure {\ref{fig:circ_evol}} gives an example of how orbits
evolve when their eccentricity is zero.

\subsubsection{Evolving through an orbital sequence.}
\label{sec:evolve}

\begin{figure}[t]
\includegraphics[width=5.3in]{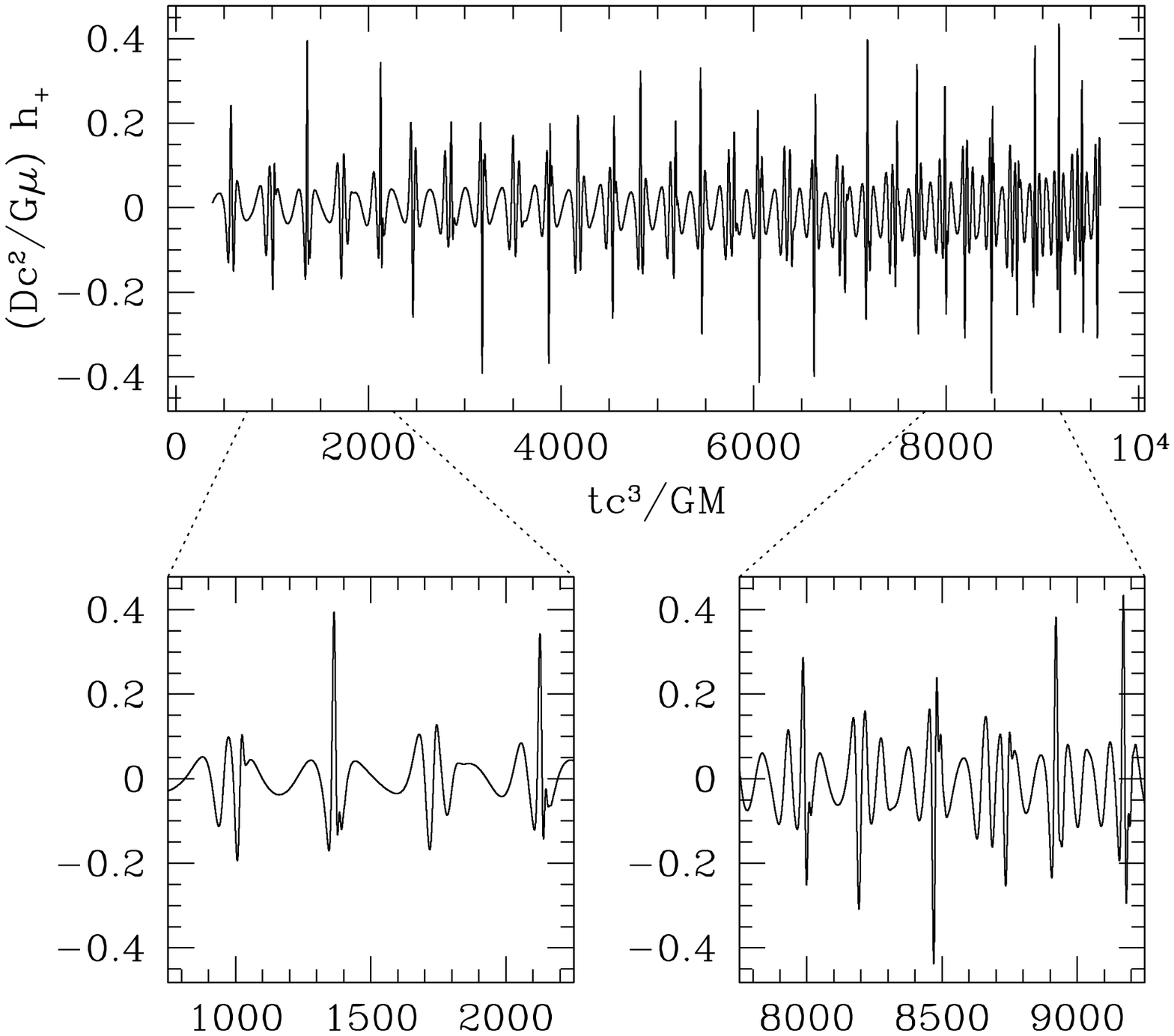}
\caption{Plus polarization of wave generated by a small body spiraling
into a massive black hole; this figure is adapted from
{\cite{pkhd08}}.  The amplitude is scaled to the source's distance $D$
and the small body's mass $\mu$; time is measure in units $c^3/GM$.
For this calculation, the binary's initial parameters are $p =
10GM/c^2$, $e = 0.5$, and $\theta_{\rm min} \simeq 61^\circ$; the
binary's mass ratio is fixed to $\mu/M = 0.016$, and the larger black
hole's spin parameter is $a = 0.5M$.  The insets show spans of length
$\Delta t \sim 1000 GM/c^3$ early and late in the inspiral.  Note the
substantial evolution of the wave's frequencies as the orbit shrinks.}
\label{fig:td_teuk}
\end{figure}

It is not too difficult to compute the sequence of orbits
[parameterized as $E(t)$, $L_z(t)$, $Q(t)$ or $r_{\rm min/max}(t)$,
$\theta_{\rm min}(t)$] that an inspiraling body passes through before
finally plunging into its companion black hole.  Once these are known,
it is straightforward to build the worldline that a small body follows
as it spirals into the black hole.  From the worldline, we can build a
source function ${\cal T}(t)$ for Eq.\ (\ref{eq:teukolsky}) and
compute the evolving inspiral waves.  Figure {\ref{fig:td_teuk}} gives
an example of a wave arising from the inspiral of a small body into a
black hole; see {\cite{pkhd08}} for details of how these waves are
computed.

\section{Post-Newtonian theory}
\label{sec:pn}

Suppose we cannot use mass ratio as an expansion parameter.  For
instance, if the members of the binary are of equal mass, then $\eta
\equiv m_1 m_2/(m_1 + m_2)^2 = 0.25$.  This is large enough that
neglect of ${\cal O}(\eta^2)$ and higher terms is problematic.  The
techniques discussed in Sec.\ {\ref{sec:pert}} will not be
appropriate.

If the mass ratio is not a good expansion parameter, the potential
$\phi \equiv GM/rc^2$ may be.  The {\it post-Newtonian} (pN) expansion
of GR results when we use $\phi$ as our expansion parameter.  We now
summarize the main concepts which underlie the pN formalism, turning
next to a discussion of the pN waveform and its interesting features.
Much of our discussion is based on {\cite{blanchet06}}.

\subsection{Basic concepts and overview of formalism}
\label{sec:pn_formal}

One typically begins the pN expansion by examining the Einstein field
equations in {\it harmonic} or deDonder coordinates (e.g.,
{\citealt{weinberg72}}, Sec.\ 7.4).  In these coordinates, one defines
\begin{equation}
h^{\mu\nu} \equiv \sqrt{-g}g^{\mu\nu} - \eta^{\mu\nu}\;,
\label{eq:h_harmonic}
\end{equation}
where $g$ is the determinant of $g_{\mu\nu}$.  This looks similar to
the flat spacetime perturbation defined in Sec.\ {\ref{sec:waveform}};
however, we do not assume that $h$ is small.  We next impose the gauge
condition
\begin{equation}
\partial_\alpha h^{\alpha\beta} = 0\;.
\end{equation}
With these definitions, the {\it exact} Einstein field equations are
\begin{equation}
\Box h^{\alpha\beta} = \frac{16\pi G}{c^4}\tau^{\alpha\beta}\;,
\label{eq:pn_efe}
\end{equation}
where $\Box = \eta^{\alpha\beta}\partial_\alpha\partial_\beta$ is the
{\it flat} spacetime wave operator.  The form of Eq.\
(\ref{eq:pn_efe}) means that the radiative Green's function we used to
derive Eq.\ (\ref{eq:lin_soln}) can be applied here as well; the
solution is simply
\begin{equation}
h^{\alpha\beta} = -\frac{4G}{c^4}\int \frac{\tau_{\alpha\beta}({\bf x}',
t - |{\bf x} - {\bf x}'|/c)}{|{\bf x} - {\bf x}'|}d^3x'\;. 
\label{eq:pn_formal_soln}
\end{equation}

Formally, Eq.\ (\ref{eq:pn_formal_soln}) is exact.  We have swept some
crucial details under the rug, however.  In particular, we never
defined the source $\tau^{\alpha\beta}$.  It is given by
\begin{equation}
\tau^{\alpha\beta} = (-g)T^{\alpha\beta} +
\frac{c^4\Lambda^{\alpha\beta}}{16\pi G}\;,
\label{eq:pn_tau_def}
\end{equation}
where $T^{\alpha\beta}$ is the usual stress energy tensor, and
$\Lambda^{\alpha\beta}$ encodes much of the nonlinear
structure of the Einstein field equations:
\begin{eqnarray}
\Lambda^{\alpha\beta} &\equiv& 16\pi(-g)t^{\alpha\beta}_{\rm LL} +
\partial_\nu h^{\alpha\mu}\partial_\mu h^{\beta\nu} -
\partial_\mu \partial_\nu h^{\alpha\beta} h^{\mu\nu}
\\
&=& N^{\alpha\beta}[h,h] + M^{\alpha\beta}[h,h,h]
+ L^{\alpha\beta}[h,h,h,h] + {\cal O}(h^5)\;.
\end{eqnarray}
On the first line, $t^{\alpha\beta}_{\rm LL}$ is the Landau-Lifshitz
pseudotensor, a quantity which (in certain gauges) allows us to
describe how GWs carry energy through spacetime (\citealt{ll75}, Sec.\
96).  On the second line, the term $N^{\alpha\beta}[h,h]$ means a
collection of terms quadratic in $h$ and its derivatives,
$M^{\alpha\beta}[h,h,h]$ is a cubic term, etc.  Our solution
$h^{\alpha\beta}$ appears on both the left- and right-hand sides of
Eq.\ (\ref{eq:pn_formal_soln}).  Such a structure can be handled very
well {\it iteratively}.  We write
\begin{equation}
h^{\alpha\beta} = \sum_{n = 1}^\infty G^n h_n^{\alpha\beta}\;.
\label{eq:iteration}
\end{equation}
The $n = 1$ term is essentially the linearized solution from Sec.\
{\ref{sec:waveform}}.  To go higher, let $\Lambda_n^{\alpha\beta}$
denote the contribution of $\Lambda^{\alpha\beta}$ to the solution
$h_n^{\alpha\beta}$.  We find
\begin{equation}
\Lambda_2^{\alpha\beta} = N^{\alpha\beta}[h_1,h_1]\;,
\end{equation}
\begin{equation}
\Lambda_3^{\alpha\beta} = M^{\alpha\beta}[h_1,h_1,h_1] +
N^{\alpha\beta}[h_2,h_1] + N^{\alpha\beta}[h_1,h_2]\;,
\end{equation}
etc.; higher contributions to $\Lambda^{ab}$ can be found by expanding
its definition and gathering terms.  By solving the equations which
result from this procedure, it becomes possible to build the spacetime
metric and describe the motion of the members of a binary and the
radiation that they emit.

\subsection{Features of the post-Newtonian binary waveform}
\label{sec:pn_waveform}

The features of the pN binary waveform are most naturally understood
by first considering how we describe the motion of the members of the
binary.  Take those members to have masses $m_1$ and $m_2$, let their
separation be $r$, and let $\mathbf{\hat r}$ point to body 1 from body
2.  Then, in the harmonic gauge used for pN theory, the acceleration
of body 1 due to the gravity of body 2 is
\begin{equation}
{\bf a} = {\bf a}_0 + {\bf a}_2 + {\bf a}_4 + {\bf a}_5 + {\bf a}_6 +
{\bf a}_7 \ldots \;.
\label{eq:pNorbitaccel}
\end{equation}
The zeroth term,
\begin{equation}
{\bf a}_0 = -\frac{G m_2}{r^2} \mathbf{\hat r},
\end{equation}
is just the usual Newtonian gravitational acceleration.  Each ${\bf
a}_n$ is a pN correction of order $(v/c)^n$.  The first such
correction is
\begin{equation}
{\bf a}_2 = \left[\frac{5G^2m_1m_2}{r^3} + \frac{4G^2m_2^2}{r^3} +
\frac{Gm_2}{r^2} \left(\frac{3}{2}({\mathbf{\hat r}}\cdot{\bf v_2})^2
- v_1^2 + 4{\bf v_1}\cdot{\bf v_2} -
2v_2^2\right)\right]\frac{\mathbf{\hat r}}{c^2}\;.
\label{eq:pN_a2}
\end{equation}
For the acceleration of body 2 due to body 1, exchange labels 1 and 2
and replace $\mathbf{\hat r}$ with $-\mathbf{\hat r}$.  Note that
${\bf a}_2$ changes the dependence of the acceleration with respect to
orbital separation.  It also shows that the acceleration of body 1
depends on its mass $m_1$.  This is a pN manifestation of the ``self
force'' discussed in Sec.\ {\ref{sec:evolve_perturbation}}.  So far,
the pN acceleration has been computed to order $(v/c)^7$.  As we go to
high order, the expressions for ${\bf a}_n$ become quite lengthy.  An
excellent summary is given in {\cite{blanchet06}}, Eq.\ (131) and
surrounding text.  (Note that the expression for ${\bf a}_6$ fills
over two pages in that paper!)

At higher order, we also find a distinctly non-Newtonian element to
binary dynamics: its members spins {\it precess} due to their motion
in the binary's curved spacetime.  If the spins are ${\bf S}_1$ and
${\bf S}_2$, one finds {\citep{th85}}
\begin{equation}
\frac{d{\bf S}_1}{dt} = \frac{G}{c^2r^3}\left[\left(2 +
\frac{3}{2}\frac{m_2}{m_1}\right)\mu\sqrt{M r}\hat{\bf L}\right]
\times{\bf S}_1 + \frac{G}{c^2r^3}\left[\frac{1}{2}{\bf S}_2 -
\frac{3}{2}({\bf S}_2\cdot\hat{\bf L})\hat{\bf L}\right] \times{\bf
S}_1\;,
\label{eq:dS1dt}
\end{equation}
\begin{equation}
\frac{d{\bf S}_2}{dt} = \frac{G}{c^2r^3}\left[\left(2 +
\frac{3}{2}\frac{m_1}{m_2}\right)\mu\sqrt{M r}\hat{\bf L}\right]
\times{\bf S}_2 + \frac{G}{c^2r^3}\left[\frac{1}{2}{\bf S}_1 -
\frac{3}{2}({\bf S}_1\cdot\hat{\bf L})\hat{\bf L}\right] \times{\bf
S}_2\;.
\label{eq:dS2dt}
\end{equation}

We now discuss the ways in which aspects of pN binary dynamics color a
system's waves.

\subsubsection{Gravitational-wave amplitudes.}
\label{sec:pn_amplitude}

\begin{figure}[t]
\includegraphics[width=5.3in]{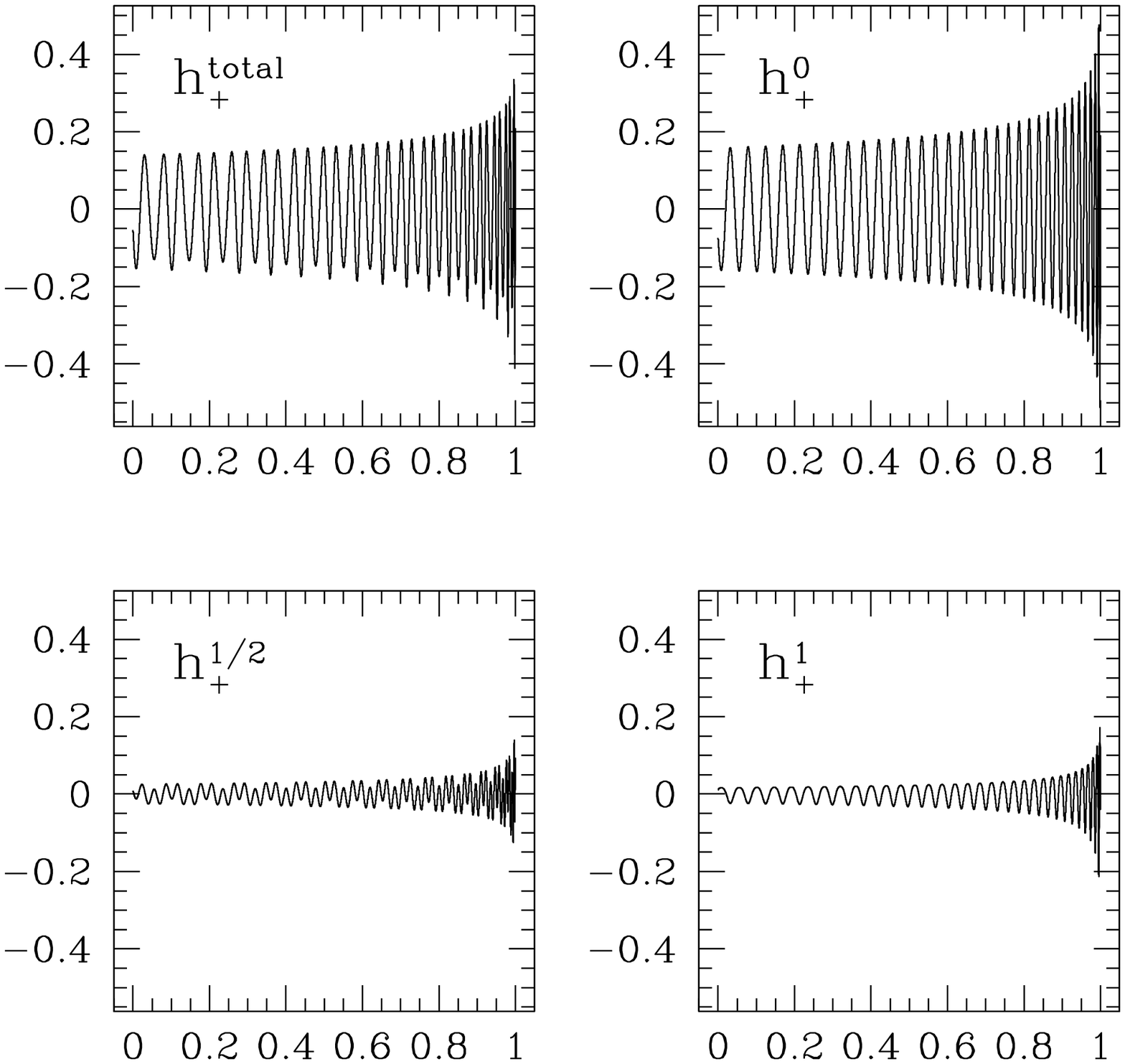}
\caption{The first three contributions to the $+$ GW polarization, and
their sum.  In all panels, we plot $(c^2D/G\mu)h_+$ versus $c^3t/GM$.
The upper left panel gives the sum [cf.\ Eq.\ (\ref{eq:hpn_sum})]
arising from $H^0_+$, $H^{1/2}_+$, and $H^1_+$; the other panels show
the individual contributions from those $H^n_+$.  Although
subdominant, the terms other than $H^0_+$ make a substantial
contribution to the total, especially at the end of inspiral (here
normalized to $c^3t/GM = 1$).}
\label{fig:hpn}
\end{figure}

Although a binary's {\it dominant} waves come from variations in its
mass quadrupole moment, Eq.\ (\ref{eq:multipolar_form}) shows us that
higher moments also generate GWs.  In the pN framework, these moments
contribute to the amplitude of a binary's waves beyond the quadrupole
form, Eq.\ (\ref{eq:h_NQ}).  Write the gravitational waveform from a
source as
\begin{equation}
h_{+,\times} = \frac{2G{\cal M}}{c^2D}\left(\frac{\pi G{\cal
M}f}{c^3}\right)^{2/3} \left[H^0_{+,\times} +
v^{1/2}H^{1/2}_{+,\times} + v H^1_{+,\times} + \ldots\right] \;,
\label{eq:hpn_sum}
\end{equation}
where $v \equiv (\pi G M f/c^3)^{1/3}$ is roughly the orbital speed of
the binary's members (normalized to $c$).  The contributions
$H^0_{+,\times}$ reproduce the waveform presented in Eq.\
(\ref{eq:h_NQ}).  The higher-order terms $H^{1/2}_{+,\times}$ and
$H^1_{+,\times}$ can be found in \cite{blanchet06}, his Eqs.\ (237)
through (241).  For our purposes, the key point to note is that these
higher-order terms introduce new dependences on the binary's orbital
inclination and its masses.  As such, measurement of these terms can
provide additional constraints to help us understand the system's
characteristics.  Figure {\ref{fig:hpn}} illustrates the three
contributions $H_0$, $H_{1/2}$, and $H_1$ to a binary's GWs.

\subsubsection{Orbital phase.}
\label{sec:pn_phase}

The motion of a binary's members about each other determines the
orbital phase.  Specializing to circular orbits, we can determine the
orbital frequency from the acceleration of the binary's members;
integrating up this frequency, we define the binary's phase $\Phi(t)$.
The first few terms of this phase are given by {\citep{bdiww95}}
\begin{eqnarray}
\Phi &=& \Phi_c - \left[\frac{c^3(t_c - t)}{5G{\cal M}}\right]^{5/8}
\left[1 + \left(\frac{3715}{8064} +
\frac{55}{96}\frac{\mu}{M}\right)\Theta^{-1/4} -\frac{3}{16}\left[4\pi
- \beta(t)\right]\Theta^{-3/8} \right.  \nonumber\\ & & \left.+
\left(\frac{9275495}{14450688} + \frac{284875}{258048}\frac{\mu}{M} +
\frac{1855}{2048}\frac{\mu^2}{M^2} + \frac{15}{64}\sigma(t)\right)
\Theta^{-1/2}\right]\;,
\label{eq:2pnPhase}
\end{eqnarray}
where
\begin{equation}
\Theta = \frac{c^3\eta}{5 G M}(t_c - t)\;.
\end{equation}
Notice that the leading term is just the Newtonian quadrupole phase,
Eq.\ (\ref{eq:phi_NQ}).  Each power of $\Theta$ connects to a higher
order in the pN; Eq.\ (\ref{eq:2pnPhase}) is taken to ``second
post-Newtonian'' order, which means that corrections of $(v/c)^4$ are
included.  Corrections to order $(v/c)^6$ are summarized in
{\cite{blanchet06}}.  In addition to the chirp mass ${\cal M}$, the
reduced mass $\mu$ enters $\Phi$ when higher order terms are included.
The high pN terms encode additional information about the binary's
masses.  At least in principle, including higher pN effects in our
wave model makes it possible to determine both chirp mass and reduced
mass, fully constraining the binary's masses.

Equation (\ref{eq:2pnPhase}) also depends on two parameters, $\beta$
and $\sigma$, which come from the binary's spins and orbit
orientation.  The ``spin-orbit'' parameter $\beta$ is
\begin{equation}
\beta = \frac{1}{2}\sum_{i = 1}^2\left[113\left(\frac{m_i}{M}\right)^2 +
75\eta\right]\frac{\hat{\bf L}\cdot{\bf S}_i}{m_i^2}\;;
\label{eq:beta_def}
\end{equation}
the ``spin-spin'' parameter $\sigma$ is
\begin{equation}
\sigma = \frac{\eta}{48m_1^2m_2^2}\left[721(\hat{\bf L}\cdot{\bf S}_1)
(\hat{\bf L}\cdot{\bf S}_2) - 247{\bf S}_1\cdot{\bf S}_2\right]\;
\label{eq:sigma_def}
\end{equation}
{\citep{bdiww95}}.  As we'll see in Sec.\ {\ref{sec:gwastro}}, these
parameters encode valuable information, especially when spin
precession is taken into account.

The $\mu \ll M$ limit of Eq.\ (\ref{eq:2pnPhase}) can be computed with
black hole perturbation theory (Sec.\ {\ref{sec:pert}}) evaluated for
circular orbits with $r \gg GM/c^2$.  The orbital phase is found by
integrating the orbital frequency.  By changing variables, one can
relate this to the orbital energy and the rate at which GWs evolve
this energy:
\begin{equation}
\Phi = \int\Omega^{\rm orb}\,dt = \int \frac{dE^{\rm
orb}/d\Omega}{dE^{\rm GW}/dt} \Omega\,d\Omega\;.
\end{equation}
The orbital energy $E^{\rm orb}$ is simple to calculate and to express
as a function of orbital frequency.  For example, for orbits of
non-rotating black holes, we have
\begin{equation}
E^{\rm orb} = \mu c^2\frac{1 - 2v^2/c^2}{\sqrt{1 -
3v^2/c^2}}\;,
\end{equation}
where $v \equiv r\Omega$.  For circular, equatorial orbits,
{\cite{msstt97}} {\it analytically} solve the Teukolsky equation as an
expansion in $v$, calculating $dE^{\rm GW}/dt$ to ${\cal
O}[(v/c)^{11}]$.  This body of work confirms, in a completely
independent way, all of the terms which do not depend on the mass
ratio $\mu/M$ in Eq.\ (\ref{eq:2pnPhase}).  The fact that these terms
are known to such high order is an important input to the effective
one-body approach described in Sec.\ {\ref{sec:eff_one_body}}.

\subsubsection{Spin precession.}
\label{sec:pn_precession}

\begin{figure}[t]
\includegraphics[width=5.3in]{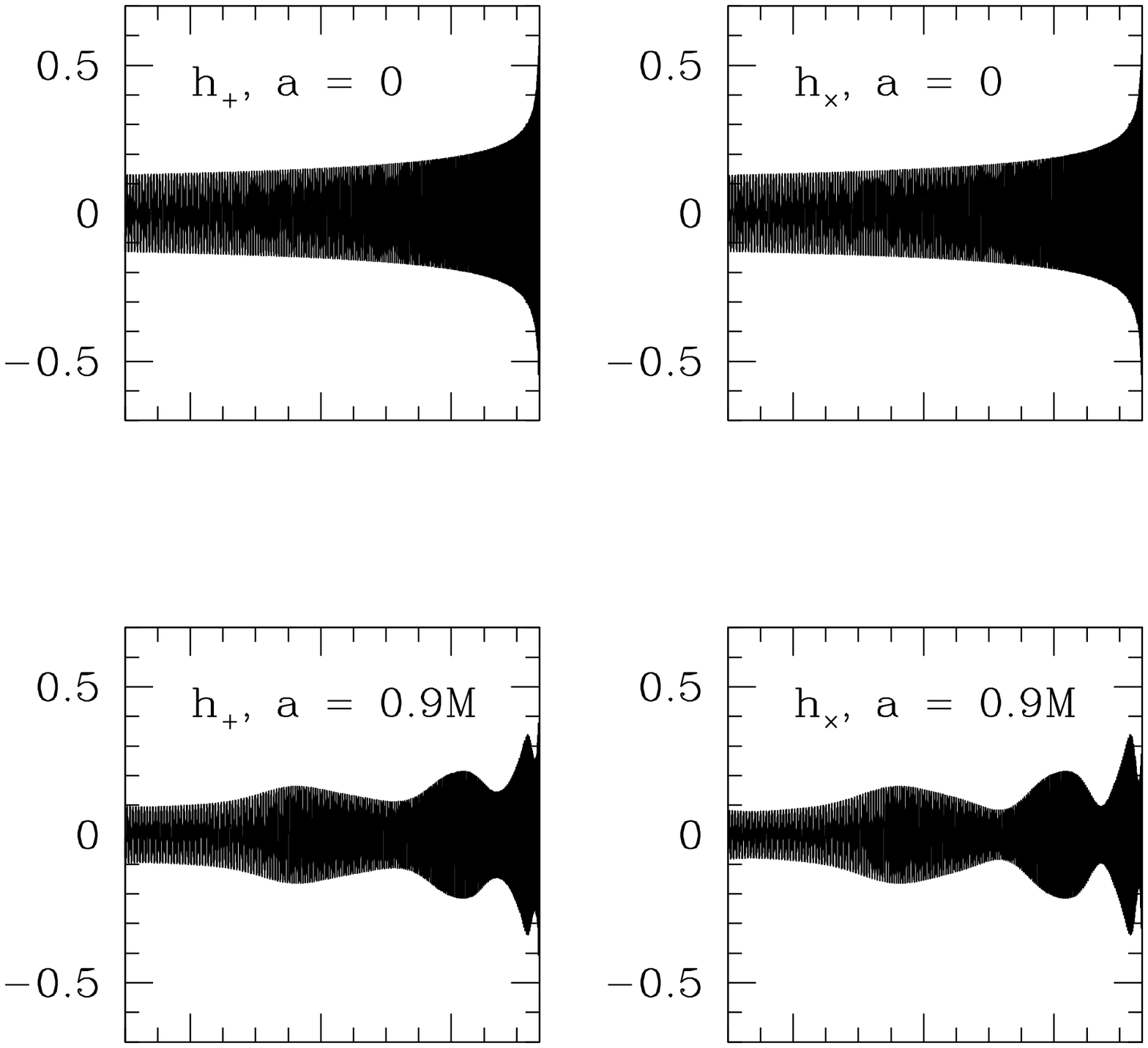}
\caption{Illustration of precession's impact on a binary's waves.  The
top panels show $h_+$ and $h_\times$ for a binary that contains
nonspinning black holes; the lower panels show the waveforms for a
binary with rapid rapidly rotating ($a = 0.9M$) holes.  The strong
amplitude modulation is readily apparent in this figure.  Less
obvious, but also included, is the frequency modulation that enters
through the spin-dependent orbital phase parameters $\beta$ and
$\sigma$ [cf.\ Eq.\ (\ref{eq:2pnPhase})].}
\label{fig:prec}
\end{figure}

Although the spin vectors ${\bf S}_1$ and ${\bf S}_2$ wiggle around
according to the prescription of Eqs.\ (\ref{eq:dS1dt}) and
(\ref{eq:dS2dt}), the system must preserve a notion of {\it global}
angular momentum.  Neglecting for a moment the secular evolution of
the binary's orbit due to GW emission, pN encodes the notion that the
total angular momentum
\begin{equation}
{\bf J} = {\bf L} + {\bf S}_1 + {\bf S}_2
\end{equation}
must be conserved.  This means ${\bf L}$ must oscillate to compensate
for the spins' dynamics, and guarantees that, when spin precession is
accounted for in our evolutionary models, the phase parameters $\beta$
and $\sigma$ become time varying.  Likewise, the inclination angle
$\iota$ varies with time.  Precession thus leads to phase and
amplitude modulation of a source's GWs.  Figure {\ref{fig:prec}}
illustrates precession's impact, showing the late inspiral waves for
binaries that are identical aside from spin.

\subsection{The effective one-body approach}
\label{sec:eff_one_body}

Because pN techniques are based on an expansion in $\phi = GM/rc^2$,
it had been thought that they would only apply for $r \gtrsim
10GM/c^2$, and that numerical relativity would be needed to cover the
inspiral past that radius, through the final plunge and merger.  This
thinking was radically changed by {\cite{bd99}}, which introduced the
{\it effective one-body} approach to two-body dynamics.  This
technique has proved to be an excellent tool for describing the late
inspiral, plunge, and merger of two black holes.  We now describe the
key ideas of this approach; our description owes much to the helpful
introductory lectures by {\cite{damour_eob08}}.

As the name suggests, the key observation of this approach is that the
motion of two bodies $(m_1, m_2)$ about one another can be regarding
as the motion of a single test body of mass $\mu = m_1m_2/(m_1 + m_2)$
in some spacetime.  One begins by examining the Hamiltonian which
gives the conservative contribution to the equations of motion.  Let
the binary's momenta be ${\bf p}_{1,2}$ and its generalized positions
${\bf q}_{1,2}$.  If we work in the center of mass frame, then the
Hamiltonian can only be a function of the {\it relative} position,
${\bf q} \equiv {\bf q}_1 - {\bf q}_2$, and can only depend on the
momentum ${\bf p} \equiv {\bf p}_1 = -{\bf p}_2$.  For example, the
conservative motion can be described to second-post-Newtonian order
[i.e., ${\cal O}(v^4/c^4)$] with the Hamiltonian
\begin{equation}
H({\bf q}, {\bf p}) = H_0({\bf p}, {\bf q}) + \frac{1}{c^2}H_2({\bf
p}, {\bf q}) + \frac{1}{c^4}H_4({\bf p}, {\bf q})\;,
\label{eq:eob_hamiltonian}
\end{equation}
where $H_0 = |{\bf p}|^2/2\mu + GM\mu/|{\bf q}|$ encodes the Newtonian
dynamics, and $H_{2,4}$ describes pN corrections to that motion.  A
binary's energy and angular momentum can be found from this
Hamiltonian without too much difficulty.

The next step is to write down an effective one-body metric,
\begin{equation}
ds^2 = -A(R) c^2 dT^2 + B(R)dR^2 + R^2(d\theta^2 +
\sin^2\theta\,d\phi^2)\;,
\label{eq:eob_metric}
\end{equation}
where $A(R) = 1 + \alpha_1(GM/Rc^2) + \alpha_2(GM/Rc^2)^2 + \ldots$; a
similar expansion describes $B(R)$.  The coefficients $\alpha_i$
depend on reduced mass ratio, $\eta = \mu/M$.  The effective problem
is then to describe the motion of a test body in the spacetime
(\ref{eq:eob_metric}).  By asserting a correspondance between certain
action variables in the pN framework and in the effective framework,
the coefficients $\alpha_i$ are completely fixed.  For example, one
finds that, as $\eta \to 0$, the metric (\ref{eq:eob_metric}) is
simply the Schwarzschild spacetime.  The effective problem can thus be
regarded as the motion of a test body around a ``deformed'' black
hole, with $\eta$ controlling the deformation.  See
{\cite{damour_eob08}} and references therein for further discussion.

One must also describe radiation reaction in the effective one-body
approach.  A key innovation introduced by Damour and coworkers [see
{\cite{dis98}} for the original presentation of this idea] is to {\it
re-sum} the pN results for energy loss due to GWs in order to obtain a
result that is good into the strong field.  In more detail, we put
\begin{equation}
\frac{dp_\phi}{dt} = -{\cal F}_\phi\;.
\end{equation}
The function ${\cal F}_\phi$ is known to rather high order in orbital
velocity $v$ by a combination of analyses in both pN theory [see,
e.g., {\cite{blanchet06}} for a review] and to analytic expansion of
results in perturbation theory {\citep{msstt97}}.  It can be written
\begin{equation}
{\cal F}(v) = \frac{32G}{5c^5}\eta r^4\Omega^5 F(v)\;,
\end{equation}
where
\begin{equation}
F(v) = 1 + \sum a_n\left(\frac{v}{c}\right)^{n/2} + \sum b_n
\log(v/c)\left(\frac{v}{c}\right)^{n/2}\;.
\end{equation}
Post-Newtonian theory allows us to compute $a_n$ including
contributions in $\mu/M$, up to $n = 7$, and shows that $b_n \ne 0$
for $n = 6$ [\citealt{blanchet06}, Eq.\ (168)].  Perturbation theory
[\citealt{msstt97}, Eq.\ (4.18)] gives us the ${\cal O}[(\mu/M)^0]$
contributions for $a_n$ up to $n = 11$, and shows that $b_n \ne 0$ for
$n = 8, 9, 10, 11$.

The resummation introduced by Damour, Iyer \& Sathyaprakash requires
factoring out a pole at $v = \hat v$ in the function $F(v)$ and then
reorganizing the remaining terms using a {\it Pad\'e approximant}:
\begin{equation}
F^{\rm rs}(v) = \left(1 - v/\hat v\right)^{-1}
P\left[(1 - v/\hat v)F(v)\right]\;.
\label{eq:Pade_form}
\end{equation}
The approximant $P$ converts an $N$-th order polynomial into a ratio
of $N/2$-th order polynomials whose small $v$ expansion reproduces the
original polynomial:
\begin{equation}
P\left[1 + \sum_{n = 1}^N c_n (v/c)^n\right] = \frac{1 + \sum_{n =
1}^{N/2} d_n (v/c)^n}{1 + \sum_{n = 1}^{N/2} e_n (v/c)^n}\;.
\end{equation}
Using this approach to define the evolution of a system due to GW
backreaction, it is not so difficult to compute the waves that a
binary generates as its members spiral together.  Indeed, by
augmenting these waves with the ``ringdown'' that comes once the
spacetime is well-described by a single black hole, the effective
one-body approach has recently had great success in matching to the
waveforms that are produced by numerical relativity simulations.  We
defer a discussion of this matching until after we have described
numerical relativity in more detail, in order that the effectiveness
of this comparison can be made more clear.

\section{Numerical relativity}
\label{sec:numrel}

Numerical relativity means the direct numerical integration of the
Einstein field equations, evolving from an ``initial'' spacetime to a
final state.  This requires rethinking some of our ideas about GR.  As
a prelude, consider Maxwell's equations, written in somewhat
non-standard form:
\begin{eqnarray}
\nabla\cdot{\bf E} = 4\pi\rho\;,&\qquad&
\nabla\cdot{\bf B} = 0\;;
\label{eq:div_eqs}\\
\frac{\partial{\bf B}}{\partial t} = -c\nabla\times{\bf E}\;,
&\qquad&
\frac{\partial{\bf E}}{\partial t} = 4\pi{\bf J} - c\nabla\times{\bf
B}\;.
\label{eq:curl_eqs}
\end{eqnarray}
These equations tell us how ${\bf E}$ and ${\bf B}$ are related
throughout spacetime.  Notice that Eqs.\ (\ref{eq:div_eqs}) and
(\ref{eq:curl_eqs}) play very different roles here.  The divergence
equations contain no time derivatives; if we imagine ``slicing''
spacetime into a stack of constant time slices, then Eq.\
(\ref{eq:div_eqs}) tells us how ${\bf E}$ and ${\bf B}$ are {\it
constrained} on each slice.  By constrast, the curl equations do
include time operators, and so tell us how ${\bf E}$ and ${\bf B}$ are
related as we {\it evolve} from slice to slice.  We turn now to
developing the Einstein equations into a form appropriate for evolving
from an initial form; our discussion has been heavily influenced by
the nicely pedagogical presentation of \cite{baumshap03}.

\subsection{Overview: From geometric equations to evolution equations}
\label{sec:nr_overview}

How do we use Eq.\ (\ref{eq:einstein}) to evolve spacetime from some
initial state?  The Einstein field equations normally treat space and
time democratically --- no explicit notion of time is built into Eq.\
(\ref{eq:einstein}).  The very question requires us to change our
thinking: ``evolving'' from an ``initial'' state requires some notion
of time.

\begin{figure}[t]
\includegraphics[width=5.2in]{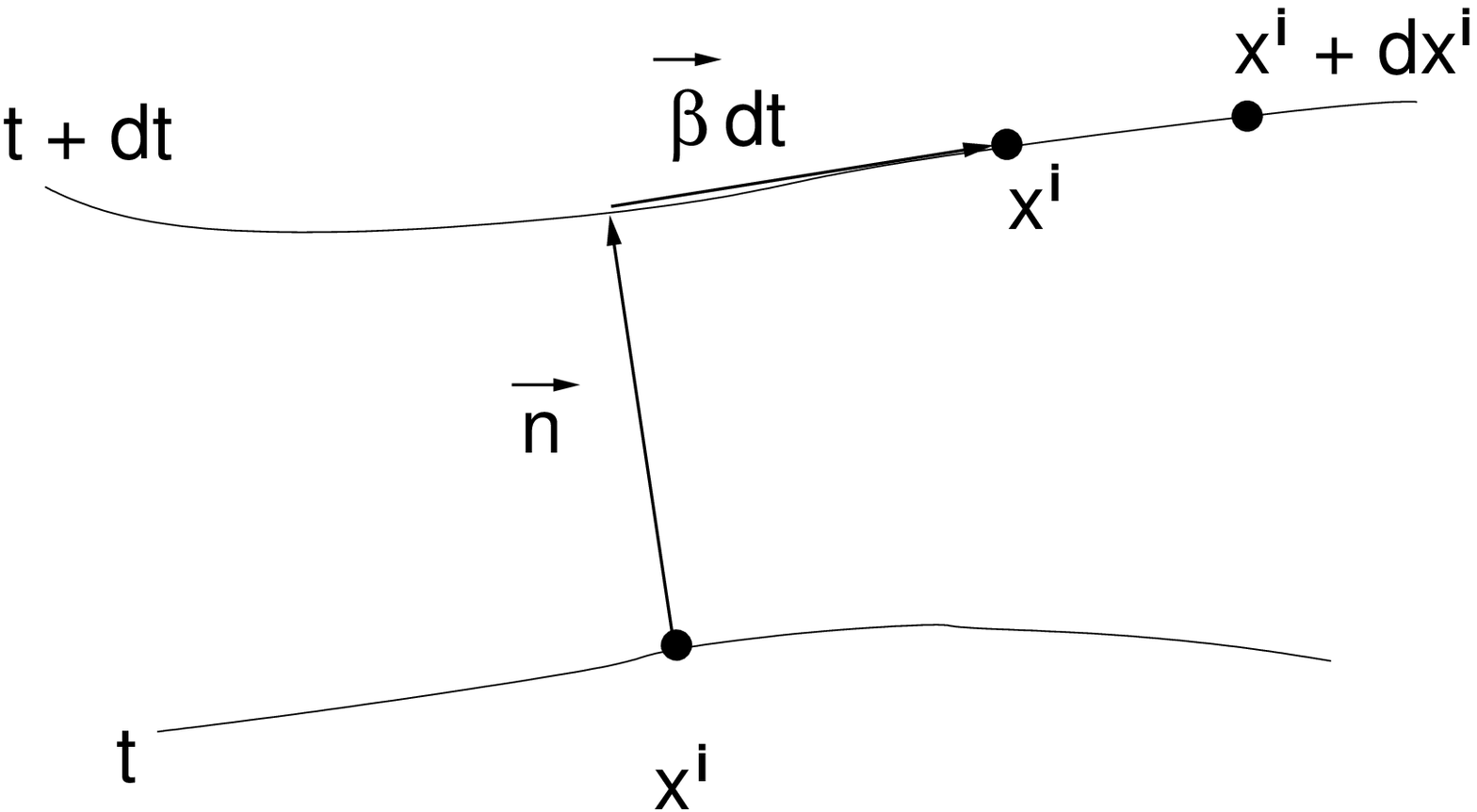}
\caption{Slicing of ``spacetime'' into ``space'' and ``time.''  The
vector $\vec n$ is normal to the slice at $t$.  An observer who moves
along $\vec n$ experiences an interval of proper time $d\tau =
\alpha\,dt$, where $\alpha$ is the {\it lapse} function.  The {\it
shift} $\vec\beta$ describes the translation of spatial coordinates
$x^i$ from slice-to-slice as seen by that normal observer.}
\label{fig:slices}
\end{figure}

Suppose that we have chosen a time coordinate, defining a way to slice
spacetime into space and time.  We must reformulate the Einstein field
equations using quantities defined solely on a given time slice.
Figure {\ref{fig:slices}} illustrates how two nearby time slices may
be embedded in spacetime.  Once time is set, we can freely choose
spatial coordinates in each slice; we show $x^i$ on both slices.

Let $\vec n$ be normal to the bottom slice.  The {\it lapse} $\alpha
\equiv d\tau/dt$ sets the proper time experienced by an observer who
moves along $\vec n$; the {\it shift} $\beta^i$ tells us by how much
$x^i$ is displaced (``shifted'') on the second slice relative to the
normal observer.  We will soon see that $\alpha$ and $\beta^i$ are
completely unconstrained by Einstein's equations.  They let us set
coordinates as conveniently as possible, generalizing the gauge
generator $\xi^\mu$ used in linearized theory (cf.\ Sec.\
\ref{sec:gwbasics}) to the strong field.

The proper spacetime separation of $x^i$ and $x^i + dx^i$ is then
\begin{equation}
ds^2 = -\alpha^2 dt^2 + g_{ij}(dx^i + \beta^i dt)(dx^j + \beta^j
dt)\;.
\end{equation}
(In this section, we will put $c = 1$; various factors become rather
unwieldy otherwise.)  We now have a form for the metric of spacetime,
a notion of constant time slices, and the normal to a slice $\vec n$.
Because we are interested in understanding quantities which ``live''
in a given slice (i.e., orthogonal to the normal $\vec n$), we build
the projection tensor $\gamma_{\mu\nu} = g_{\mu\nu} + n_\mu n_\nu$.
This tensor is just the metric for the geometry in each slice.  We can
choose coordinates so that $\gamma_{tt} = \gamma_{ti} = 0$, and
$\gamma_{ij} = g_{ij}$; we will assume these from now on.

We now have enough pieces to see how to build the field equations in
this formalism: We take Eq.\ (\ref{eq:einstein}) and project
components parallel and orthogonal to $\vec n$.  Consider first the
component that is completely parallel to $\vec n$:
\begin{equation}
G_{\alpha\beta} n^\alpha n^\beta = 8\pi G T_{\alpha\beta} n^\alpha
n^\beta \quad\longrightarrow\quad
R + K^2 - K_{ij} K^{ij} = 16\pi G\rho\;.
\label{eq:hamilton}
\end{equation}
[See {\cite{baumshap03}} for a detailed derivation of Eq.\
(\ref{eq:hamilton}).]  In this equation, $R$ is the Ricci scalar for
the 3-metric $\gamma_{ij}$, $\rho = T_{\alpha\beta} n^\alpha n^\beta$,
and
\begin{eqnarray}
K_{ij} &\equiv& -{\gamma_i}^\alpha{\gamma_j}^\beta \nabla_\alpha
n_\beta
\nonumber\\
&=& \frac{1}{2\alpha}\left(-\partial_t\gamma_{ij} + D_i\beta_j +
D_j\beta_i\right)
\label{eq:extrinsic}
\end{eqnarray}
is the {\it extrinsic curvature}.  (The operator $D_i$ is a covariant
derivative for the metric $\gamma_{ij}$.)  It describes the portion of
the curvature which is due to the way that each constant time slice is
embedded in the full spacetime.  Equation (\ref{eq:hamilton}) is known
as the {\it Hamiltonian constraint}.  [See {\cite{baumshap03}} for
details of how to go from the first line of (\ref{eq:hamilton}), which
is a definition, to the second line, which is more useful here.]
Notice that it contains no time derivatives of $K_{ij}$.  This
equation is thus a {\it constraint}, relating data on a given
timeslice.

Next, components parallel to $\vec n$ on one index and orthogonal on
the other:
\begin{equation}
G_{\alpha\beta} n^\alpha {\gamma_i}^\beta = 8\pi G T_{\alpha\beta}
n^\alpha {\gamma_i}^\beta \quad \longrightarrow \quad D_j{K^j}_i - D_i
K = 8\pi G j_i
\label{eq:momentum}
\end{equation}
The matter current $j_i = -T_{\alpha\beta}n^\alpha{\gamma_i}^\beta$.
Equation (\ref{eq:momentum}) is the {\it momentum constraint}; notice
it also has no time derivatives of $K_{ij}$.

Finally, project completely orthogonal to $\vec n$:
\begin{eqnarray}
G_{\alpha\beta} {\gamma_i}^\alpha {\gamma_j}^\beta &=& 8\pi G
T_{\alpha\beta} {\gamma_i}^\alpha {\gamma_j}^\beta \quad
\longrightarrow
\nonumber\\
\partial_t K_{ij} &=& -D_i D_j\alpha + \alpha\left[R_{ij} - 2K_{ik}{K^k}_j +
K K_{ij} - 8\pi G\alpha(\mbox{matter})\right]
\nonumber\\
& + & \beta^k D_k K_{ij} + K_{ik} D_j\beta^k + K_{kj} D_i\beta^k\;.
\label{eq:evolution}
\end{eqnarray}
[We have abbreviated a combination of projections of the stress-energy
tensor as ``matter.''  Interested readers can find more details in
{\cite{baumshap03}}.]  Combining Eqs.\ (\ref{eq:evolution}) with
(\ref{eq:extrinsic}) gives us a full set of {\it evolution equations}
for the metric and the extrinsic curvature, describing how the
geometry changes as we evolve from time slice to time slice.

The field equations sketched here are the {\it ADM} equations
{\citep{adm62}}.  Today, most groups work with modified versions of
these equations; a particularly popular version is the {\it BSSN}
system, developed by {\cite{bs99}}, building on foundational work by
{\cite{sn95}}.  In BSSN, one rewrites the spacetime metric as
\begin{equation}
{\tilde\gamma}_{ij} = e^{-4\phi}\gamma_{ij}\;,
\label{eq:conformal}
\end{equation}
where $\phi$ is chosen so that $e^{12\phi} = {\rm det}(\gamma_{ij})$.
With this choice, ${\rm det}({\tilde\gamma}_{ij}) = 1$.  Roughly
speaking, the decomposition (\ref{eq:conformal}) splits the geometry
into ``transverse'' and ``longitudinal'' degrees of freedom
(encapsulated by ${\tilde\gamma}_{ij}$ and $\phi$, respectively.)  One
similarly splits the extrinsic curvature into ``longitudinal'' and
``transverse'' parts by separately treating its trace and its
trace-free parts:
\begin{equation}
A_{ij} = K_{ij} - \frac{1}{3}\gamma_{ij}K\;.
\end{equation}
It is convenient to conformally rescale $A_{ij}$, using ${\tilde
A}_{ij} = e^{-4\phi}A_{ij}$.  One then develops evolution equations
for $\phi$, ${\tilde\gamma}_{ij}$, $K$, and ${\tilde A}_{ij}$.  See
{\cite{bs99}} for detailed discussion.

\subsection{The struggles and the breakthrough}
\label{sec:nr_breakthrough}

Having recast the equations of GR into a form that lets us evolve
forward in time, we might hope that simulating the merger of two
compact bodies would now not be too difficult.  In addition to the
equations discussed above, we need a few additional pieces:

\begin{enumerate}

\item {\it Initial data}; i.e., a description of the metric and
extrinsic curvature of a binary at the first moment in a simulation.
Ideally, we might hope that this initial data set would be related to
an earlier inspiral of widely separated bodies, (e.g.,
{\citealt{samaya06}}, and {\citealt{ytob06}}).  However, any method
which can produce a bound binary with specified masses $m_1$, $m_2$
and spins ${\bf S}_1$, ${\bf S}_2$ should allow us to simulate a
binary (although it may be ``contaminated'' by having the wrong GW
content at early times).

\item {\it Gauge or coordinate conditions}; i.e., an algorithm by
which the lapse $\alpha$ and shift $\beta^i$ are selected.  Because
these functions are not determined by the Einstein field equations but
are instead freely specified, they can be selected in a way that is as
convenient as possible.  Such wonderful freedom can also be horrible
freedom, as one could choose gauge conditions which obscure the
physics we wish to study, or not facilitate a stable simulation.

\item {\it Boundary conditions.} If the simulation contains black
holes, then they should have an event horizon from which nothing comes
out.  Unfortunately, we cannot know where horizons are located until
the full spacetime is built (though we have a good estimate, the
``apparent horizon,'' which can be computed from information on a
single time slice).  They also contain singularities.  Hopefully,
event horizons will prevent singular fields from contaminating the
spacetime.  The computation will also have an outer boundary.  Far
from the binary, the spacetime should asymptote to a flat (or
Robertson-Walker) form, with a gentle admixture of outgoing GWs.

\end{enumerate}

How to choose these ingredients has been active research for many
years.  Early on, some workers were confident it was just a matter of
choosing the right combination of methods and the pieces would fall
into place.  The initial optimism was nicely encapsulated by the
following quote from {\cite{300yrs}}, p.\ 379:

\begin{quote}

$\ldots$ numerical relativity is likely to give us, in the next five
years or so, a detailed and highly reliable picture of the final
coalescence and the wave forms it produces, including the dependence
on the holes' masses and angular momenta.

\end{quote}

For many years, Thorne's optimism seemed misplaced.  Despite having a
detailed understanding of the principles involved, it seemed simply
not possible to evolve binary systems for any interesting length of
time.  In most cases, the binary's members could complete a fraction
of an orbit and then the code would crash.  As Joan Centrella has
emphasized in several venues, by roughly 2004 ``People were beginning
to say `numerical relativity cannot be done'{''} (J.\ Centrella,
private communication).

A major issue with many of these simulations appeared to be {\it
constraint violating modes}.  These are solutions of the system
$(\partial_t\gamma_{ij}, \partial_t K_{ij})$ that do not satisfy the
constraint equations (\ref{eq:hamilton}) and (\ref{eq:momentum}).  As
with the Maxwell equations, one can prove that a solution which
satisfies the constraints initially will continue to satisfy them at
later times {\it in the continuum limit} [cf.\ Sec.\ IIIC of
{\cite{pretorius07}}].  Unfortunately, numerical relativity does not
work in the continuum limit; and, even more unfortunately, constraint
violating modes generically tend to be unstable {\citep{ls02}}.  This
means that small numerical errors can ``seed'' the modes, which then
grow unstably and swamp the true solution.  The challenge was to keep
the seed instabilities as small as possible and then to prevent them
from growing.  In this way, {\cite{btj04}} were able to compute one
full orbit of a black hole binary before their simulation crashed.

It was thus something of a shock when {\cite{pretorius05}}
demonstrated a binary black hole simulation that executed a full
orbit, followed by merger and ringdown to a Kerr black hole, with no
crash apparent.  Pretorius used a formulation of the Einstein
equations based on coordinates similar to the de Donder coordinates
described in Sec.\ {\ref{sec:pn}}.  These are known as ``generalized
harmonic coordinates''; see {\cite{pretorius07}} for detailed
discussion.  Because his success came with such a radically different
system of equations, it was suspected that the ADM-like equations
might have to be abandoned.  These concerns were allayed by near
simultaneous discoveries by numerical relativity groups then at the
University of Texas in Browsnville {\citep{campanellietal06}} and the
Goddard Space Flight Center {\citep{bakeretal06}}.  The Campanelli et
al.\ and Baker et al.\ groups both used the BSSN formalism described
at the end of Sec.\ {\ref{sec:nr_overview}}.  To make this approach
work, they independently discovered a gauge condition that allows the
black holes in their simulations to move across their computational
grid.  (Earlier calculations typically fixed the coordinate locations
of the black holes, which means that the coordinates effectively
co-rotated with the motion of the binary.)  It was quickly shown that
this approach yielded results that agreed excellently with Pretorius'
setup {\citep{bcpz07}}.

Implementing the so-called ``moving puncture'' approach was
sufficiently simple that the vast majority of numerical relativity
groups were able to immediately apply it to their codes and begin
evolving binary black holes.  These techniques have also
revolutionized our ability to model systems containing neutron stars
(\citealt{su06}, \citealt{etienne08}, \citealt{bgr08},
\citealt{skyt09}).  In the past four years, we have thus moved from a
state of being barely able to model a single orbit of a compact binary
system in full GR to being able to model nearly arbitrary binary
configurations.  Though Thorne's 1987 prediction quoted above was far
too optimistic on the timescale, his prediction for how well the
physics of binary coalescence would be understood appears to be
exactly correct.

\subsection{GWs from numerical relativity and effective one body}
\label{sec:nr_eob}

Prior to this breakthrough, the effective one-body approach gave the
only strong-field description of GWs from the coalescence of two black
holes.  Indeed, these techniques made a rather strong prediction: The
coalescence waveform should be fairly ``boring,'' in the sense that we
expect the frequency and amplitude to chirp up to the point at which
the physical system is well modeled as a single deformed black hole.
Then, it should rapidly ring down to a quiescent Kerr state.

Such a waveform is indeed exactly what numerical simulations find, at
least for the cases that have been studied so far.  It has since been
found that predictions from the effective one-body formalism give an
outstanding description of the results from numerical relativity.
There is some freedom to adjust how one matches effective one-body
waves to the numerical relativity output [e.g., the choice of pole
$\hat v$ in Eq.\ (\ref{eq:Pade_form})]; {\cite{betal07}} and
{\cite{dnhhb08}} describe how to do this matching.

\begin{figure}[t]
\includegraphics[width=2.55in]{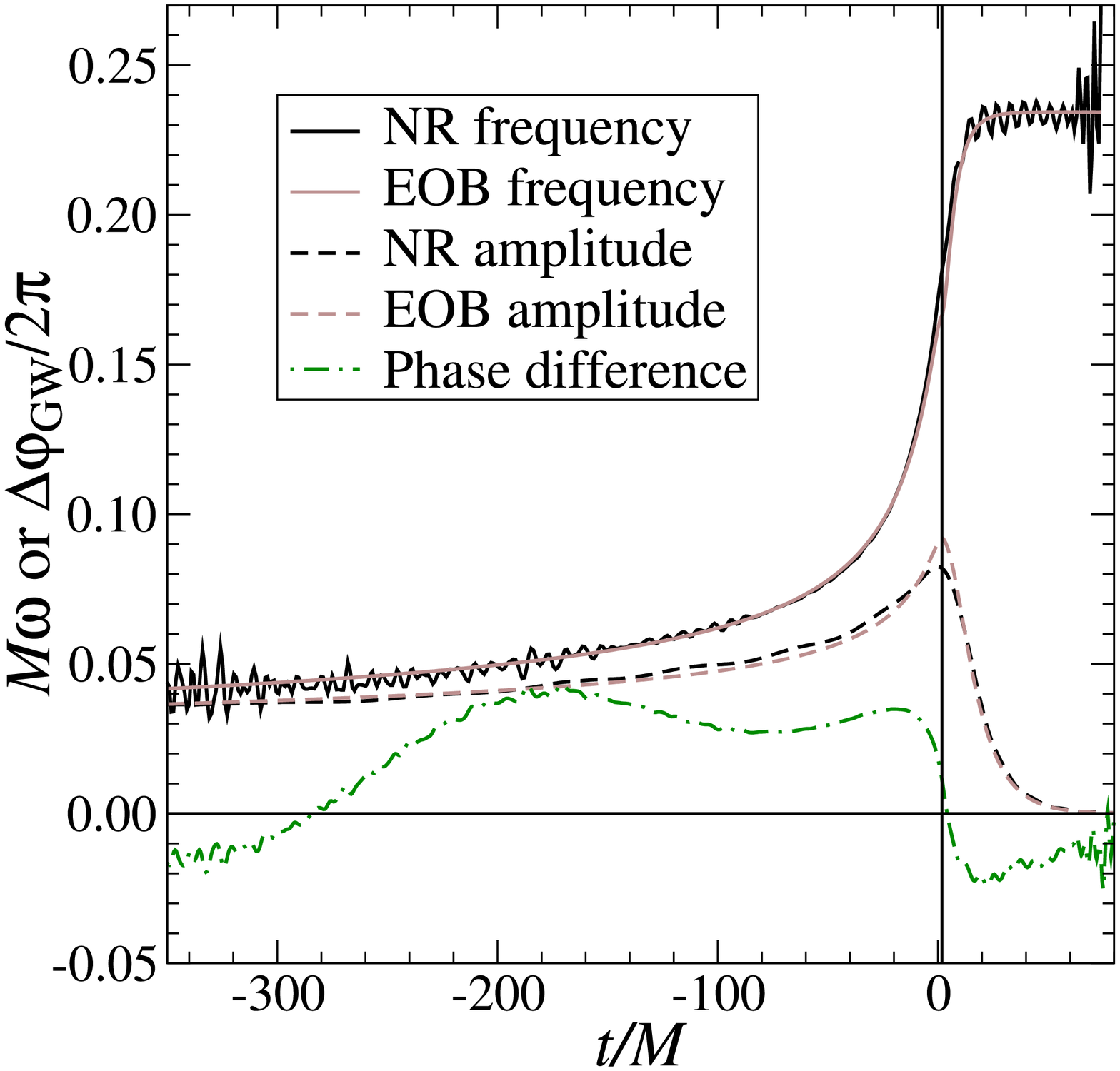}
\hskip 1pc
\includegraphics[width=2.55in]{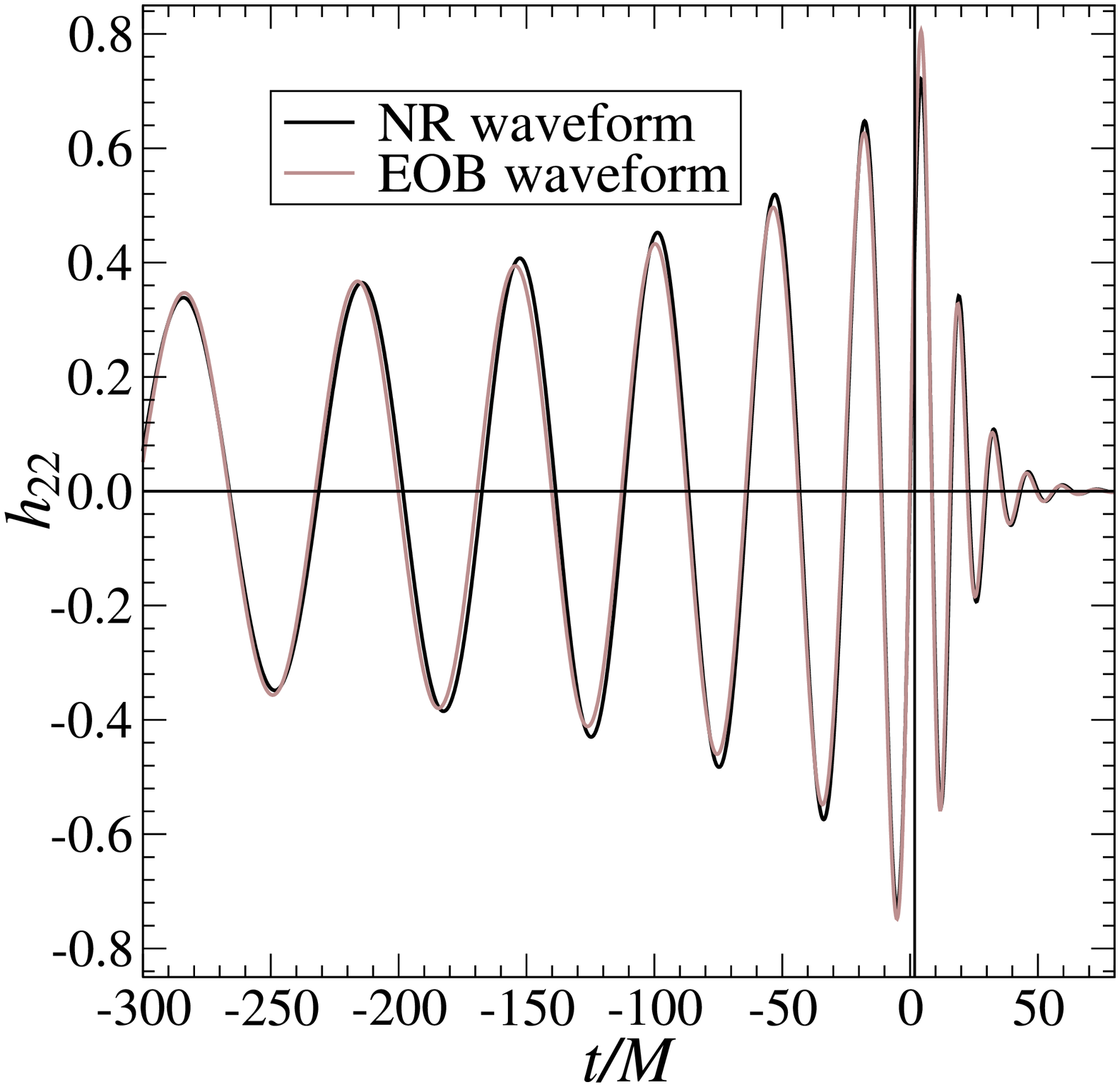}
\caption{Left panel: Comparison between the numerical relativity
computed frequency and phase and the effective one-body frequency and
phase.  Right panel: Gravitational waveform computed by those two
methods.  These plots are for the coalescence of two non-spinning
black holes with a mass ratio $m_2/m_1 = 4$.  Figure kindly provided
to the author by Alessandra Buonanno, taken from \cite{betal07}.  Note
that $G = c = 1$ in the labels to these figures.}
\label{fig:eobnr}
\end{figure}

Figure {\ref{fig:eobnr}}, taken from Buonanno et al., gives an example
of how well the waveforms match one another.  Over the entire span
computed, the two waveforms differ in phase by only a few hundredths
of a cycle.  The agreement is so good that one can realistically
imagine ``calibrating'' the effective one-body waveforms with a
relatively small number of expensive numerical relativity
computations, and then densely sampling the binary parameter space
using the effective one-body approach.

\section{Gravitational-wave recoil}
\label{sec:recoil}

That GWs carry energy and angular momentum from a binary, driving its
members to spiral together as described in Sec.\
{\ref{sec:newtonian_waves}}, is widely appreciated.  Until recently,
it was not so well appreciated that the waves can carry {\it linear}
momentum as well.  If the binary and its radiation pattern are
asymmetric, then that radiation carries a net flux of momentum given
by
\begin{equation}
\frac{dp^i}{dt} = \frac{R^2}{c}\int d\Omega\, T^{00}\,n^i\;,
\end{equation}
where $T^{00}$ is the energy-flux component of the Isaacson tensor
(\ref{eq:isaacson_tmunu}), $n^i$ is the $i$-th component of the radial
unit vector, and the integral is taken over a large sphere ($R \to
\infty$) around the source.  Recent work has shown that the
contribution to this ``kick'' from the final plunge and merger of
coalescing black holes can be particularly strong.  We now summarize
the basic physics of this effect, and survey recent results.

{\cite{bekenstein73}} appears to have first appreciated that the
momentum flux of GWs could have interesting astrophysical
consequences.  {\cite{fitchett83}} then estimated the impact this flux
could have on a binary.  An aspect of the problem which Fitchett's
analysis makes very clear is that the recoil comes from the beating of
different multipolar contributions to the recoil: If one only looks at
the quadrupole part of the GWs [cf.\ Eq.\ (\ref{eq:multipolar_form})],
the momentum flux is zero.  Fitchett's analysis included octupole and
current-quadrupole radiation, and found
\begin{equation}
v_{\rm kick} \simeq 1450\,{\rm km/sec}\;\frac{f(q)}{f_{\rm max}}\,
\left(\frac{GM_{\rm tot}/c^2}{R_{\rm term}}\right)^4\;,
\label{eq:kickmag}
\end{equation}
where $f(q) = q^2(1 - q)/(1 + q)^5$ gives the dependence on mass ratio
$q = m_1/m_2$.  This function has a maximum at $q \simeq 0.38$.  The
radius $R_{\rm term}$ describes when wave emission cuts off; for
systems containing black holes this will scale with the total mass.
Thus, the recoil does not depend on total mass, just mass ratio.

Fitchett's analysis is similar to our discussion in Sec.\
{\ref{sec:newtonian_waves}} in that Newtonian dynamics are
supplemented with multipolar wave emission.  Because the effect is
strongest when $R_{\rm term}$ is smallest, it was long clear that a
proper relativistic analysis was needed to get this kick correct.
Indeed, a prescient analysis by \cite{rr89} suggested that binaries
containing rapidly spinning black holes were likely to be especially
interesting; as we shall see in a few moments, they were absolutely
correct.

{\cite{fhh04}} provided the first estimates of recoil which did the
strong-field physics more-or-less correctly.  They used the
perturbative techniques described in Sec.\ {\ref{sec:pert}}, arguing
that one can extrapolate from the small mass ratio regime to $q \sim
0.2$ or so with errors of a few tens of percent.  Unfortunately, their
code at that time did not work well with plunging orbits, so they had
large error bars.  They did find, however, that the maximum recoil
probably fell around $v_{\rm kick} \simeq (250 \pm 150)$ km/sec, at
least if no more than one body spins, and if the spin and orbit are
aligned.  {\cite{bqw05}} revisited this treatment, with a particular
eye on the final plunge waves, using the pN methods outlined in Sec.\
{\ref{sec:pn}}.  Their results were consistent with Favata et al., but
reduced the uncertainty substantially, finding a maximum recoil
$v_{\rm kick} \simeq (220 \pm 50)$ km/sec.

These numbers stood as the state-of-the-art in black hole recoil for
several years, until numerical relativity's breakthrough (Sec.\
{\ref{sec:numrel}}) made it possible to study black hole mergers
without any approximations.  The Favata et al.\ and Blanchet et al.\
numbers turn out to agree quite well with predictions for the merger
of non-spinning black holes; {\cite{gsbhh07}} find the maximum kick
for non-spinning merger comes when $q = 0.38$, yielding $v_{\rm kick}
= (175 \pm 11)$ km/sec.  When spin is unimportant, kicks appear to be
no larger than a few hundred km/sec.

When spin {\it is} important, the kick can be substantially larger.
Recent work by {\cite{ghsbh07}} and {\cite{clz07}} shows that when the
holes have large spins and those spins are aligned just right (equal
in magnitude, antiparallel to each other, and orthogonal to the
orbital angular momentum), the recoil can be a few {\it thousand}
km/sec.  Detailed parameter exploration is needed to assess how much
of this maximum is actually achieved; early work on this problem is
finding that large kicks (many hundreds to a few thousand of km/sec)
can be achieved for various spin orientations as long as the spins are
large (\citealt{tm07}, \citealt{pollney_etal07}); recent work by
\cite{bkn08} shows how the recoil can depend on spin and spin
orientation, suggesting a powerful way to organize the calculation to
see how generic such large kicks actually are.  That the maximum is so
much higher than had been appreciated suggests that substantial
recoils may be more common than the Favata et al.\ and Blanchet et
al.\ calculations led us to expect {\citep{sb07}}.

Many recent papers have emphasized that kicks could have strong
astrophysical implications, ranging from escape of the black hole from
its host galaxy to shocks in material accreting onto the large black
hole.  The first possible detection of black hole recoil was recently
announced {\citep{kzl08}}.  As that claim is assessed, we anticipate
much activity as groups continue to try to identify a signature of a
recoiling black hole.

\section{Astronomy with gravitational waves}
\label{sec:gwastro}

Direct GW measurement is a major motivator for theorists seeking to
understand how binary systems generate these waves.  The major
challenge one faces is that GWs are extremely weak; as we derive in
detail in this section, a wave's strain $h$ sets the change in length
$\Delta L$ per length $L$ in the arms of a GW detector.  Referring to
Eq.\ (\ref{eq:h_NQ}), we now estimate typical amplitudes for binary
sources:
\begin{eqnarray}
h_{\rm amp} &\simeq& \frac{2G{\cal M}}{c^2 D}\left(\frac{\pi G {\cal
M} f}{c^3}\right)^{2/3}
\nonumber\\
&\simeq& 10^{-23}\,\left(\frac{2.8\,M_\odot}{M}\right)^{5/3}
\left(\frac{f}{100\,{\rm Hz}}\right)^{2/3} \left(\frac{200\,{\rm
Mpc}}{D}\right)
\nonumber\\
&\simeq& 10^{-19}\,\left(\frac{2\times10^6\,M_\odot}{M}\right)^{5/3}
\left(\frac{f}{10^{-3}\,{\rm Hz}}\right)^{2/3} \left(\frac{5\,{\rm
Gpc}}{D}\right)\;.
\label{eq:h_fiducial}
\end{eqnarray}
On the last two lines, we have specialized to a binary whose members
each have mass $M/2$, and have inserted fiducial numbers corresponding
to targets for ground-based detectors (second line) and space-based
detectors (third line).

In the remainder of this section, we summarize the principles behind
modern interferometric GW detectors.  Given that these principles are
likely to be novel for much of the astronomical community, we present
this material in some depth.  Note that \cite{finn08} has recently
flagged some important issues in the ``standard'' calculation of an
interferometer's response to GWs (which, however, do not change the
final results).  For the sake of brevity, we omit these issues and
recommend his article for those wishing a deeper analysis.  We then
briefly describe existing and planned detectors, and describe how one
measures a binary's signal with these instruments.  This last point
highlights why theoretical modeling has been so strongly motivated by
the development of these instruments.

\subsection{Principles behind interferometric GW antennae}
\label{sec:interferometers}

As a simple limit, treat the spacetime in which our detector lives as
flat plus a simple GW propagating down our coordinate system's
$z$-axis:
\begin{equation}
ds^2 = -c^2dt^2 + (1 + h)dx^2 + (1 - h)dy^2 + dz^2\;,
\label{eq:detector_spacetime}
\end{equation}
where $h = h(t - z)$.  We neglect the influence of the earth (clearly
important for terrestrial experiments) and the solar system (which
dominates the spacetime of space-based detectors).  Corrections
describing these influences can easily be added to Eq.\
(\ref{eq:detector_spacetime}); we neglect them, as they represent
influences that vary on much longer timescales than the GWs.

\begin{figure}[t]
\includegraphics[width=5in]{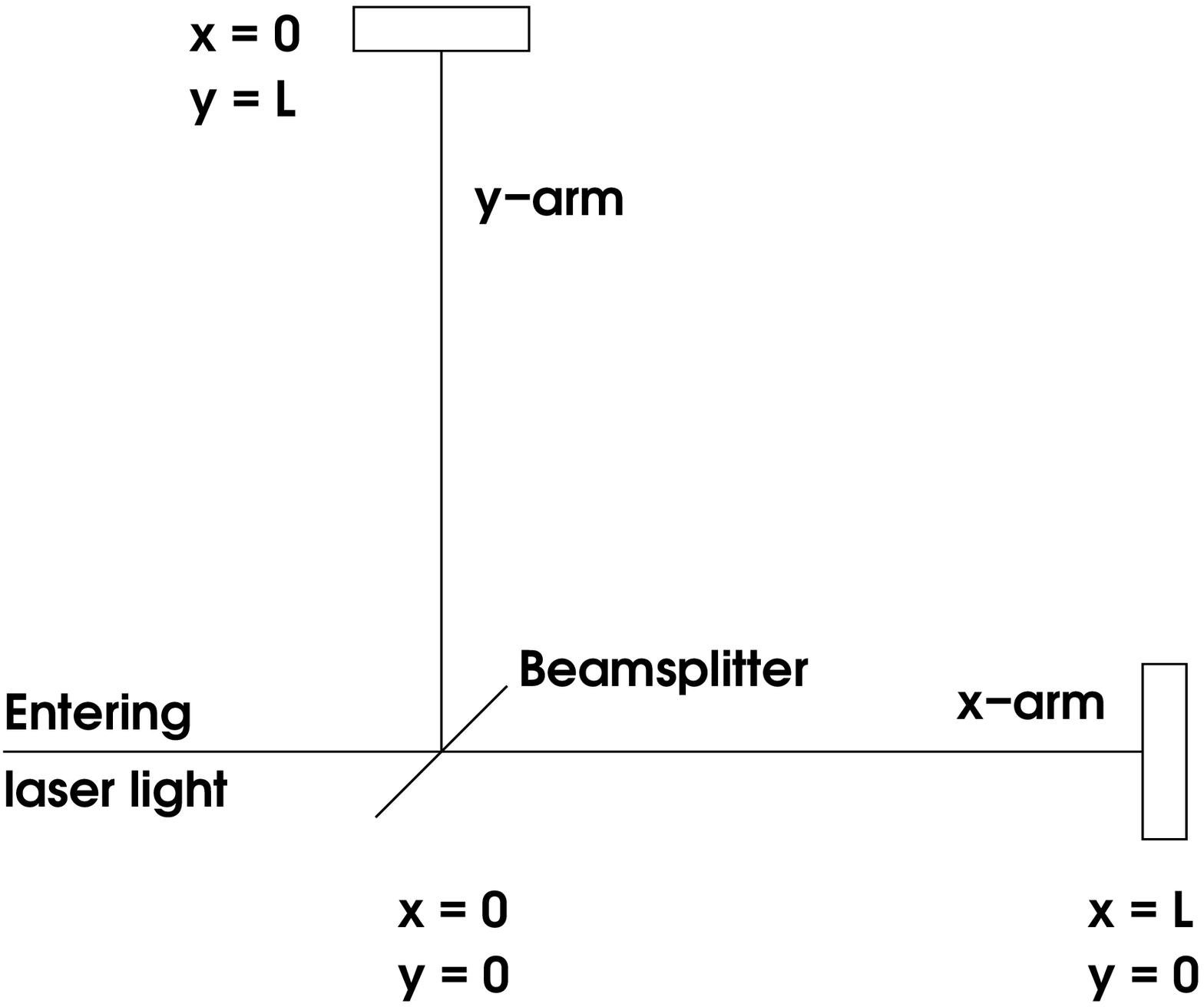}
\caption{Schematic of an interferometer that could be used to detect
GWs.  Though real interferometers are vastly more complicated, this
interferometer topology contains enough detail to illustrate the
principle by which such measurements are made.}
\label{fig:interf}
\end{figure}

Figure {\ref{fig:interf}} sketches an interferometer that can measure
a GW.  Begin by examining the geodesics describing the masses at the
ends of the arms, and the beam splitter at the center.  Take these
objects to be initially at rest, so that $(dx^\mu/d\tau)_{\rm before}
\doteq (c,0,0,0)$.  The GW shifts this velocity by an amount of order
the wave strain: $(dx^\mu/d\tau)_{\rm after} = (dx^\mu/d\tau)_{\rm
before} + {\cal O}(h)$.  Now examine the geodesic equation:
\begin{equation}
\frac{d^2x^j}{d\tau^2} + {\Gamma^j}_{\alpha\beta}
\frac{dx^\alpha}{d\tau}\frac{dx^\beta}{d\tau} = 0\;.
\end{equation}
All components of the connection are ${\cal O}(h)$.  Combining this
with our argument for how the GW affects the various velocities, we
have
\begin{equation}
\frac{d^2x^j}{d\tau^2} + {\Gamma^j}_{00}
\frac{dx^0}{d\tau}\frac{dx^0}{d\tau} + {\cal O}(h^2) = 0\;.
\end{equation}
Now,
\begin{equation}
{\Gamma^j}_{00} = \frac{1}{2}g^{jk}\left(\partial_0 g_{k0} +
\partial_0 g_{0k} - \partial_k g_{00}\right) = 0
\end{equation}
as the relevant metric components are constant.  Thus,
\begin{equation}
\frac{d^2x^j}{d\tau^2} = 0 \;.
\end{equation}
{\it The test masses are unaccelerated to leading order in the GW
amplitude $h$.}

This seems to say that the GW has no impact on the masses.  However,
the geodesic equation describes motion {\it with respect to a
specified coordinate system}.  These coordinates are effectively
``comoving'' with the interferometer's components.  This is
convenient, as the interferometer's components remain fixed in our
coordinates.  Using this, we can show that the {\it proper} length of
the arms does change.  For instance, the $x$-arm has a proper length
\begin{equation}
D_x = \int_0^L \sqrt{g_{xx}}\,dx = \int_0^L \sqrt{1 + h}\,dx \simeq
\int_0^L \left(1 + \frac{h}{2}\right)dx = L\left(1 +
\frac{h}{2}\right)\;.
\label{eq:proper_length}
\end{equation}
Likewise, the $y$-arm has a proper length $D_y = L(1 - h/2)$.

This result tells us that the armlengths as measured by a ruler will
vary with $h$.  One might worry, though, that the ruler will vary,
cancelling the measurement.  This does not happen because rulers are
not made of freely-falling particles: The elements of the ruler are
{\it bound} to one another and act against the GW.  The ruler will
feel {\it some} effect from the GW, but it will be far smaller than
the variation in the separation.

The ruler used by the most sensitive current and planned detectors is
based on laser interferometry.  We now briefly outline how a GW
imprints itself on the phase of the interferometer sketched in Fig.\
{\ref{fig:interf}}.  For further discussion, we recommend
{\citet{faraoni07}}; we also recommend \cite{finn08} for more detailed
discussion.  In the interferometer shown, laser light enters from the
left, hits the beam splitter, travels down both the $x$- and $y$-arms,
bounces back, and is recombined at the beam splitter.  Begin with
light in the $x$-arm, and compute the phase difference between light
that has completed a round trip and light that is just entering.  The
phase of a wavefront is constant as it follows a ray through
spacetime, so we write this difference as
\begin{equation}
\Delta\Phi_x \equiv \Phi(T_{\rm round-trip}) - \Phi(0) = \omega_{\rm
proper} \times(\mbox{proper round-trip travel time})\;,
\label{eq:phase_change_schematic}
\end{equation}
where $\omega_{\rm proper}$ is the laser frequency measured by an
observer at the beam splitter.  Detecting a GW is essentially
precision timing, with the laser acting as our clock.

Consider first the proper frequency.  Light energy as measured by some
observer is $E = -\vec p\cdot\vec u$, where $\vec p$ is the light's
4-momentum and $\vec u$ is the observer's 4-velocity.  We take this
observer to be at rest, so
\begin{equation}
E = -p_t = -g_{tt}p^t = p^t\;.
\end{equation}
Put $p^\mu = \hat p^\mu + \delta p^\mu$, where $\hat p^\mu$ describes
the light in the absence of a GW, and $\delta p^\mu$ is a GW shift.
In the $x$-arm, $\hat p^\mu \doteq \hbar\omega(1,\pm1,0,0)$; the signs
correspond to before and after bounce.  We compute the shift $\delta
p^\mu$ using the geodesic equation (\ref{eq:geodesic}):
\begin{equation}
\frac{d\delta p^\mu}{d\chi} + {\Gamma^\mu}_{\alpha\beta}\hat p^\alpha
\hat p^\beta = 0\;.
\end{equation}
(We use $d\hat p^\mu/d\chi = 0$ in the background to simplify.)  Focus
on the $\mu = t$ component.  The only nontrivial connection is
${\Gamma^t}_{xx} = \partial_t h/2$.  Using in addition the facts that
$p^\mu = dx^\mu/d\chi$ plus $\hat p^t = \pm \hat p^x$ reduces the
geodesic equation to
\begin{equation}
\frac{d\delta p^t}{dt} = -\frac{1}{2}\hat p^t \partial_t h\;.
\label{eq:nullgeod}
\end{equation}
Integrating over a round trip, we find
\begin{equation}
\delta p^t = -\frac{\hat p^t}{2}\left[h(T_{\rm round-trip}) - h(0)\right]\;.
\label{eq:deltapt}
\end{equation}
So, we finally find the proper frequency:
\begin{eqnarray}
\omega_{\rm proper} \equiv E/\hbar &=&\omega\left(1 +
\frac{1}{2}\left[h(0) - h(T_{\rm round-trip})\right]\right)
\nonumber\\ &\simeq& \omega\;.
\label{eq:omega_proper}
\end{eqnarray}
On the second line of Eq.\ (\ref{eq:omega_proper}), we take $T_{\rm
round-trip}$ to be much smaller than the wave period.  This is an
excellent approximation for ground-based interferometers; the exact
result must be used for high-frequency response in space.

Turn next to the proper round-trip time.  The metric
(\ref{eq:detector_spacetime}) shows us that proper time measured at
fixed coordinate is identical to the coordinate time $t$.  For light
traveling in the $x$-arm, $0 = -c^2dt^2 + (1 + h)dx^2$, so
\begin{equation}
dx = \pm c\,dt\left(1 - \frac{h}{2}\right) + {\cal O}(h^2)\;.
\label{eq:dtdx}
\end{equation}
Now integrate over $x$ from $0$ to $L$ and back, and over $t$ from $0$
to $T_{\rm round-trip}$:
\begin{eqnarray}
2L &=& c T_{\rm round-trip} - \frac{c}{2}\int_0^{\rm T_{\rm
round-trip}}h\,dt
\nonumber\\
&=& c T_{\rm round-trip} - \frac{c}{2}\int_0^{2L/c}h\,dt + {\cal
O}(h^2)\;.
\end{eqnarray}
We thus find
\begin{eqnarray}
T_{\rm round-trip} &=& \frac{2L}{c} + \frac{1}{2}\int_0^{2L/c}h\,dt
\nonumber\\
&\simeq& \frac{2L}{c}\left(1 + \frac{1}{2}h\right)\;.
\label{eq:roundtrip}
\end{eqnarray}
The second line describes a wave which barely changes during a round
trip.

The total phase change is found by combining Eqs.\
(\ref{eq:phase_change_schematic}), (\ref{eq:omega_proper}) and
(\ref{eq:roundtrip}):
\begin{eqnarray}
\Delta\Phi_x &=& \omega\left(\frac{2L}{c} + \frac{1}{2}\int_0^{2L/c}h
dt\right)\left(1 + \frac{1}{2}\left[h(0) - h(2L/c)\right]\right)
\nonumber\\
&\simeq&  \frac{2\omega L}{c}\left(1 + \frac{1}{2}h\right)\;.
\label{eq:DeltaPhi_x}
\end{eqnarray}
The second line is for a slowly varying wave, and the first is exact
to order $h$.  We will use the slow limit in further calculations.
Repeating for the $y$-arm yields
\begin{equation}
\Delta\Phi_y = \frac{2\omega L}{c}\left(1 - \frac{1}{2}h\right)\;.
\end{equation}
Notice that this GW acts {\it antisymmetrically} on the arms.  By
contrast, any laser phase noise will be {\it symmetric}: because the
same laser state is sent into the arms by the beam splitter, we have
$\Delta\Phi_x^{\rm Noise} = \Phi_y^{\rm Noise}$.  We take advantage of
this by reading out light produced by {\it destructive} interference
at the beamsplitter:
\begin{eqnarray}
\Delta\Phi^{\rm Read-out} &=& \Delta\Phi_x - \Delta\Phi_y
= \left(\Delta\Phi_x^{\rm GW} + \Delta\Phi_x^{\rm Noise}\right) -
\left(\Delta\Phi_y^{\rm GW} + \Delta\Phi_y^{\rm Noise}\right)
\nonumber\\
&=& 2\Delta\Phi_x^{\rm GW}\;.
\end{eqnarray}
{\it An L-shaped interferometer is sensitive only to the GW, not to
laser phase noise.}  This is the major reason that this geometry is
used; even if an incident wave is oriented such that the response of
the arms to the GW is not asymmetric, one is guaranteed that phase
noise will be cancelled by this configuration.

From basic principles we turn now to a brief discussion of current and
planned detectors.  Our goal is not an in-depth discussion, so we
refer readers interested in these details to excellent reviews by
\citet{hr00} (which covers in detail the characteristics of the
various detectors) and \citet{td05} (which covers the interferometry
used for space-based detectors).

\subsection{Existing and planned detectors}
\label{sec:detectors}

When thinking about GW detectors, a key characteristic is that the
frequency of peak sensitivity scales inversely with armlength.  The
ground-based detectors currently in operation (and undergoing or about
to undergo upgrades) are sensitive to waves oscillating at 10s --
1000s of Hertz.  Planned space-based detectors will have sensitivities
at much lower frequencies, ranging from $10^{-4}$ -- 0.1 Hz
(corresponding to waves with periods of tens of seconds to hours).

The ground-based detectors currently in operation are LIGO ({\it Laser
Interferometer Gravitational-wave Observatory}), with antennae in
Hanford, Washington and Livingston, Louisiana; Virgo near Pisa, Italy;
and GEO near Hanover, Germany.  The LIGO interferometers each feature
4-kilometer arms, and have a peak sensitivity near 100 Hz.  Virgo is
similar, with 3-kilometer arms and sensitivity comparable to the LIGO
detectors.  GEO (or GEO600) has 600-meter arms; as such, its peak
sensitivity is at higher frequencies than LIGO and Virgo.  By using
advanced interferometry techniques, it is able to achieve sensitivity
competitive with the kilometer-scale instruments.  All of these
instruments will be upgraded over the course of the next few years,
installing more powerful lasers, and reducing the impact of local
ground vibrations.  The senstivity of LIGO should be improved by
roughly a factor of ten, and the bandwidth increased as well.  See
{\cite{fritschel03}} for detailed discussion.

There are also plans to build additional kilometer-scale instruments.
The detector AIGO ({\it Australian International Gravitational
Observatory}) is planned as a detector very similar to LIGO and Virgo,
but in Western Australia {\citep{mcc02}}.  This location, far from the
other major GW observatories, has great potential to improve the
ability of the worldwide GW detector network to determine the
characteristics of GW events {\citep{ssmf06}}.  In particular, AIGO
should be able to break degeneracies in angles that determine a
source's sky position and polarization, greatly adding to the
astronomical value of GW observations.  The Japanese GW community,
building on their experience with the 300-meter TAMA interferometer,
hopes to build a 3-kilometer {\it underground} instrument.  Dubbed
LCGT ({\it Large-scale Cryogenic Gravitational-wave Telescope}), the
underground location takes advantage of the fact that local ground
motions tend to decay fairly rapidly as we move away from the earth's
surface.  They plan to use cryogenic cooling to reduce noise from
thermal vibrations.

\begin{figure}[t]
\includegraphics[width=5.3in]{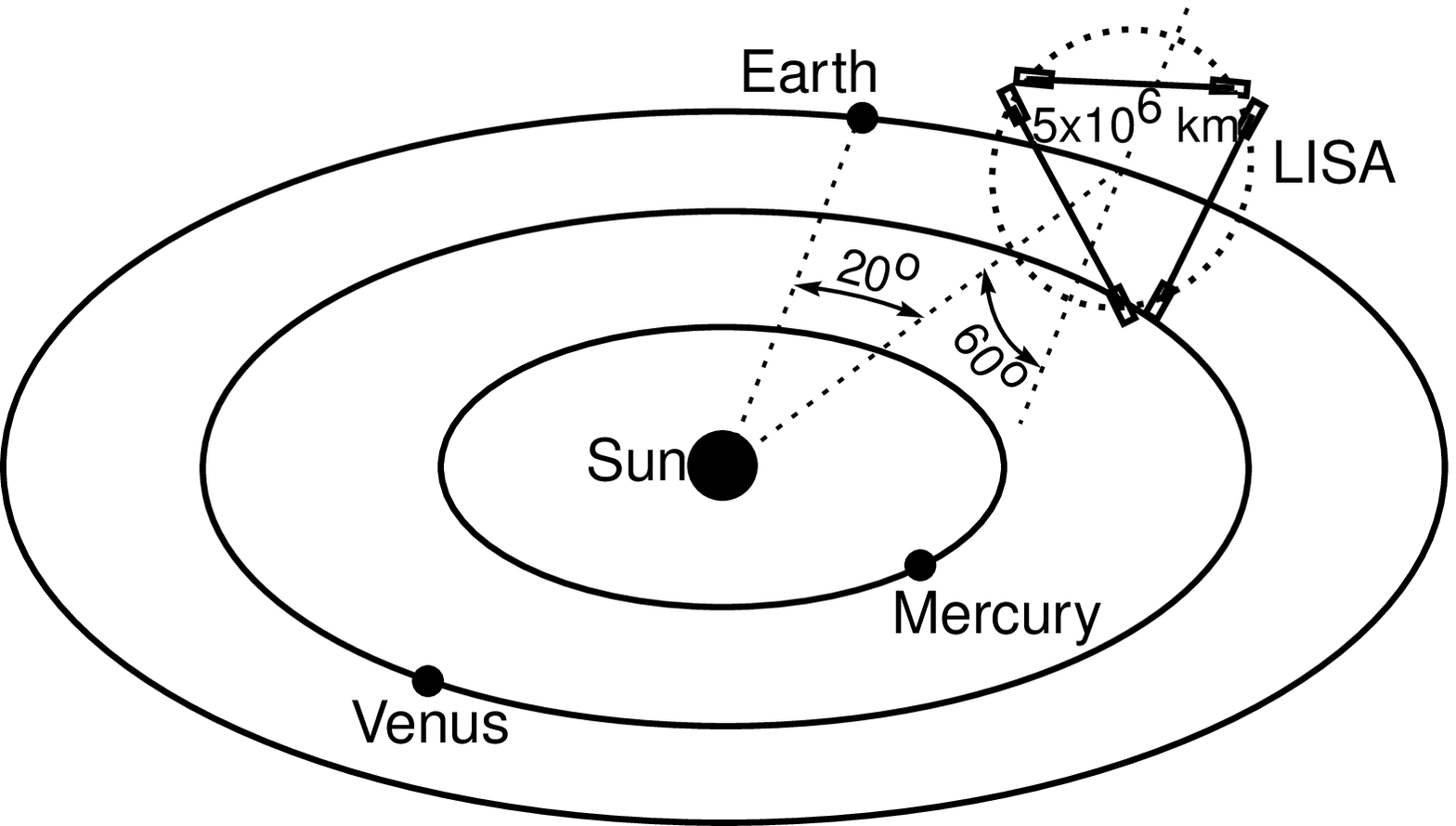}
\caption{Schematic of the LISA constellation in orbit about the sun.
Each arm of the triangle is $5\times10^6$ km; the centroid of the
constellation lags the Earth by $20^\circ$, and its plane is inclined
to the ecliptic by $60^\circ$.  Note that the spacecraft orbit freely;
there is no formation flying in the LISA configuration.  Instead, each
spacecraft is in a slightly eccentric, slightly inclined orbit; their
individual motions preserve the near-equilateral triangle pattern with
high accuracy for a timescale of decades.}
\label{fig:lisa_orb}
\end{figure}

In space, the major project is LISA ({\it Laser Interferometer Space
Antenna}), a 5-million kilometer interferometer under development as a
joint NASA-ESA mission.  LISA will consist of three spacecraft placed
in orbits such their relative positions form an equilateral triangle
whose centroid lags the earth by $20^\circ$, and whose plane is
inclined to the ecliptic by $60^\circ$; see Fig.\
{\ref{fig:lisa_orb}}.  Because the spacecraft are free, they do not
maintain this constellation precisely; however, the variations in
armlength occur on a timescale far longer than the periods of their
target waves, and so can be modeled out without too much difficulty.
The review by {\citet{td05}} discusses in great detail how one does
interferometry on such a baseline with time-changing armlengths.  LISA
is being designed to target waves with periods of several hours to
several seconds, a particularly rich band for signals involving black
holes that have $10^5\, M_\odot \lesssim M \lesssim 10^7\, M_\odot$;
the LISA {\it Pathfinder}, a testbed for some of the mission's
components, is scheduled for launch in the very near future
{\citep{vitale05}}.

Somewhat smaller than LISA, The Japanese GW community has proposed
DECIGO ({\it DECI-hertz Gravitational-wave Observatory}), a space
antenna to target a band at roughly $0.1$ Hz.  This straddles the peak
sensitivities of LISA and terrestrial detectors, and may thus act as a
bridge for signals that evolve from one band to the other.  See
{\cite{decigo}} for further discussion.

It's worth nothing that, in addition to the laser interferometers
discussed here, there have been proposals to measure GWs using atom
interferometry.  A particularly interesting proposal has been
developed by \cite{dimo08}.  Sources of noise in such experiments are
quite different than in the case of laser interferometers, and may
usefully complement the existing suite of detectors in future
applications.

\subsection{Measuring binary signals}
\label{sec:measurement}

The central principle guiding the measurement of GWs is that their
weakness requires {\it phase coherent} signal measurement.  This is
similar to how one searches for a pulsar in radio or x-ray data.  For
pulsars, one models the signal as a sinusoid with a phenomenological
model for frequency evolution:
\begin{equation}
\Phi(t ; \Phi_0, f_0, \dot f_0, \ddot f_0, \ldots) = \Phi_0 +
2\pi\left(f_0 t + \frac{1}{2}\dot f_0 t^2 + \frac{1}{6}\ddot f_0 t^3 +
\ldots \right)\;.
\end{equation}
The cross correlation of a model, $\cos[\Phi]$, with data is maximized
when the parameters $(\Phi_0,f_0,\dot f_0,\ddot f_0,\ldots)$
accurately describe a signal's phase.  For a signal that is $N$ cycles
long, it is not too difficult to show that the cross-correlation is
enhanced by roughly $\sqrt{N}$ when the ``template'' matches the data.

For binary GWs, one similarly cross-correlates models against data,
looking for the template which maximizes the correlation.  [Given the
signal weakness implied by Eq.\ (\ref{eq:h_fiducial}), the
cross-correlation enhancement is sure to be crucial for measuring
these signals in realistic noise.]  Imagine, for example, that the
rule given by Eq.\ (\ref{eq:phi_NQ}) accurately described binary
orbits over the band of our detectors.  We would then expect a model
based on
\begin{equation}
\Phi(t;\Phi_c, t_c, {\cal M}) = \Phi_c -
\left[\frac{c^3(t_c - t)}{5G{\cal M}}\right]^{5/8}
\end{equation}
to give a large correlation when the coalescence phase $\Phi_c$,
coalescence time $t_c$, and chirp mass ${\cal M}$ are chosen well.  As
we have discussed at length, Eq.\ (\ref{eq:phi_NQ}) is not a good
model for strong-field binaries.  The need to faithfully track what
nature throws at us has been a major motivation for the developments
in perturbation theory, pN theory, and numerical relativity discussed
here.

When one determines that some set of parameters maximizes the
correlation, that set is an estimator for the parameters of the
binary.  More formally, the cross-correlation defines a {\it
likelihood function}, which gives the probability of measuring some
set of parameters from the data {\citep{finn92}}.  By examining how
sharply the likelihood falls from this maximum, one can estimate how
accurately data determines parameters.  For large cross-correlation
(large signal-to-noise ratio), this is simply related to how the wave
models vary with parameters.  Let $\theta^a$ represent the $a$-th
parameter describing a waveform $h$.  If $\langle h | h \rangle$
denotes the cross-correlation of $h$ with itself, then define the
covariance matrix
\begin{equation}
\Sigma^{ab} = \left(\left\langle\frac{\partial h}{\partial\theta^a}
\biggl| \frac{\partial h}{\partial\theta^b}\right\rangle\right)^{-1}
\label{eq:covariance_matrix}
\end{equation}
(where the $-1$ power denotes matrix inverse).  Diagonal elements of
this matrix are $1-\sigma$ parameter errors; off-diagonal elements
describe parameter correlations.  See {\cite{finn92}} for derivations
and much more detailed discussion.  In the discussion that follows, we
use Eq.\ (\ref{eq:covariance_matrix}) to drive the discussion of how
model waveforms are used to understand how well observations will be
able to determine the properties of GW-generating systems.

\subsection{What we learn by measuring binary GWs}
\label{sec:observables}

\subsubsection{Overview}

Given all that we have discussed, what can we learn by observing
compact binary merger in GWs?  To set the context, consider again the
``Newtonian'' waveform:
\begin{eqnarray}
h_+ &=& -\frac{2G{\cal M}}{c^2D}\left(\frac{\pi G{\cal
M}f}{c^3}\right)^{2/3}(1 + \cos^2\iota) \cos2\Phi_N(t)\;,
\nonumber\\
h_\times &=& -\frac{4G{\cal M}}{c^2D}\left(\frac{\pi G{\cal
M}f}{c^3}\right)^{2/3}\cos\iota \sin2\Phi_N(t)\;;
\nonumber\\
\Phi_N(t) &=& \Phi_c - \left[\frac{c^3(t_c - t)}{5G{\cal
      M}}\right]^{5/8}\;.
\label{eq:h_NQ2}
\end{eqnarray}
A given interferometer measures a combination of these two
polarizations, with the weights set by the interferometers
antenna response functions:
\begin{equation}
h_{\rm meas} = F_+(\theta,\phi,\psi)h_+ +
F_\times(\theta,\phi,\psi)h_\times\;.
\label{eq:h_meas}
\end{equation}
The angles $(\theta,\phi)$ give a sources position on the sky; the
angle $\psi$ (in combination with $\iota$) describes the orientation
of the binary's orbital plane with respect to a detector.  See
\cite{300yrs}, Eqs.\ (103) -- (104) for further discussion.

Imagine that Eq.\ (\ref{eq:h_NQ2}) accurately described GWs in nature.
By matching phase with the data, measurement of the GW would determine
the chirp mass ${\cal M}$.  Calculations using Eq.\
(\ref{eq:covariance_matrix}) to estimate measurement error
{\citep{fc93}} show that ${\cal M}$ should be determined with
exquisite accuracy, $\Delta{\cal M}/{\cal M} \propto 1/N_{\rm cyc}$,
where $N_{\rm cyc}$ is the number of GW cycles measured in our band.

The amplitude of the signal is determined with an accuracy
$\Delta{\cal A}/{\cal A} \sim 1/\mbox{SNR}$.  This means that, for a
given GW antenna, a combination of the angles $\theta$, $\phi$,
$\iota$, $\psi$, and the source distance $D$ are measured with this
precision; however, those parameters are {\it not} invidually
determined.  A single interferometer cannot break the distance-angle
correlations apparent in Eqs.\ (\ref{eq:h_NQ2}) and (\ref{eq:h_meas}).
Multiple detectors (which will each have their own response functions
$F_+$ and $F_\times$) are needed to measure these source
characteristics.  This is one reason that multiple detectors around
the globe are being built. (For LISA, the constellation's motion
around the sun makes $F_+$ and $F_\times$ effectively time varying.
The modulation imposed by this time variation means that the single
LISA antenna can break these degeneracies, provided that a source is
sufficiently long-lived for the antenna to complete a large fraction
of an orbit.)

As we have discussed, Eq.\ (\ref{eq:h_NQ2}) does not give a good
description of GWs from strong-field binaries.  Effects which this
``Newtonian gravity plus quadrupole waves'' treatment misses come into
play.  Consider the pN phase function, Eq.\ (\ref{eq:2pnPhase}).  Not
only does the chirp mass ${\cal M}$ influence the phase; so too does
the binary's reduced mass $\mu$ and its ``spin-orbit'' and
``spin-spin'' parameters $\beta$ and $\sigma$.  The pN phasing thus
encodes more detail about the binary's characteristics.
Unfortunately, these parameters may be highly correlated.  For
example, \cite{cf94} show that when precession is neglected, errors in
a binary's reduced mass and spin-orbit parameter are typically $90\%$
or more correlated with each other.  This is because the time
dependence of their contributions to the phase is not very different.

\begin{figure}[t]
\includegraphics[width=5.3in]{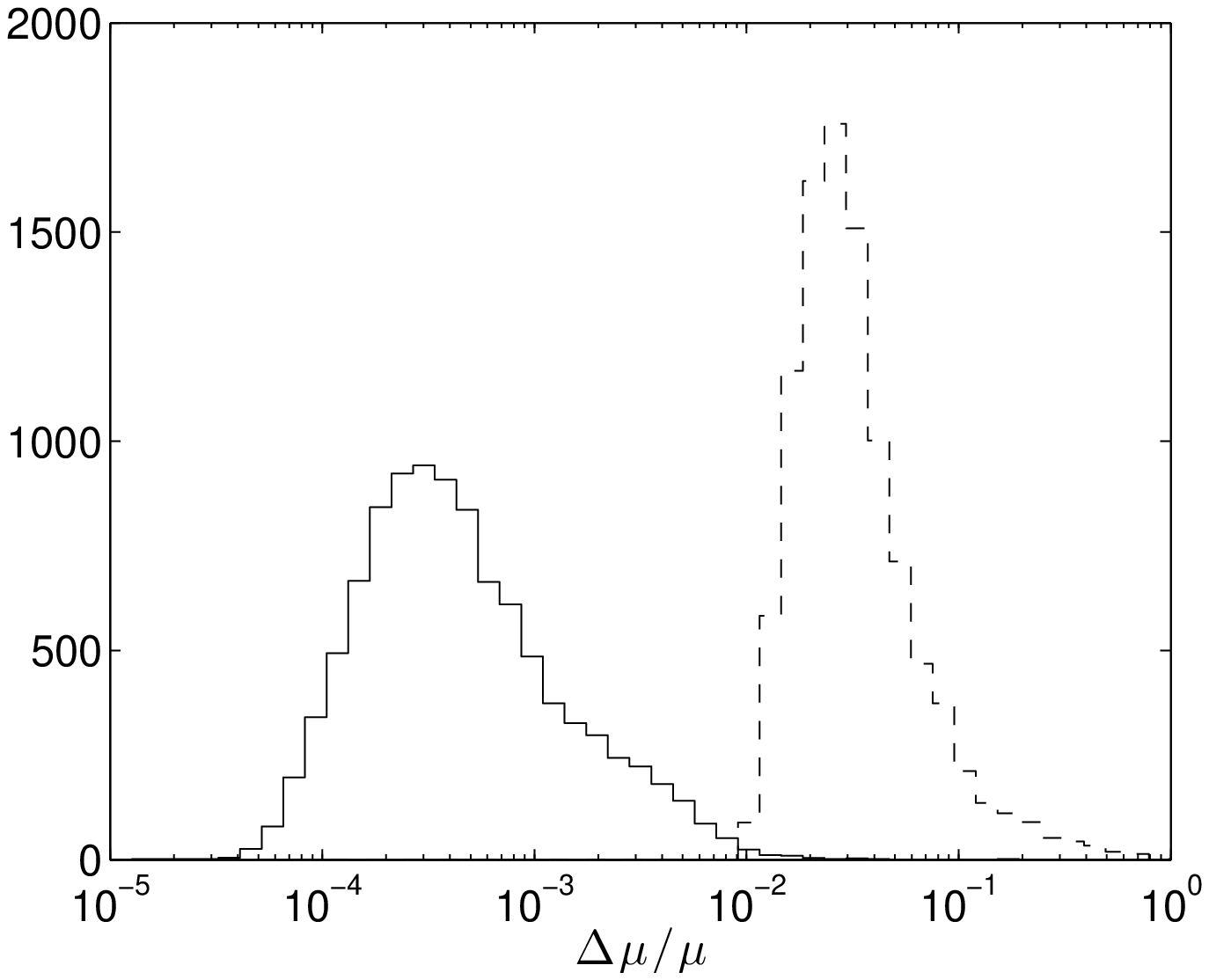}
\caption{Accuracy with which reduced mass $\mu$ is measured by LISA
for binaries at $z = 1$ with masses $m_1 = 3 \times 10^5\,M_\odot$,
$m_2 = 10^6\,M_\odot$.  The two curves come from a Monte-Carlo
simulation in which the sky is populated with $10^4$ binaries whose
positions, orientations, and spins have been randomly chosen.
Horizontal axis is the logarithmic error $\Delta\mu/\mu$; vertical
axis is the number of binaries that fall in an error bin.  The dashed
line neglects spin precession effects; note that the distribution
peaks at an error $\Delta\mu/\mu \simeq 0.03$.  The solid line
includes spin precession; note that the peak error is smaller by
roughly two orders of mangnitude.}
\label{fig:mucomp}
\end{figure}

These correlations can be broken when we make our models more
complete.  As we've discussed, the spin precession equations
(\ref{eq:dS1dt}) and (\ref{eq:dS2dt}) cause $\beta$ and $\sigma$ to
oscillate.  This modulates the waves' phase; as first demonstrated by
\cite{v04} and then examined in greater depth by \cite{lh06}, the
modulations break parameter degeneracies and improve our ability to
characterize the system whose waves we measure.  Figure
(\ref{fig:mucomp}), taken from \cite{lh06}, shows this effect for LISA
measurements of coalescing binary black holes.  Accounting for
precession improves the accuracy with which the reduced mass is
measured by roughly two orders of magnitude.  We similarly find that
the members' spins can be determined with excellent accuracy.  GW
measurements will be able to map the mass and spin distributions of
coalescing binaries.

Equation (\ref{eq:h_NQ2}) is also deficient in that only the leading
quadrupole harmonic is included.  As the discussion in Sec.\
{\ref{sec:pn_amplitude}} demonstrates, that is just one harmonic among
many that contribute to a binary's GWs.  Recent work
(\citealt{aissv07}, \citealt{ts08}, \citealt{pc08}) has looked at how
our ability to characterize a source improves when those ``higher
harmonics'' are included.  Typically, one finds that these harmonics
improve our ability to determine a binary's orientation $\iota$.  This
is largely because each harmonic has a slightly different functional
dependence on $\iota$, so each encodes that information somewhat
differently than the others.  The unique functional dependence of each
harmonic on $\iota$ in turn helps break degeneracies between that
angle and the source distance $D$.

\subsubsection{``Bothrodesy'': Mapping black hole spacetimes}
\label{sec:bothros}

Extreme mass ratio captures may allow a unique GW measurement: We may
use them to ``map'' the spacetimes of large black holes and test how
well they satisfy the (rather stringent) requirements of GR.  As
discussed in Sec.\ {\ref{sec:pert}}, an extreme mass ratio inspiral is
essentially a sequence of orbits.  Thanks to the mass ratio, the small
body moves through this sequence slowly, spending a lot of time
``close to'' any orbit in the sequence.  Also thanks to the mass
ratio, each orbit's properties are mostly determined by the larger
body.  In analogy to {\it geodesy}, the mapping of earth's gravity
with satellite orbits, one can imagine {\it bothrodesy}\footnote{This
name was coined by Sterl Phinney, and comes from the word $\beta
o\theta\!\rho o\varsigma$, which refers to a sacrificial pit in
ancient Greek.  This author offers an apology to speakers of modern
Greek.}, the mapping of a black hole's gravity by studying the orbits
of inspiraling ``satellites.''

In more detail, consider first Newtonian gravity.  The exterior
potential of a body of radius $R$ can be expanded in a set of
multipole moments:
\begin{equation}
\Phi_N = -\frac{GM}{r} + G\sum_{l = 2}^\infty
\left(\frac{R}{r}\right)^{l + 1} M_{lm} Y_{lm}(\theta,\phi)\;.
\label{eq:earth_pot}
\end{equation}
Studying orbits allows us to map the potential $\Phi_N$, and thus to
infer the moments $M_{lm}$.  By enforcing Poisson's equation in the
interior, $\nabla^2\Phi_N = 4\pi G\rho$, and then matching at the
surface $R$, one can relate the moments $M_{lm}$ to the distribution
of matter.  In this way, orbits allow us to map in detail the
distribution of matter in a body like the earth.

Bothrodesy applies the same basic idea to a black hole.  The spacetime
of any stationary, axisymmetric body can be described by a set of
``mass moments'' $M_l$, similar to the $M_{lm}$ of Eq.\
(\ref{eq:earth_pot}); and a set of ``current moments'' $S_l$ which
describe the distribution of mass-energy's {\it flow}.  What makes
this test powerful is that the moments of a black hole take a simple
form: for a Kerr black hole (\ref{eq:kerr_metric}) with mass $M$ and
spin parameter $a$,
\begin{equation}
M_l + i S_l = M(ia)^l\;.
\label{eq:kerr_moments}
\end{equation}
A black hole has a mass moment $M_0 = M$ and a current moment $S_1 =
aM$ (i.e., the magnitude of its spin is $aM$, modulo factors of $G$
and $c$). {\it Once those moments are known, all other moments are
fixed if the Kerr solution describes the spacetime.}  This is a
restatement of the ``no hair'' theorem {\citep{carter71, robinson75}}
that a black hole's properties are set by its mass and spin.

The fact that an object's spacetime (and hence orbits in that
spacetime) is determined by its multipoles, and that the Kerr moments
take such a simple form, suggests a simple consistency test: Develop
an algorithm for mapping a large object's moments by studying orbits
of that object, and check that the $l \ge 2$ moments satisfy Eq.\
(\ref{eq:kerr_moments}).  \cite{ryan95} first demonstrated that such a
measurement can in principle be done, and \cite{brink08} has recently
clarified what must be done for such measurements to be done in
practice.  \cite{ch04} took the first steps in formulating this
question as a null experiment (with the Schwarzschild solution as the
null hypothesis).  \cite{gb06} formulated a similar approach
appropriate to Kerr black holes, and Vigeland (Vigeland \& Hughes, in
preparation) has recently extended the Collins \& Hughes formalism in
that direction.

A robust test of the Kerr solution is thus a very likely outcome of
measuring waves from extreme mass ratio captures.  If, however,
testing metrics is not your cup of tea, precision black hole metrology
may be: In the process of mapping a spacetime, one measures with
exquisite accuracy both the mass and the spin of the large black hole.
\cite{bc04} have found that in most cases these events will allow us
to determine both the mass and the spin of the large black hole with
$0.1\%$ errors are better.

\subsubsection{Binary inspiral as a standard ``siren.''}
\label{sec:siren}

A particularly exciting astronomical application of binary inspiral
comes from the fact that the GWs depend on, and thus directly encode,
distance to a source.  Binary inspiral thus acts as a standard candle
(or ``standard siren,'' so named because it is often useful to regard
GWs as soundlike), with GR providing the standardization.
{\cite{schutz86}} first demonstrated the power of GW observations of
merging binaries to pin down the Hubble constant; {\cite{markovic93}}
and {\cite{fc93}} analyzed Schutz's argument in more detail, in
addition assessing how well other cosmological parameters could be
determined.  More recently, {\cite{hh05}} have examined what can be
done if a GW merger is accompanied by an ``electromagnetic''
counterpart of some kind.  We now describe how inspiral waves can
serve as a standard siren.

Imagine that we measure a nearby source, so that cosmological redshift
can be neglected.  The measured waveform generically has a form
\begin{equation}
h = \frac{G M(m_i)}{c^2 r}{\cal A}(t)\cos\left[\Phi(t; m_i, {\bf
S}_i)\right]\;,
\label{eq:waveform_generic}
\end{equation}
where $m_i$ are the binary's masses, ${\bf S}_i$ are its spins, and
$M(m_i)$ is a function of the masses with dimension mass.  For
example, for the Newtonian quadrupole waveform (\ref{eq:h_NQ}), this
function is the chirp mass, $M(m_i) = {\cal M} = (m_1m_2)^{3/5}/(m_1 +
m_2)^{1/5}$.  The function ${\cal A}(t)$ is a slowly varying,
dimensionless function which depends most strongly on parameters such
as the source inclination $\iota$.

Now place this source at a cosmological distance.  Careful analysis
shows that the naive Euclidean distance measure $r$ should be the {\it
proper motion distance} $D_M$ (\citealt{carroll}, Chap.\ 8); see,
e.g., {\cite{fc93}} for a derivation.  Also, all timescales which
characterize the source will be redshifted: If $\tau$ is a timescale
characterizing the source's internal dynamics, $\tau \to (1 + z)\tau$.

What is the phase $\Phi$ for this cosmological binary?  Because it
evolves solely due to gravity, any parameter describing the binary's
dynamics enters as a timescale.  For example, a mass parameter becomes
a time: $m \to \tau_m \equiv Gm/c^3$.  This time suffers cosmological
redshift; the mass that we infer by measuring it is likewise
redshifted: $m_{\rm meas} = (1 + z)m_{\rm local}$.  Spin variables
pick up a squared reshift factor: $S_{\rm meas} = (1 + z)^2 S_{\rm
local}$.  This tells us is that redshift ends up {\it degenerate} with
other parameters: A binary with masses $m_i$ and spins ${\bf S}_i$ at
redshift $z$ has a phase evolution that looks just like a binary with
$(1 + z)m_i$, $(1 + z)^2{\bf S}_i$ in the local universe.  So, if we
put our source at redshift $z$, Eq.\ (\ref{eq:waveform_generic})
becomes
\begin{equation}
h = \frac{GM(m_i)}{c^2 D_M}{\cal A}(t)\cos\left[\Phi(t; (1 + z)m_i, (1 +
z)^2{\bf S}_i)\right]\;.
\label{eq:eq:waveform_generic_z1}
\end{equation}
Recall that proper motion distance is related to luminosity distance
by $D_M = D_L/(1 + z)$.  Because we don't measure masses but rather
$(1 + z)$ times masses, it makes sense to adjust the amplitude and put
\begin{equation}
h = \frac{G(1 + z)M(m_i)}{c^2 D_L}{\cal A}(t)\cos\left[\Phi(t; (1 +
z)m_i, (1 + z)^2{\bf S}_i)\right]\;.
\label{eq:eq:waveform_generic_z}
\end{equation}
The key point here is that measurements {\it directly encode the
luminosity distance to a source}, $D_L$; however, they do {\it not}
tell us anything about a source's redshift $z$.  In this sense GW
measurements of merging binaries can be distance probes that are
highly {\it complementary} to most other astronomical distance
measures.  Indeed, analyses indicate that the distance should be
measured to $\sim 10 - 20\%$ accuracy using ground-based instruments
(e.g., \citealt{cf94}), and to $\sim 1 - 5\%$ from space
(\citealt{lh06}, \citealt{aissv07}, \citealt{ts08}, \citealt{pc08}).

Suppose that we measure GWs from a merging compact binary, allowing us
to measure $D_L$ with this accuracy.  {\it If} it is possible to
measure the source's redshift [either from the statistical properties
of the distribution of events, as emphasized by \cite{schutz86} and
\cite{cf93}, or by direct association with an ``electromagnetic''
event \citep{hh05}], {\it then} one may be able to accurately
determine both distance and redshift for that event --- a potentially
powerful constraint on the universe's cosmography with completely
different systematic properties than other standard candles.  An
example of an event which may constitute such a standard siren is a
short-hard gamma-ray burst.  Evidence has accumulated recently
consistent with the hypothesis that at least some short-hard bursts
are associated with NS-NS or NS-BH mergers (e.g., \citealt{fox05},
\citealt{nakar06}, \citealt{perley08}).  Near simultaneous measurement
of a GW signal with a short-hard burst is a perfect example of what
can be done as these detectors reach maturity and inaugurate GW
astronomy.

\section*{Acknowledgments}

I thank Daniel Kennefick for helpful discussion about the history of
this field, Thomas Baumgarte and Stuart Shapiro for teaching me most
of what I know about the foundations of numerical relativity, Vicky
Kalogera and Fred Rasio for helping me untangle some of the literature
on rate estimates for compact binary mergers, Alessandra Buonanno for
providing figures and background for the material on the effective
one-body approach, Plamen Fiziev for pointing out that Chandrasekhar's
massive tome develops Kerr perturbations in the language of metric
variables, and Daniel Holz and Samaya Nissanke for providing
particularly thorough comments on an early draft of this paper.  Some
of this material was presented at the 2008 Summer School in Cosmology
at the Abdus Salam International Center for Theoretical Physics, in
Trieste, Italy; I thank the organizers of that school for the
invitation and for the opportunity to develop and organize this
material.  The work I have discussed here owes a particular debt to my
collaborators Neil Cornish, Steve Drasco, Marc Favata, \'Eanna
Flanagan, Joel Franklin, Daniel Holz, Gaurav Khanna, and Samaya
Nissanke; as well as to current and former graduate students Nathan
Collins, Ryan Lang, Stephen O'Sullivan, Pranesh Sundararajan, and
Sarah Vigeland.  Finally, I thank Deepto Chakrabarty for five years of
teasing, which inspired me to insert the factors of $G$ and $c$
included in the equations here.  My research in gravitational waves
and compact binaries is supported by NSF Grant PHY-0449884 and NASA
Grant NNX08AL42G.  Some of the work discussed here was also supported
by NASA Grant NNG05G105G and the MIT Class of 1956 Career Development
Fund.

\bibliography{araa}

\end{document}